\documentclass[11pt,a4paper]{article}
\pdfoutput=1
\usepackage{jcappub}
\usepackage{bm}
\usepackage{graphicx}
\usepackage{feynmp}
\usepackage[labelfont={bf,scriptsize},textfont={scriptsize},justification=justified,format=hang]{caption,subfig}
\usepackage{color}

\definecolor{forestgreen}{rgb}{0.133,0.545,0.133}
\renewcommand{\boxed}[1]{{\color{forestgreen}#1}}

\usepackage[T1]{fontenc}

\RequirePackage{ifpdf}
\ifpdf
\else 
\fi

\renewcommand{\d}{\mathrm{d}}
\newcommand{\D}{\mathrm{D}}
\newcommand{\im}{\mathrm{i}}
\renewcommand{\geq}{\geqslant}
\renewcommand{\leq}{\leqslant}
\newcommand{\grad}{\nabla}

\newcommand{\Mp}{M_{\mathrm{P}}}
\newcommand{\EulerGamma}{\gamma_{\mathrm{E}}}
\newcommand{\R}{\mathcal{R}}

\newcommand{\LO}{\text{\textsc{lo}}}
\newcommand{\NLO}{\text{\textsc{nlo}}}
\newcommand{\DRGE}{\textsc{drge}}
\newcommand{\RGE}{\textsc{rge}}

\newcommand{\F}{\text{\textsc{f}}}

\newcommand{\hard}{{\text{hard}\ast}}

\newcommand{\Umatrix}{\mathcal{U}}

\newcommand{\Gmetric}{\bm{\mathrm{G}}}
\newcommand{\umetric}{\bm{\mathrm{w}}}
\newcommand{\Pimetric}{\bm{\Pi}}
\newcommand{\RiemannMetric}{\bm{\mathrm{R}}}

\newcommand{\kfact}{k_{\F}}

\newcommand{\vect}[1]{\bm{\mathrm{{#1}}}}
\newcommand{\e}[1]{\mathrm{e}^{{#1}}}

\DeclareMathOperator{\TimeOrder}{\mathsf{T}}
\DeclareMathOperator{\Or}{O}
\DeclareMathOperator{\NewRe}{Re}

\renewcommand{\Re}{\NewRe}

\renewcommand{\baselinestretch}{1.05}

\begin{document}

\title{The $\delta N$ formula is the dynamical renormalization group}

\author{Mafalda Dias,$^1$}
\author{Raquel H. Ribeiro,$^{2,3}$}
\author{and David Seery$^1$}

\affiliation{$^1$ Astronomy Centre, University of Sussex, Falmer, Brighton,
BN1 9QH, UK}

\affiliation{$^2$ Department of Physics, Case Western Reserve University, 
10900 Euclid Ave, Cleveland,\\ OH~44106, USA}
\affiliation{$^3$ Department of Applied Mathematics and Theoretical Physics,
University of Cambridge, Wilberforce Road, Cambridge CB3 0WA, UK}
\emailAdd{M.Dias@sussex.ac.uk}
\emailAdd{RaquelHRibeiro@case.edu}
\emailAdd{D.Seery@sussex.ac.uk}

\abstract{We derive the `separate universe' method
for the inflationary bispectrum,
beginning directly from
a field-theory calculation.
We work to tree-level in quantum effects but to all orders in the
slow-roll expansion,
with masses accommodated perturbatively.
Our method provides a systematic basis to account for
novel sources of time-dependence
in inflationary correlation functions,
and has immediate applications.
First, we use our result to obtain the correct matching prescription
between the `quantum' and `classical' parts of the
separate universe computation.
Second, we elaborate on the 
application of this method in situations where its validity is not clear.
As a by-product of our calculation we give the leading slow-roll
corrections to the three-point function of field fluctuations on
spatially flat hypersurfaces in a canonical, multiple-field model.}


\maketitle

\begin{fmffile}{drg-diags}

\newpage
\section{Introduction}
\label{sec:introduction}

Since Maldacena's computation of the three-point function
produced by an epoch of single-field
inflation~\cite{Maldacena:2002vr},
considerable effort has been invested in
obtaining
correlation functions for more complex scenarios.
In part this investment has been motivated by the increasing sophistication
of observational 
cosmology.
But an equally important influence has been the
drive to understand
the implications of our inflationary
theories, especially in
examples with many
interacting degrees of
freedom.

The tool used to
extract predictions from inflationary scenarios is
quantum field theory,
in which each observable is constructed from correlation functions.
Quantum field theory incorporates the effect of fluctuations on all
scales, making each correlation function potentially sensitive
to both short- and long-distance effects.
Over very short distances this implies that
they must
be defined carefully by a renormalization prescription, as in
flat spacetime.
But the most interesting outcome of our investment
in studying correlation functions has been the
realization that they have a rich infrared structure
which is very different to the case of flat spacetime.
Some aspects of this structure were reviewed in
Refs.~\cite{Seery:2010kh,Bartolo:2007ti,Pimentel:2012tw}.

Nontrivial structure
in correlation functions
is often associated with the existence of kinematic hierarchies.
A 
paradigmatic
example from particle physics
is a hadronic jet carrying some large energy
$E$
but small invariant mass $m_{\text{jet}}$.
Another is electroweak processes at large momentum transfer
$s \gg M_Z^2$. (Here $s$ is the Mandelstam variable
and $M_Z$ is the mass of the $Z$ boson.)
The kinematic region where ratios such as
$E/m_{\text{jet}}$ or $s/M_Z^2$ become large
is
called the \emph{Sudakov region}.
In this region logarithms of large ratios
invalidate perturbation theory,
and
to obtain even a qualitative description requires resummation.

Analogues of these effects are responsible for the rich
infrared structure
of inflationary correlation functions.
Viewed as a sum of Feynman diagrams,
a generic $n$-point function
is characterized by the comoving momenta carried by each propagator.
At tree-level
these are linear combinations $\vect{q}_i$
of the external momenta $\{ \vect{k}_1, \ldots, \vect{k}_n \}$.
Another scale
is the comoving Hubble length $\tau \sim (aH)^{-1}$
where $a$ is the scale factor
and $H = \dot{a}/a$ is the Hubble rate.
Sudakov-like logarithms involving ratios of these scales
appear in each correlation function.%
	\footnote{Despite the appealing
	analogy, the logarithms which appear in inflationary
    correlation functions are \emph{not} Sudakov logarithms in
    the following 
    sense: in scattering
    calculations the Kinoshita--Lee--Nauenberg theorem
    implies that sufficiently inclusive
    observables are infrared safe, due to cancellation between virtual
    diagrams and real soft or collinear
    emission~\cite{Kinoshita:1962ur,Lee:1964is}.
    Sudakov effects arise in
    comparatively
    exclusive observables because
    exclusion of
    phase space regions prevents complete cancellation.
    The Sudakov logarithms measure this
    large remainder.
    In inflation the background fields themselves have time
    dependence 
    and there is no meaningful sense
    in which these logarithms arise as a cancellation between
    real and virtual effects.
    For this reason we shall refer to them
    as \emph{Sudakov-like logarithms}.}

When there are no hierarchies
among the $|\vect{q}_i|$ and $\tau$,
each of these scales
will be comparable to
some reference scale $k_\ast$.
If the $n$-point functions are
expressed in terms of background quantities evaluated at
the horizon-crossing time for $k_\ast$,
the Sudakov-like logarithms are small
and the correlation functions are
simple.
However,
if large hierarchies exist 
it will not
be possible to find a $k_\ast$ for which all
logarithms are small and the correlation functions develop a
complex structure.
Indeed,
we will see that
it is necessary to
carry out a resummation,
just as for hadronic jets and high-energy electroweak
observables.
Understanding the details of this enhanced structure is
particularly important because
correlation functions with large hierarchies
are valuable observables:
for example,
$n$-point functions with
$q_i/q_j \gg 1$
can be used to
determine
nonlinear and stochastic halo bias~\cite{Baumann:2012bc}.
But despite their significance, a systematic treatment of these
effects has not appeared in the literature.

In this paper we begin a systematic treatment of
correlation functions
containing
Sudakov-like logarithms.
We focus on logarithms containing the comoving
Hubble scale $\tau$,
which may all be rewritten in the form $\ln |k_\ast \tau|$
(perhaps at the cost of introducing other logarithms of the form
$\ln q_i/k_\ast$).
In this form they coincide with
the `time-dependent' logarithms $\ln |k_\ast \tau|$
described in Ref.~\cite{Seery:2010kh};
see also Ref.~\cite{Pimentel:2012tw}.
We will return to the remaining
`scale' and `shape' logarithms
(involving $\ln q_i/k_\ast$ and
$\ln q_i / q_j$, respectively~\cite{Burrage:2011hd})
in a future publication.
We work to tree-level in quantum effects but
to \emph{all orders} in the slow-roll expansion.
In particular,
this means that masses are accommodated perturbatively.
We will comment on the implications of this choice as we develop
our argument in \S\S\ref{sec:factorization}--\ref{sec:applications}
below.

\paragraph{Separate-universe formulae.}
It has been understood for some time that logarithms of
the form $\ln |k_\ast \tau|$
describe time evolution of each $n$-point
function~\cite{Falk:1992sf,Zaldarriaga:2003my,Seery:2007wf,Burgess:2009bs}.
But, although the necessity to account for
time evolution is well-known,
the precise role of the $\ln |k_\ast \tau|$ terms has
received comparatively little attention.
Most discussion of time dependence are framed in terms of the
`separate universe' approach to perturbation
theory~\cite{Lyth:1984gv,Starobinsky:1986fxa,Sasaki:1995aw,Lyth:2005fi}.
On very large scales this gives a description of a perturbed
universe by patching together unperturbed solutions
with different initial conditions.
One implementation follows
Sasaki \& Tanaka~\cite{Sasaki:1998ug}, who
argued that the superhorizon limit of
perturbation theory
could be obtained from Jacobi fields of the
background phase space.
(A similar interpretation was recently given in~Ref.\cite{Seery:2012sx}.)
Related discussions of perturbation theory on large
scales were given in Refs.~\cite{Wands:2000dp,Rigopoulos:2003ak,Lyth:2004gb}.
On the basis of these arguments one can
determine the evolution of correlation functions
by tracking the evolution of these Jacobi fields and averaging
over a suitable ensemble of stochastic initial conditions.

This approach gives a reliable description of effects occurring
on superhorizon scales, but it cannot describe subhorizon physics
or phenomena which occur near the epoch of horizon exit.
For example, this leads to some ambiguity about the
precise ensemble of initial conditions which should be used
when computing correlation functions,
to which we will return in {\S\ref{sec:matching}}.
In this paper we argue that
the separate-universe method can be reproduced
by resummation of Sudakov-like logarithms.
Among other benefits,
this provides precise information about the role of
horizon-crossing effects which cannot be captured
from the superhorizon limit
of perturbation theory.

After constructing suitable Jacobi fields
and setting up an ensemble of
initial conditions $\delta \phi_i(\vect{k}_1)$,
the separate-universe
approach we have just described
asserts that
the two- and three-point functions for a set of light scalar fields
during inflation satisfy~\cite{Lyth:2005fi}
\begin{subequations}
\begin{align}
	\langle
		\delta \phi_\alpha(\vect{k}_1)
		\delta \phi_\beta(\vect{k}_2)
	\rangle
	& \supseteq
		\frac{\partial \phi_\alpha}{\partial \phi_i^\ast}
		\frac{\partial \phi_\beta}{\partial \phi_j^\ast}
		\langle
			\delta \phi_i(\vect{k}_1)
			\delta \phi_j(\vect{k}_2)
		\rangle_\ast
	\label{eq:deltaN-2pf}
	\\ \nonumber
	\langle
		\delta \phi_\alpha(\vect{k}_1)
		\delta \phi_\beta(\vect{k}_2)
		\delta \phi_\gamma(\vect{k}_3)
	\rangle
	& \supseteq
		\frac{\partial \phi_\alpha}{\partial \phi_i^\ast}
		\frac{\partial \phi_\beta}{\partial \phi_j^\ast}
		\frac{\partial \phi_\gamma}{\partial \phi_k^\ast}
		\langle
			\delta \phi_i(\vect{k}_1)
			\delta \phi_j(\vect{k}_2)
			\delta \phi_k(\vect{k}_3)
		\rangle_\ast
	\\ \nonumber &
		\quad \mbox{} +
		\frac{\partial^2 \phi_\alpha}{\partial \phi_i^\ast \partial \phi_j^\ast}
		\frac{\partial \phi_\beta}{\partial \phi_k^\ast}
		\frac{\partial \phi_\gamma}{\partial \phi_m^\ast}
		\int \frac{\d^3 q}{(2\pi)^3} \;
		\langle
			\delta \phi_i(\vect{k}_1 - \vect{q})
			\delta \phi_k(\vect{k}_2)
		\rangle_\ast
		\langle
			\delta \phi_j(\vect{q})
			\delta \phi_m(\vect{k}_3)
		\rangle_\ast
	\\ &
		\quad \mbox{} + \text{cyclic permutations} .
	\label{eq:deltaN-3pf}
\end{align}
\end{subequations}
We
have adopted the notation of Refs.~\cite{Seery:2012sx,Elliston:2012he},
which will be used throughout this paper. On the left-hand side, each
correlation function is evaluated at some late time $t$
and transforms as a tensor in the
tangent space
associated with the field-space coordinate
$\phi_\alpha(t)$.
Indices in this tangent space are labelled
$\{ \alpha, \beta, \ldots \}$.
On the right-hand side each correlation function is evaluated
at the horizon-crossing time $t_\ast$ for the reference scale $k_\ast$,
and transforms as a tensor in the
tangent space associated with the field-space coordinate
$\phi_i(t_\ast)$.
Indices in this tangent
space are labelled
$\{ i, j, \ldots \}$.
Where the field-space manifold is curved it is essential to preserve
this
distinction~\cite{Elliston:2012he}.
In this paper we will work with a flat field-space metric
except in~\S\ref{sec:nontrivial-metric}
but this index convention remains convenient.
The bitensors $\partial \phi_\alpha / \partial \phi^\ast_i$,
$\partial^2 \phi_\alpha / \partial \phi^\ast_i \partial \phi^\ast_j$
measure variation of the inflationary trajectory
under a change of initial conditions at the
horizon-crossing time for the reference scale,
and
the symbol `$\supseteq$' indicates that subleading corrections
from `loop' diagrams have been ignored
together with disconnected contributions.

Eqs.~\eqref{eq:deltaN-2pf}--\eqref{eq:deltaN-3pf} make a number
of strong
assertions. First, Eq.~\eqref{eq:deltaN-2pf} asserts that, at all times
and accounting for
large contributions from all orders in the slow-roll
expansion,
the two-point function can be written as a linear
combination of its values at some arbitrary
earlier time with coefficients
$\partial \phi_\alpha / \partial \phi_i^\ast$.
This implies that
(at least when decaying modes have become negligible)
all time-dependent Sudakov-like logarithms can be
factorized into these coefficients.
Here and below, `to all orders in the slow-roll expansion'
is a statement about a perturbative expansion in terms of slow-roll
parameters evaluated at the horizon-crossing time for $k_\ast$.

Second, Eq.~\eqref{eq:deltaN-3pf} asserts that any time-dependent
Sudakov-like logarithms appearing in the three-point function
exhibit a similar
factorizable structure.
In principle
the three-point function could
have arbitrary dependence on
the external momenta $k_i$~\cite{Maldacena:2002vr,Seery:2005gb}.
Therefore, at each order, the Sudakov-like logarithms could
enter with different functions of the $k_i$.
But Eq.~\eqref{eq:deltaN-3pf} asserts that,
to all orders in the slow-roll expansion,
almost all Sudakov-like logarithms
factorize into
the coefficients $\partial \phi_\alpha / \partial \phi_i^\ast$
which appear in the two-point function~\eqref{eq:deltaN-2pf}.
Exceptions are allowed
only for
Sudakov-like logarithms which multiply functions of the $k_i$
obtainable from a product of two-point functions.
For massless fields, or where masses are taken into account
perturbatively, these are the `local' momentum combinations
$(k_2^3 k_3^3)^{-1}$ and its permutations.
Even for these exceptions, Eq.~\eqref{eq:deltaN-3pf} requires that the
resummation of Sudakov-like logarithms produces a result which
is related in a nontrivial way to the factorizable coefficients
already present in
the two-point function.

Each of these assertions is a complex statement,
applicable to all orders in the slow-roll expansion,
about the
structure of infrared divergences
which can be produced by the underlying quantum field theory.
Although these results are implied
by the perturbation-theory results described
above, it is nontrivial to see how they
are reproduced by the Sudakov-like logarithms which
arise from the underlying quantum field theory.
In this paper we explain how this occurs
by deducing the time-dependent structure of each correlation
function directly from the Sudakov-like divergences it
contains.
We will see that this provides a practical and convenient
means to compute the evolution of each $n$-point function
in cases where the separate universe picture is complicated
to apply, or an untrustworthy guide for our intuition.

\paragraph{Summary.}
The Sudakov-like time-dependent logarithms
in inflationary correlation functions
are qualitatively similar to those occurring in particle physics,
and therefore can be understood using the same tools.
In~\S\ref{sec:resummation} we sketch the
main steps,
focusing on a physical interpretation of the 
`factorizable' structure.
In~\S\ref{sec:factorization} we begin to develop this argument in
detail.
The strategy breaks into two parts.
\begin{itemize}
	\item First, in~\S\ref{sec:factorization-thm}
	we prove a \emph{factorization theorem}
	which provides constraints
	on the terms in each correlation function which can be
	logarithmically enhanced.
	This factorization theorem plays the same role as a proof
	of renormalizability in applying renormalization-group arguments
	to the ultraviolet behaviour of correlation functions:
	it guarantees that all large logarithms can be assembled into
	a finite number of `renormalized' functions.
	\item
	Second, in~\S\ref{sec:threepf},
	we derive
	a \emph{renormalization group equation} (`{\RGE}')
	which can be used to determine these unknown `renormalized'
	functions.
	Once the renormalization group equations have been solved
	the full correlation function can be reconstructed.
	It will turn out that
	the {\RGE}s are equivalent to the transport equations
	used to determine $\partial \phi_\alpha / \partial \phi_i^\ast$
	and their higher derivatives~\cite{Yokoyama:2007uu,Yokoyama:2007dw,
	Yokoyama:2008by,Mulryne:2009kh,Mulryne:2010rp,Seery:2012sx,Elliston:2012he}.
\end{itemize}
In~\S\ref{sec:applications}
we illustrate our method using examples.
When justified from
the superhorizon limit of perturbation theory
the separate-universe approach requires
a set of initial conditions,
usually assumed to be generated
from quantum fluctuations
which become classical after passing outside the horizon.
Implicit in this point of view is a `matching'
between quantum and classical calculations.
Several authors have noted that the precise choice of matching time
is ambigious, which could lead to
uncertainties in an accurate
calculation~\cite{Polarski:1994rz,Leach:2000yw,Nalson:2011gc}.
In \S\ref{sec:matching}
we revisit this question using the {\RGE} approach,
in which
there is no need to invoke
a matching from quantum to classical evolution because the entire calculation
takes place within the framework of quantum field theory.%
	\footnote{Note that this is purely a statement about the calculation
	of correlation functions. The need to understand how fluctuations
	decohere (presumably solving the Schr\"{o}dinger's cat problem)
	remains.}
We show how this gives a unique prescription which resolves
the matching ambiguity.

In~\S\ref{sec:nontrivial-metric} we illustrate the practical
utility of the {\RGE} method
by using it to determine evolution equations for
covariant
versions of the coefficients
$\partial \phi_\alpha / \partial \phi_i^\ast$
and their higher-order generalizations
in the presence of a nontrivial
field-space metric.
These coefficients are not straightforward to
determine using classical intuition,
making the renormalization-group approach simple and reliable.
This example could be adapted to more complex cases, perhaps including
the effect of loop corrections,
where the usual classical separate-universe arguments do not
function or are prohibitively difficult to apply.

In \S\ref{sec:discussion} we conclude with
a short discussion of our main results.
We provide details of the next-order calculation of the three-point
correlation function (which is required to deduce an appropriate
renormalization group equation for the coefficient functions)
in Appendix~\ref{appendix:3pf}.

\paragraph{Notation and conventions.}
Throughout this paper, we adopt units in which $c = \hbar = 1$.
Our index conventions for scalar fields are described
below Eqs.~\eqref{eq:deltaN-2pf}--\eqref{eq:deltaN-3pf}.
We work with the action
\begin{equation}
	S = \frac{1}{2} \int \d^4 x \; \sqrt{-g} \, \Big\{
		\Mp^2 R - \partial_a \phi_\alpha \partial^a \phi_\alpha - 2V 
	\Big\} , 
	\label{eq:action}
\end{equation}
where $V = V(\phi)$ is an arbitrary potential
and Latin indices $a$, $b$, {\ldots}, are Lorentz indices contracted with
the spacetime metric $g_{ab}$.
Except where explicitly indicated,
scalar field indices are contracted using the flat field-space metric
$\delta_{\alpha\beta}$.
We will generally set the reduced
Planck mass $\Mp \equiv (8\pi G)^{-1/2}$ to unity.

\section{Why is resummation necessary?}
\label{sec:resummation}

It is not at all clear from inspection of
Eqs.~\eqref{eq:deltaN-2pf}--\eqref{eq:deltaN-3pf}
that they involve
a resummation of
time-dependent Sudakov-like logarithms.
In this section and the next we explain why resummation is necessary
and develop a heuristic approach to these equations.
In~\S\ref{sec:drge}
we will consider the same problem from a
more formal viewpoint, that of the dynamical renormalization group.

To quadratic order, the action for fluctuations of the scalar fields
$\phi_\alpha$ in~\eqref{eq:action},
measured on uniform-expansion hypersurfaces,
can be written
\begin{equation}
	S = \frac{1}{2} \int \d^3 x \, \d t \; a^3 \left(
		\delta \dot{\phi}_\alpha
		\delta \dot{\phi}_\alpha
		-
		\frac{1}{a^2}
		\partial_i \delta \phi_\alpha
		\partial_i \delta \phi_\alpha
		-
		m_{\alpha\beta}
		\delta \phi_\alpha
		\delta \phi_\beta
	\right) ,
	\label{eq:quadratic-action}
\end{equation}
where
an overdot denotes differentiation with respect to 
cosmic time.
The mass-mixing matrix $m_{\alpha\beta}$
can be written explicitly in
terms of slow-roll parameters~\cite{Mukhanov:1985rz,Sasaki:1986hm,Sasaki:1995aw}
\begin{equation}
	m_{\alpha\beta} = V_{\alpha\beta} -
		\frac{1}{a^3} \frac{\d}{\d t} \left(
			\frac{a^3}{H} \dot{\phi}_\alpha \dot{\phi}_\beta
		\right)
		=
		V_{\alpha\beta}
		- 3 \dot{\phi}_\alpha \dot{\phi}_\beta
		- \varepsilon \dot{\phi}_\alpha \dot{\phi}_\beta
		- \frac{2}{H} \dot{\phi}_{(\alpha} \ddot{\phi}_{\beta)} ,
	\label{eq:mass-matrix}
\end{equation}
where $\varepsilon = - \dot{H}/H^2$ is the usual slow-roll
parameter.
Inflation occurs whenever $\varepsilon < 1$.
At lowest order in slow-roll
the mass-matrix can be
related to the `expansion tensor' $u_{\alpha\beta}$
introduced in Ref.~\cite{Seery:2012sx},
\begin{equation}
	u_{\alpha\beta} = - \frac{m_{\alpha\beta}}{3H^2} .
	\label{eq:expansion-tensor}
\end{equation}

\paragraph{Two-point function.}
It is convenient to work in conformal time, defined
by $\tau = \int^t_{\infty} \d t' / a(t')$.
For the two-point function only a single 
hierarchy
can be generated, measured by the ratio of $k$
to the comoving Hubble scale $aH \approx -\tau^{-1}$.
Therefore $|k\tau| \approx 1$ at horizon crossing.
Working to next-order in the slow-roll expansion
and evaluating the two-point function at least a little after
horizon-crossing we find
\begin{equation}
	\begin{split}
	\langle
		\delta & \phi_{\alpha}(\vect{k}_1)
		\delta \phi_{\beta}(\vect{k}_2)
	\rangle_\tau
	=
	(2\pi)^3
	\delta(\vect{k}_1 + \vect{k}_2)
	\frac{H_\ast^2}{2k^3}
	\\
	& \times \left\{
		\delta_{\alpha\beta} \left[
			1
			+ 2 \varepsilon_\ast \left(
				1 - \EulerGamma - \ln \frac{2k}{k_\ast}
			\right)
		\right]
		+
		2 u_{\alpha\beta}^\ast \left[
			2
			- \EulerGamma
			- \ln (-k_\ast \tau)
			- \ln \frac{2k}{k_\ast}
		\right]
		+ \cdots
	\right\} ,
	\end{split}
	\label{eq:twopf-fixed}
\end{equation}
where `$\cdots$' indicates higher-order corrections which we have
not computed.
We have set $k = k_1 = k_2$ to be the common magnitude of the
external momenta
and dropped contributions which decay like positive powers of $\tau$.
The scale $k_\ast$ determines 
a horizon-crossing time
around which we have chosen to Taylor-expand background time-dependent
quantities such as $H$.
Evaluation at the horizon-crossing time for $k_\ast$ is denoted
by a subscript `$\ast$'.
At this stage
its precise assigment is arbitrary and can be chosen to suit our own
convenience.
The terms involving $\ln k/k_\ast$
and $\ln (-k_\ast\tau)$
are our first examples of
Sudakov-like logarithms.

This result has appeared in various forms in the literature.
It was given in the single-field case by
Stewart \& Lyth~\cite{Stewart:1993bc}, who worked directly in terms
of the conserved comoving-gauge curvature perturbation $\R$.
The corresponding result for
field fluctuations in the uniform-expansion gauge
was given by Nakamura \& Stewart~\cite{Nakamura:1996da},
who accounted for multiple fields
and a nontrivial field-space metric.
Higher-order corrections were given by Gong \& Stewart,
who developed an algorithmic approach to compute them
using Green's functions of the Mukhanov--Sasaki
equation~\cite{Gong:2001he,Gong:2002cx}.%
	\footnote{Eq.~\eqref{eq:twopf-fixed} of this paper agrees with Eq.~(43)
	of Gong \& Stewart \cite{Gong:2002cx},
	although the time-dependent terms appear different due to the
	way Gong \& Stewart selected their time of evaluation.

	Results restricted to a two-field model were quoted by
	Byrnes \& Wands~\cite{Byrnes:2006fr}
	and Lalak~et~al.~\cite{Lalak:2007vi},
	who worked with an explicit adiabatic--isocurvature basis.
	Their results neglect the
	time-dependent term $\ln(-k_\ast \tau)$ and therefore disagree with
	our Eq.~\eqref{eq:twopf-fixed}.
	Formulae compatible with~\eqref{eq:twopf-fixed}
	were given by
	Avgoustidis~et~al.~\cite{Avgoustidis:2011em}.}

\paragraph{Callan--Symanzik equation.}
The reference scale $k_\ast$
is not physical, and therefore the correlation functions cannot depend
on it.
Invariance under a change in $k_\ast$ is
expressed by the
Callan--Symanzik equation.
Taking all background quantities to be functions of the fields
$\phi_\alpha$, this equation can be written
\begin{equation}
	\left(
		\frac{\partial}{\partial \ln k_\ast}
		+
		\frac{\partial \phi_\lambda^\ast}{\partial \ln k_\ast}
		\frac{\partial}{\partial \phi_\lambda^\ast}
	\right)
	\langle
		\delta \phi_{\alpha_1}(\vect{k}_1)
		\cdots
		\delta \phi_{\alpha_n}(\vect{k}_n)
	\rangle
	= 0 .
	\label{eq:CS}
\end{equation}
The differential operator in brackets $( \cdots )$ is simply
the total derivative $\d / \d \ln k_\ast$.
Specializing~\eqref{eq:CS} to the two-point function yields,
to lowest-order in a slow-roll expansion,
\begin{equation}
	\frac{\d \ln H_\ast}{\d \ln k_\ast}
	+ \varepsilon_\ast
	= 0 .
	\label{eq:CS-on-shell}
\end{equation}
As above, a subscript `$\ast$' indicates evaluation at the horizon-crossing
point for $k_\ast$. Since~\eqref{eq:CS-on-shell} holds for any
$k_\ast$ it expresses the condition $\varepsilon = - \dot{H}/H^2$
which is satisfied for any change of the background
fields $\phi_\alpha$ which
themselves
satisfy the background equations of motion.
This is sometimes expressed by saying that the change should be `on-shell'.
Eq.~\eqref{eq:CS-on-shell} guarantees that we are free to change
$k_\ast$ provided that we compensate
by adjusting all background quantities according to their classical
equations of motion. The appearance of classical equations is a consequence
of our restriction to tree-level processes.%
	\footnote{One might have expected to recover the equation of motion
	$3 H \dot{\phi}_\alpha = - V_{,\alpha}$ for each field.
	This does not happen because~\eqref{eq:twopf-fixed} does not depend
	on the fields individually,
	but only through their aggregate contribution
	to $H$.
	Eq.~\eqref{eq:CS-on-shell} can be regarded as the
	equation of motion for $H$.
	In a model with a single field $\phi$
	it is equivalent to the classical
	equation of motion $\d\phi/\d N =-\sqrt{2\varepsilon}$.}

\paragraph{Time evolution.}
We now return to Eq.~\eqref{eq:twopf-fixed} and consider its behaviour
for different values of $k$ and $\tau$.

An expression such as~\eqref{eq:twopf-fixed} is said to be computed using
\emph{fixed order} perturbation theory, because the calculation is
carried to a predetermined order in the slow-roll expansion.
When $k$ is comparable to $aH$,
Eq.~\eqref{eq:CS-on-shell} shows that
we can choose $k_\ast$ to be approximately
their common magnitude provided we evaluate background quantities
near the horizon-crossing time for $k$.
Then $\ln(-k_\ast \tau)$ and $\ln k/k_\ast$
will both be negligible,
and a fixed-order expression such as Eq.~\eqref{eq:twopf-fixed}
is a good approximation.
It is dominated by its lowest-order term
$H_\ast^2/2k^3$.

More than a few e-folds
after horizon-crossing the hierarchy $k/aH = |k\tau|$
becomes exponentially
small.
No matter how we choose $k_\ast$ it is no longer possible
to make
both $\ln k/k_\ast$ and $\ln (-k_\ast \tau)$ negligible.
Therefore
we must accept the 
appearance of large, growing contributions
which cannot be absorbed into background quantities
by a choice of evaluation time.
Growing terms of this kind are potentially hazardous.
They are sometimes described as divergences, because on its own
$\ln(-k_\ast\tau)$ becomes unboundedly large as $\tau \rightarrow 0$.
This is simply an artefact of perturbation theory,
in the same way that the Taylor expansion of any function
\begin{displaymath}
	f(x) = f(x_0) + f'(x_0) (x - x_0) + \cdots
\end{displaymath}
appears to diverge when $|x - x_0| \gg 1$.
In reality, when $\varepsilon_\ast \ln (-k_\ast \tau) \sim 1$
we must sum
an infinite number of terms
from a fixed-order
expression such as~\eqref{eq:deltaN-2pf}
before we can determine even its qualitative behaviour.
Therefore, unless
some principle or symmetry enforces a precise cancellation,
fixed-order perturbation theory can never provide
an adequate description.
This obligation to deal with
an infinite sequence of terms
of the form $[\varepsilon_\ast \ln (-k_\ast \tau)]^n$
was emphasized by Weinberg~\cite{Weinberg:2006ac},
who
went on to speculate that the methods of the renormalization
group could be used to perform the resummation,
in analogy with the case of Sudakov effects in QCD.
In this paper we show that this is indeed the case.

We conclude that any
expression which is valid to
arbitrarily late times, such as~\eqref{eq:deltaN-2pf},
must resum an infinite number of terms.
For example,
consider the well-studied model of double quadratic
inflation~\cite{Silk:1986vc,Polarski:1994rz,GarciaBellido:1995qq,
Langlois:1999dw}.
The action is
\begin{equation}
	S = -\dfrac{1}{2} \int \d^4 x \; \sqrt{-g} \, \Big\{
		R
		+ \partial_a \varphi \partial^a \varphi
		+ \partial_a \chi \partial^a \chi
		+ m_{\varphi}^2 \varphi^2 + m_{\chi}^2 \chi^2
	\Big\} ,
	\label{eq:double-quadratic}
\end{equation}	
where $\varphi$ and $\chi$ are two light scalar fields
with $m_\varphi$, $m_\chi < H$.
We focus on the dimensionless two-point function
$\Sigma_{\alpha\beta}$, defined by
\begin{equation}
	\langle
		\delta \phi_{\alpha}(\vect{k}_1)
		\delta \phi_{\beta}(\vect{k}_2)
	\rangle_\tau
	=
	(2\pi)^3
	\delta(\vect{k}_1 + \vect{k}_2)
	\frac{\Sigma_{\alpha\beta}}{2k^3} ,
	\label{eq:factorized-2pf}
\end{equation}
but similar remarks
apply to higher $n$-point functions.

\begin{figure}[h]
	\begin{center}
		\includegraphics[width=9cm]{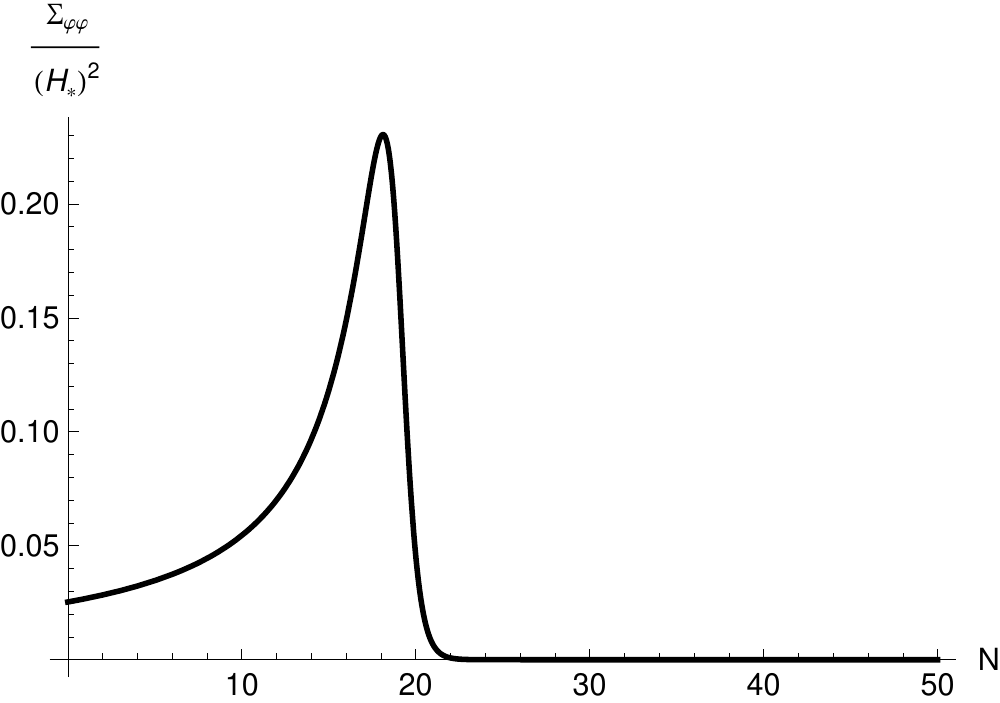}
		\caption{Super-horizon
		evolution of the two-point function $\Sigma_{\varphi\varphi}$
		in the double-quadratic model~\eqref{eq:double-quadratic},
		normalized by the Hubble scale at horizon crossing.
		This evolution is for the scale which left the horizon
		(at $N=0$)
		fifty e-folds before the end of inflation
		(at $N=50$).
		The initial conditions were $\varphi_{\text{init}}=8.2 \Mp$
		and $\chi_{\text{init}}=12.9 \Mp$ with $m_{\varphi}/m_{\chi}=9$.}
		\label{fig:dqpowerspectrum}
	\end{center}
\end{figure}

In Fig.~\ref{fig:dqpowerspectrum}
we present
a typical super-horizon evolution of 
$\Sigma_{\varphi\varphi}$
calculated using the formula~\eqref{eq:deltaN-2pf}.
There is a clearly defined peak in the vicinity of $N
= -\ln(-k_\ast \tau) \simeq 20$,
caused by a turn of the trajectory in field space~\cite{Vernizzi:2006ve},
after
which $\Sigma_{\varphi\varphi}$ asymptotes to a constant.
Inflation is ongoing through the entire evolution, with $\varepsilon$
growing to a maximum value $\sim 0.26$ near the peak.
Fig.~\ref{fig:dqpowerspectrum} shows clearly that
$\Sigma_{\alpha\beta}$ can exhibit very rapid
evolution.
The linear approximation
involving
$\varepsilon_\ast \ln (-k_\ast \tau)$
is already poor for $N \gtrsim 10$ and is
totally incorrect (even qualitatively) for $N \gtrsim 20$.
If the initial time had been chosen closer to the peak,
the linear approximation could easily have been invalidated
within $\lesssim 1$ e-fold.

The flat asymptote for $N \gtrsim 20$ is associated with
convergence to an `adiabatic limit' in which
the uniform-density gauge curvature perturbation
$\zeta$ is
conserved~\cite{Elliston:2011dr}.
To describe this asymptotic plateau certainly requires
resummation of an infinite number of powers of $N = - \ln(-k_\ast \tau)$.
In any finite combination the highest power of $N$
must eventually dominate, leading to power-law growth at large $N$.
It is only in an infinite sum that sufficiently many terms
remain available at high-order to balance the growing
contribution at low orders,
enabling $\Sigma_{\varphi\varphi}$ to be
constant at late times.

\section{Towards a resummation prescription}
\label{sec:factorization}

We have identified growth of
the logarithmic `divergence' in~\eqref{eq:twopf-fixed}
with the infrared region where $k/aH \ll 1$.
It represents the cumulative effect of
interactions which operate over very many Hubble times.
We would like to separate the physics of these
interactions from the creation of fluctuations in each
$k$-mode, which takes place
over only a few Hubble times.
The disparity of these timescales implies that,
when they are created,
the fluctuations do not depend on details
of the inflationary model in which they are embedded.
It is the systematic separation
of short-distance, model-independent `hard' physics
and long-distance, model-dependent `soft' effects
which
produces the
factorization in Eqs.~\eqref{eq:deltaN-2pf}--\eqref{eq:deltaN-3pf}.

In this section we illustrate how to perform and interpret
factorization of the two-point 
function, and indicate how to extend the method
to higher $n$-point functions.

\subsection{Factorization of the two-point function}
\label{sec:subfactorization}

We work with the dimensionless two-point function
$\Sigma_{\alpha\beta}$ defined in Eq.~\eqref{eq:factorized-2pf}.
The formula~\eqref{eq:twopf-fixed} includes terms at lowest-order
in the slow-roll expansion and 
$\Or(\varepsilon)$ corrections.
We describe these, respectively, as `lowest-order' ($\LO$) and
`next-lowest order' (or next-order, $\NLO$) terms.
Working to $\NLO$, we identify
\begin{equation}
	\Sigma_{\alpha\beta}^{\NLO}
	\supseteq
		H_\ast^2
		\left(
			\delta_{\alpha\beta}
			+
			2 r_{\alpha\beta}^\ast
			- 2 u^\ast_{\alpha\beta} \ln (-k_\ast \tau)
			- 2 M^\ast_{\alpha\beta} \ln \frac{2k}{k_\ast}
		\right) .
	\label{eq:twopf-basic}
\end{equation}
The constant matrices
$r_{\alpha\beta}$
and
$M_{\alpha\beta}$
are
\begin{subequations}
\begin{align}
	r^\ast_{\alpha\beta} & = \varepsilon_\ast \delta_{\alpha \beta}
		( 1 - \EulerGamma )
		+
		u^\ast_{\alpha\beta}
		( 2 - \EulerGamma )
	\label{eq:r-def}
	\\
	M^\ast_{\alpha\beta} & = \varepsilon_\ast \delta_{\alpha\beta}
		+ u^\ast_{\alpha\beta} .
	\label{eq:M-def}
\end{align}
\end{subequations}

\paragraph{Heuristic treatment.}
In what follows, we assume that $\Sigma_{\alpha\beta}$ contains
at most
powers of $\ln k$, but not powers of $k$ itself.
This is a consequence of approximate scale invariance.
Our aim is to separate
the long- and short-distance contributions to~\eqref{eq:twopf-basic}.
It was explained above that when $k/aH \ll 1$
we cannot
make a choice of $k_\ast$ which
eliminates all large logarithms.
In effect, we do not have enough free scales
to absorb every large contribution:
it is only
possible to use $k_\ast$ to eliminate contributions from \emph{one}
of $\ln k/k_\ast$ and $\ln(-k_\ast\tau)$.
We choose
$k_\ast = k$ to eliminate $\ln k/k_\ast$ and
\
introduce a new, arbitrary `factorization scale' $\kfact$.
(However, we continue to denote quantities evaluated at the horizon-crossing
time
for $k_\ast$ with a subscript `$\ast$'.)
Ultimately, we will see that the factorization scale
describes the boundary between
long- and short-distance physics.
We write
\begin{equation}
\begin{split}
	\Sigma^{\NLO}_{\alpha\beta}
	& \supseteq H_\ast^2
		\Big(
			\delta_{\alpha\beta}
			+ 2 \tilde{r}^\ast_{\alpha\beta}
			- 2 u^\ast_{\alpha\beta} \Big[
				\ln \frac{k}{\kfact} + \ln (-\kfact \tau)
			\Big]
		\Big) \\
	& = H_\ast^2
		\Big( \delta_{\alpha i} - u^\ast_{\alpha i} \ln (-\kfact \tau) \Big)
		\Big( \delta_{\beta j} - u^\ast_{\beta j} \ln (-\kfact \tau) \Big)
		\Big(
			\delta_{ij}
			+ 2 \tilde{r}^\ast_{ij}
			- 2 u^\ast_{ij} \ln \frac{k}{\kfact}
		\Big)
		,
\end{split}
\label{eq:twopf-nlo-factorized}
\end{equation}
where $\tilde{r}_{\alpha\beta} = r_{\alpha\beta} - M_{\alpha\beta} \ln 2$.
In the first line $\kfact$ cancels out.
The second line is valid to next-order, which is the accuracy to which
we are working.
Based on the structure of~\eqref{eq:twopf-nlo-factorized},
we introduce a `form-factor' $\Gamma_{\alpha i}$,
which is a function only of $\tau$ and the factorization scale
$\kfact$,
\begin{equation}
	\Sigma_{\alpha\beta}
	= H_{\ast}^2 \Gamma_{\alpha i} \Gamma_{\beta j}
	\Big(
		\delta_{ij}
		+ 2 \tilde{r}^\ast_{ij}
		- 2 u^\ast_{ij} \ln \frac{k}{\kfact}
	\Big) .
	\label{eq:twopf-factorization-ansatz}
\end{equation}
We will see that this form-factor modifies the
lowest-order
functional dependence of $\Sigma_{\alpha\beta}$
on background quantities,
giving the two-point function an enhanced structure which is not visible
in fixed-order perturbation theory.
Modifications of this type
are an inevitable consequence of large contributions which
cannot be absorbed into a change of evaluation point for the background
quantities already present 
in a fixed-order
expression such as~\eqref{eq:twopf-fixed} or~\eqref{eq:twopf-basic}.

Since $\Sigma_{\alpha\beta}$ must be independent of
$\kfact$ it is possible to write a Callan--Symanzik equation
similar to~\eqref{eq:CS}. For~\eqref{eq:twopf-factorization-ansatz}
to be independent of $\kfact$ requires
\begin{equation}
	\frac{\d \Gamma_{\alpha i}}{\d \ln \kfact}
	= - \Gamma_{\alpha j} u_{ji} .
	\label{eq:twopf-backwards-rge}
\end{equation}
In Eqs.~\eqref{eq:twopf-nlo-factorized}--\eqref{eq:twopf-backwards-rge}
we have made use of the index convention described in~\S\ref{sec:introduction}.
The indices $\alpha$, $\beta$, {\ldots} label the tangent space
at the late time $\tau$, whereas indices $i$, $j$, {\ldots} label
the tangent space at the horizon-crossing time for the
factorization scale $\kfact$.
Note that in~\eqref{eq:twopf-backwards-rge} we have rewritten the time of
evaluation for $u_{ij}$ as this horizon-crossing time.
This is formally acceptable because the error we
incur is $\Or(\varepsilon^2)$ and therefore below the precision to which
we are working.
In~\S\ref{sec:drge} below
we will see how to
give a more mathematically satisfactory formulation of this argument.
In this section
the discussion is not intended to be rigorous,
but to highlight the physical meaning
of factorization.

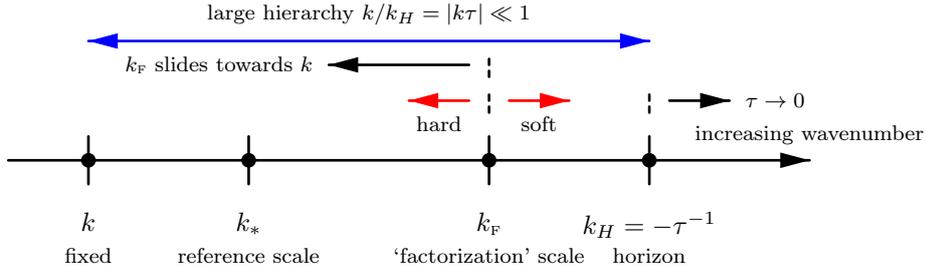
\begin{figure*}

	\small
	
	\hfill
	\begin{fmfgraph*}(300,90)
		\fmfcmd{%
 				style_def end_arrow expr p =
 				cdraw p;
 				cfill (harrow (p, 1));
 				enddef;}
 		 	\fmfcmd{%
 				style_def double_arrow expr p =
 				cdraw p;
 				cfill (harrow (reverse p, 1));
 				cfill (harrow (p, 1));
 				enddef;}
		\fmfpen{thin}
		\fmfipair{a,b}				
		\fmfipair{eta,muf,k,kstar}	
		\fmfiequ{a}{(0,0.5h)}
		\fmfiequ{b}{(w,0.5h)}
		\fmfi{end_arrow}{a .. b}
		\fmfiequ{k}{point 0.1*length(a .. b) of (a .. b)}
		\fmfiequ{kstar}{point 0.3*length(a .. b) of (a .. b)}
		\fmfiequ{muf}{point 0.6*length(a .. b) of (a .. b)}
		\fmfiequ{eta}{point 0.8*length(a .. b) of (a .. b)}
		\fmfiv{decoration.shape=circle,decoration.size=0.05h,decoration.filled=full}{eta}
		\fmfiv{decoration.shape=circle,decoration.size=0.05h,decoration.filled=full}{muf}
		\fmfiv{decoration.shape=circle,decoration.size=0.05h,decoration.filled=full}{kstar}
		\fmfiv{decoration.shape=circle,decoration.size=0.05h,decoration.filled=full}{k}
		\fmfi{plain}{eta shifted (0,0.1h) .. eta shifted (0,-0.1h)}
		\fmfi{plain}{muf shifted (0,0.1h) .. muf shifted (0,-0.1h)}
		\fmfi{plain}{kstar shifted (0,0.1h) .. kstar shifted (0,-0.1h)}
		\fmfi{plain}{k shifted (0,0.1h) .. k shifted (0,-0.1h)}
		\fmfiv{label={$k_H = -\tau^{-1}$},label.angle=-90}{eta shifted (0,-0.15h)}
		\fmfiv{label={$\kfact$},label.angle=-90}{muf shifted (0,-0.15h)}
		\fmfiv{label={$k_\ast$},label.angle=-90}{kstar shifted (0,-0.15h)}
		\fmfiv{label={$k$},label.angle=-90}{k shifted (0,-0.15h)}
		\fmfi{double_arrow,foreground=blue,label={\scriptsize large hierarchy $k/k_H = |k\tau| \ll 1$},label.side=right}{eta shifted (0,0.5h) .. k shifted (0,0.5h)}
		\fmfi{end_arrow,foreground=red,label={\scriptsize hard},label.side=left}{muf shifted (-0.025w,0.25h) .. muf shifted (-0.1w,0.25h)}
		\fmfi{end_arrow,foreground=red,label={\scriptsize soft},label.side=right}{muf shifted (0.025w,0.25h) .. muf shifted (0.1w,0.25h)}
		\fmfi{dashes}{muf shifted (0,0.2h) .. muf shifted (0,0.3h)}
		\fmfi{dashes}{muf shifted (0,0.35h) .. muf shifted (0,0.45h)}
		\fmfi{end_arrow}{muf shifted (-0.025w,0.4h) .. muf shifted (-0.2w,0.4h)}
		\fmfi{dashes}{eta shifted (0,0.2h) .. eta shifted (0,0.3h)}
		\fmfi{end_arrow}{eta shifted (0.025w,0.25h) .. eta shifted (0.1w,0.25h)}
		\fmfiv{label={\scriptsize increasing wavenumber},label.angle=90}{b}
		\fmfiv{label={\scriptsize `factorization' scale},label.angle=-90}{muf shifted (0,-0.3h)}
		\fmfiv{label={\scriptsize reference scale},label.angle=-90}{kstar shifted (0,-0.3h)}
		\fmfiv{label={\scriptsize horizon},label.angle=-90}{eta shifted (0,-0.3h)}
		\fmfiv{label={\scriptsize fixed},label.angle=-90}{k shifted (0,-0.3h)}
		\fmfiv{label={\scriptsize $\kfact$ slides towards $k$},label.angle=180}{muf shifted (-0.2w,0.4h)}
		\fmfiv{label={\scriptsize $\tau \rightarrow 0$},label.angle=0}{eta shifted (0.1w,0.25h)}
	\end{fmfgraph*}
	\hfill
	\mbox{}
	
	\caption{\label{fig:scales}%
	Scales appearing in factorization of the two-point function.
	At late times $\tau \rightarrow 0$, generating a large hierarchy
	between the horizon size $k_H = -\tau^{-1}$ and the fixed comoving
	scale $k$. The floating scale $\kfact$ acts as a boundary
	between the `hard' (short-distance,
	$k/aH \gtrsim 1$) and `soft' (long-distance, $k / aH \lesssim 1$)
	parts of the calculation.
	Beginning with $\kfact = k_H$,
	all soft effects are included in the fixed-order
	part of the calculation.
	The `renormalization group equation' \eqref{eq:twopf-backwards-rge}
	allows us to slide
	$\kfact$ towards $k$,
	successively absorbing long-distance effects in the
	Sudakov-like form-factor $\Gamma_{\alpha i}$.
	Eventually, only `hard' effects dominated by comoving momenta
	$\sim k$ are included in the fixed-order part of the calculation.}
	
\end{figure*}

Eq.~\eqref{eq:twopf-backwards-rge} is
the `backwards' evolution equation
introduced by
Yokoyama et~al.~\cite{Yokoyama:2007uu,Yokoyama:2007dw,Yokoyama:2008by}
for the separate-universe coefficient
$\partial \phi_\alpha / \partial \phi_i$.
This discussion shows that it can equally be regarded as
a kind of renormalization-group equation
for the `Sudakov-like' form-factor $\Gamma_{\alpha i}$.
Its solution requires an initial condition.
This can be deduced from~\eqref{eq:twopf-nlo-factorized},
which gives
$\Gamma_{\alpha i} = \delta_{\alpha i}$
when $|\kfact \tau| \sim 1$.
In combination
with~\eqref{eq:twopf-backwards-rge} the initial condition enables us
to give a physical interpretation of the factorization procedure
(see Fig.~\ref{fig:scales}).
Our starting point is a fixed-order expression such
as~\eqref{eq:twopf-basic}.

We begin with $|\kfact \tau| \sim 1$.
With this choice
the form-factors $\Gamma_{\alpha i} = \delta_{\alpha i}$ are trivial
and contain no information,
whereas the fixed-order piece involving $\ln k/\kfact$
accounts for
contributions from both long- and short-distance
effects.
These are effects arising from the regions
$k/aH \gtrsim 1$ and $k/aH \lesssim 1$, respectively.
We now use~\eqref{eq:twopf-backwards-rge} to systematically
move $\kfact$ closer to $k$.
This adjustment moves 
all effects
generated up to the point where $\kfact/aH \lesssim 1$
into
$\Gamma_{\alpha i}$.
Effects generated when $\kfact/aH \gtrsim 1$
remain within the fixed-order expression.
Since~\eqref{eq:twopf-backwards-rge} merely expresses that
$\Sigma_{\alpha\beta}$ is independent of $\kfact$ this does not
change its numerical value.
We terminate this process when $\kfact \sim k$,
at which point all soft effects
generated when $k / aH \lesssim 1$
are absorbed into the Sudakov-like form-factor.
Conversely,
the hard 
factor in~\eqref{eq:twopf-nlo-factorized}
receives contributions only from
times when $k/aH \sim 1$.
Because this hard factor is uncontaminated by any other scale,%
	\footnote{For this statement to be strictly
	correct we must choose $k_\ast$ to remove all other sources
	of large logarithms,
	in the same way that in scattering calculations we
	must choose the renormalization scale to do the same.}
fixed-order perturbation theory can be used to obtain a good approximation.

\paragraph{Infrared safety.}
In the special case that the Sudakov-like logarithms vanish
order-by-order there are no soft effects to absorb in
$\Gamma_{\alpha i}$,
and $\Gamma_{\alpha i} \approx \delta_{\alpha i}$
even for $\kfact \sim k$. In these circumstances, the hard factor
automatically
receives contributions only from $k/aH \sim 1$.
In Ref.~\cite{Elliston:2012he}, correlation functions
with this property were described
as `infrared safe', by analogy with the same situation
in scattering calculations.
Infrared safe quantities decouple from
the complex, cumulative interactions of soft modes
and depend only on whatever physics dominates the hard subprocess.
Observables built from such correlation functions
have the advantage of being theoretically `clean',
but the disadvantage that they probe
phenomena operating only over a narrow window of wavenumbers.

This discussion makes clear
that the form-factor $\Gamma_{\alpha i}$ is the differential coefficient
$\partial \phi_\alpha / \partial \phi_i$,
and therefore
that the factorization scale $\kfact$ corresponds to the initial
time `$\ast$' used to construct the separate universe
formulae~\eqref{eq:deltaN-2pf}--\eqref{eq:deltaN-3pf}.
Its arbitrariness corresponds to the arbitrary location of this
initial time-slice.
The hard, fixed-order contribution corresponds to the
two-point function $\langle \delta \phi^\alpha \delta \phi^\beta \rangle_\ast$.
As we explained above, this
receives contributions only
from times when $k \sim aH$ and describes the hard
subprocess by which fluctuations are created.
The creation process is rapid
in comparison with the subsequent evolution,
so separation of scales implies that
its form is universal.
All details of this discussion have precise parallels in
the factorization of Sudakov-like form-factors in
QCD~\cite{Collins:1989gx,Brock:1993sz,Collins:2011zzd,Dissertori:2003pj}.

\subsection{Factorization of $n$-point functions with  $n \geq 3$}
\label{sec:separate-universe}

With this background it is possible to consider higher $n$-point functions.
We begin with the three-point function.

\begin{figure*}
	
	\small

	\vspace{7mm}
    \hfill
    \subfloat[][Propagator\label{fig:propagator}]{
    	\begin{fmfgraph*}(70,60)
    		\fmfpen{thin}
    		\fmfleft{l}
    		\fmfright{r}
    		\fmf{dashes}{l,v}
    		\fmf{plain}{v,r}
    		\fmfv{decor.shape=hexacross,decor.size=0.1w}{v}
    		\fmfv{label={\scriptsize $\alpha$}}{l}
    		\fmfv{label={\scriptsize $\beta$}}{r}
    	\end{fmfgraph*}
    }
    \hspace{3cm}
    \subfloat[][Three-point function\label{fig:3pf}]{
 		\begin{fmfgraph*}(70,60)
    		\fmfpen{thin}
	    	\fmftop{t}
	    	\fmfbottom{b1,b2}
	    	\fmf{dashes}{t,ta}
	    	\fmf{dashes}{b1,ba}
	    	\fmf{dashes}{b2,bb}
	    	\fmf{plain}{ta,v}
	    	\fmf{plain}{ba,v}
	    	\fmf{plain}{bb,v}
	    	\fmfv{decor.shape=hexacross,decor.size=0.1w}{ta}
	    	\fmfv{decor.shape=hexacross,decor.size=0.1w}{ba}
	    	\fmfv{decor.shape=hexacross,decor.size=0.1w}{bb}
	    	\fmfv{label={\scriptsize $\alpha$ [time $\tau$]}}{b1}
	    	\fmfv{label={\scriptsize $\beta$ [time $\tau$]}}{b2}
	    	\fmfv{label={\scriptsize $\gamma$ [time $\tau$]}}{t}
	    	\fmfv{label={\scriptsize [time $\eta$]},label.angle=30}{v}
	    \end{fmfgraph*}
	}
    \hspace{3cm}
    \subfloat[][Vertex integral\label{fig:amp-3pf}]{
    	\begin{fmfgraph*}(70,60)
    		\fmfpen{thin}
	    	\fmftop{t}
	    	\fmfbottom{b1,b2}
	    	\fmf{phantom}{t,ta}
	    	\fmf{phantom}{b1,ba}
	    	\fmf{phantom}{b2,bb}
	    	\fmf{plain}{ta,v}
	    	\fmf{plain}{ba,v}
	    	\fmf{plain}{bb,v}
	    	\fmfv{decor.shape=hexacross,decor.size=0.1w}{ta}
	    	\fmfv{decor.shape=hexacross,decor.size=0.1w}{ba}
	    	\fmfv{decor.shape=hexacross,decor.size=0.1w}{bb}
	    	\fmfv{label={\scriptsize [time $\eta$]},label.angle=30}{v}
	    \end{fmfgraph*}
	}
	\hfill
    \mbox{}

    \caption{\label{fig:amputated}%
    Structure of the three-point function as
  	a `vertex integral' over
    individual wavefunction factors.
    Fig.~\ref{fig:propagator} depicts the propagator, which is a product
    of one conjugated wavefunction (dashed line)
    and one unconjugated wavefunction (solid line).
    The cross denotes contraction of these factors by matrix
    multiplication.
    In Fig.~\ref{fig:3pf}, three `internal' factors participate in
    an integral over the vertex whereas the three `external'
    factors do not.
    Note that the time at the vertex, $\eta$,
    is always earlier than the external time $\tau$.
    In Fig.~\ref{fig:amp-3pf} we show the vertex integral with these
    external factors removed.}

\end{figure*}
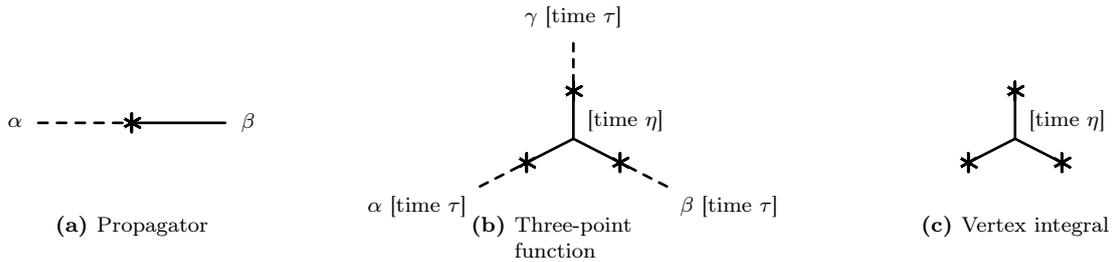

\paragraph{Removal of external wavefunctions.}
For reasons that will become clear below,
it is helpful to discuss
the $n$-point functions
separately from their external wavefunction factors.
In scattering calculations
we sometimes discuss `amputated' diagrams,
meaning that
the entirety of each external propagator is stripped away.
In a nonequilibrium or time-dependent setting
this cannot be done because
half of the propagator participates in a nontrivial
integral over the temporal position of the vertex to which it is attached.
In Appendix~\ref{appendix:propagator} we show that
the Feynman propagator takes the form%
	\footnote{Computation of expectation values requires the `in--in'
	or `Schwinger' formalism, in which all field degrees of freedom
	are doubled and therefore there are four possible
	time-ordered two point functions. Strictly, Eq.~\eqref{eq:propagator}
	represents the time-ordered $++$ two point function.
	For details, see Appendix~\ref{appendix:3pf} or
	Refs.~\cite{Chen:2010xka,Koyama:2010xj}.}
[cf. Eq.~\eqref{eq:propagator-wavefunctions}]
\begin{equation}
	\langle
		\TimeOrder
		\delta \phi_\alpha(\vect{k}_1,\tau)
		\delta \phi_\beta(\vect{k}_2,\tau')
	\rangle
	= (2\pi)^3 \delta(\vect{k}_1 + \vect{k}_2)
	\times
	\begin{cases}
		w^\ast_{\alpha \gamma}(k,\tau) w_{\gamma \beta}(k,\tau') &
			\tau < \tau' \\
		w_{\alpha \gamma}(k,\tau) w^\ast_{\gamma \beta}(k,\tau') &
			\tau' < \tau
	\end{cases}
	,
	\label{eq:propagator}
\end{equation}
where $k = |\vect{k}_1| = |\vect{k}_2|$ is the common magnitude of
the momenta on each external line,
the operator $\TimeOrder$ denotes time ordering and
the matrix-valued mode functions $w_{\alpha\beta}$
corresponding to each half of the propagator are contracted with each other.
In Fig.~\ref{fig:propagator} we depict the conjugated mode
$w^\ast_{\alpha\beta}$ as a dashed half-line,
and the unconjugated mode $w_{\alpha\beta}$ by a solid half-line.
The index contraction is denoted by a cross joining the dashed
and solid halves.
With these conventions, the three-point function can be depicted
as in Fig.~\ref{fig:3pf}.
The external, conjugated mode functions (dashed lines)
are evaluated at some time $\tau$ which is taken
to be later than the time $\eta$ associated with the internal vertex.
Finally in Fig.~\ref{fig:amp-3pf}
we show the 3-point function with these external, conjugated wavefunction
factors removed.
The rules of the `in--in' formulation
of quantum field theory (see Appendix~\ref{appendix:3pf})
show that
the full three-point function should be obtained by adding
Fig.~\ref{fig:3pf} and its complex conjugate.
Therefore the three-point function has the structure
\begin{equation}
	\langle
		\delta \phi_\alpha(\vect{k}_1)
		\delta \phi_\beta(\vect{k}_2)
		\delta \phi_\gamma(\vect{k}_3)
	\rangle_\tau
	= 2\Re \Big[
		w^\ast_{\alpha \lambda}(k_1,\tau)
		w^\ast_{\beta \mu}(k_2,\tau)
		w^\ast_{\gamma \nu}(k_3,\tau)
		I_{\lambda \mu \nu}(\tau,k_1,k_2,k_3)
	\Big] ,
	\label{eq:3pf-structure}
\end{equation}
where the vertex integral $I_{\lambda\mu\nu}$
corresponds to Fig.~\ref{fig:amp-3pf}
and is determined using the methods of the in--in formalism.
It represents the cumulative amplitude for three-body interactions up to
time $\tau$,
and 
is given by an integral over the internal wavefunctions
(which measure the probability of three
particles interacting at a point),
weighted by geometrical factors
(which measure the volume in which the interaction can take place)
and terms representing the detailed structure of
each three-body interaction.

\paragraph{External wavefunction divergences.}
The discussion in~\S\ref{sec:subfactorization} shows that although each
wavefunction $w_{\alpha\beta}(k,\tau)$ contains `divergent' logarithms
as $\tau \rightarrow 0$, these are absorbed into
the form-factor $\Gamma_{\alpha i}$ in the combination
$\Gamma_{\alpha j}(\tau,\kfact) w_{ji}(k,-\kfact^{-1})$.
(Here, the wavefunction $w_{ji}$ is evaluated at the horizon-crossing time
for the factorization scale $\kfact$.)
In concrete terms,
although the `divergent' logarithms in $w_{ij}(k,-\kfact^{-1})$
now appear as powers of $\ln \kfact$,
the combination
$\Gamma_{\alpha j}(\tau,\kfact) w_{ji} (k,-\kfact^{-1})$ is
independent of $\kfact$.
Therefore these `divergent' logarithms cancel.
Meanwhile, the overall $\tau$-dependence is determined by the solution to
the renormalization group equation~\eqref{eq:twopf-backwards-rge}
and does not suffer from the appearance of `divergent'
terms.
The same applies to 
$w^\ast_{\alpha\beta}$
provided we arrange for the phase of $w_{\alpha\beta}$ to be
constant at late times.

We now generalize this to the three-point function.
Consider the combination
\begin{equation}
	\Gamma_{\alpha i}(\tau,\kfact)
	\Gamma_{\beta j}(\tau,\kfact)
	\Gamma_{\gamma k}(\tau,\kfact)
	\langle
		\delta \phi^i(\vect{k}_1)
		\delta \phi^j(\vect{k}_2)
		\delta \phi^k(\vect{k}_3)
	\rangle_{-\kfact^{-1}} 
	\label{eq:3pf-absorb-external}
\end{equation}
which appears as one component of the separate-universe
formula~\eqref{eq:deltaN-3pf}.
Comparison with Fig.~\ref{fig:3pf} and Eq.~\eqref{eq:3pf-structure}
shows that, by construction, we expect
each form-factor
$\Gamma$
to absorb the `divergent' $\ln \kfact$ terms
from its corresponding external wavefunction.
But if the vertex integral $I_{\lambda\mu\nu}$ also contributes
$\ln \kfact$ terms
these
can
not be absorbed by the
$\Gamma$ factors.
Such large, `divergent' contributions would remain,
leaving residual $\kfact$-dependence in~\eqref{eq:3pf-absorb-external}.

Eq.~\eqref{eq:deltaN-3pf} introduces a new form-factor
$\partial^2 \phi_\alpha / \partial \phi_i^\ast \partial \phi_j^\ast$,
which provides a means by
which these $\ln \kfact$ terms can be absorbed.
However, as explained in~\S\ref{sec:introduction},
it imposes stringent conditions
(to all orders in the slow-roll expansion)
on the structure of any $\ln \kfact$ terms which appear in
the vertex integral $I_{\lambda\mu\nu}$---these must all be proportional
to the momentum combination $k_1^{-3} k_2^{-3}$ (or its permutations)
which can be generated by a product of two-point functions.
Our key task in applying the renormalization-group formalism to
the three-point function will be to prove a `factorization theorem'
which guarantees that $I_{\lambda\mu\nu}$ \emph{does} possess
this structure.

\paragraph{Higher $n$-point functions.}
Essentially the same
discussion can be given for the four- and higher $n$-point
functions, for which the separate universe method gives
a structure similar to~\eqref{eq:deltaN-3pf}.
Each $n$-point function contains a term like~\eqref{eq:3pf-absorb-external}
\cite{Seery:2006js,Byrnes:2006vq},
which absorbs $\ln \kfact$ contributions from the external wavefunctions.
Once this is done,
a finite number of form-factors are available to absorb
$\ln \kfact$ contributions coming from integration over the
vertices of the diagram, analogous to the vertex integral
$I_{\lambda\mu\nu}$.
The number of possible form-factors typically increases with $n$,
although they are not all independent.

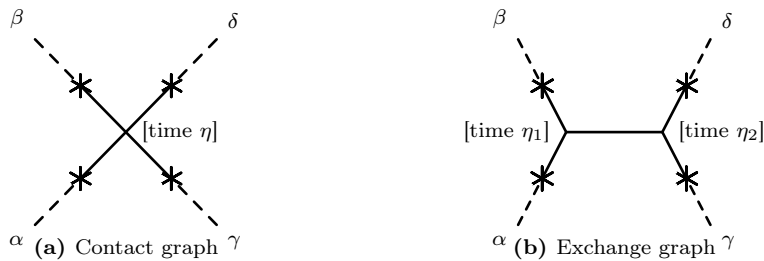
\begin{figure*}
	
	\small

	\vspace{7mm}
    \hfill
    \subfloat[][Contact graph\label{fig:4pf-contact}]{
    	\begin{fmfgraph*}(85,70)
    		\fmfleft{l1,l2}
    		\fmfright{r1,r2}
    		\fmf{dashes}{l1,l1a}
    		\fmf{dashes}{l2,l2a}
    		\fmf{dashes}{r1,r1a}
    		\fmf{dashes}{r2,r2a}
    		\fmf{plain}{l1a,v}
    		\fmf{plain}{l2a,v}
    		\fmf{plain}{r1a,v}
    		\fmf{plain}{r2a,v}
    		\fmfv{decor.shape=hexacross,decor.size=0.1w}{l1a}
    		\fmfv{decor.shape=hexacross,decor.size=0.1w}{l2a}
    		\fmfv{decor.shape=hexacross,decor.size=0.1w}{r1a}
    		\fmfv{decor.shape=hexacross,decor.size=0.1w}{r2a}
    		\fmfv{label={\scriptsize [time $\eta$]},label.angle=0}{v}
    		\fmfv{label={\scriptsize $\alpha$}}{l1}
    		\fmfv{label={\scriptsize $\beta$}}{l2}
    		\fmfv{label={\scriptsize $\gamma$}}{r1}
    		\fmfv{label={\scriptsize $\delta$}}{r2}
    	\end{fmfgraph*}
    }
    \hspace{3cm}
    \subfloat[][Exchange graph\label{fig:4pf-exchange}]{
    	\begin{fmfgraph*}(90,70)
    		\fmfleft{l1,l2}
    		\fmfright{r1,r2}
    		\fmf{dashes}{l1,l1a}
    		\fmf{dashes}{l2,l2a}
    		\fmf{dashes}{r1,r1a}
    		\fmf{dashes}{r2,r2a}
    		\fmf{plain}{l1a,v1}
    		\fmf{plain}{l2a,v1}
    		\fmf{plain}{r1a,v2}
    		\fmf{plain}{r2a,v2}
    		\fmf{plain,tension=0.5}{v1,v2}
    		\fmfv{decor.shape=hexacross,decor.size=0.1w}{l1a}
    		\fmfv{decor.shape=hexacross,decor.size=0.1w}{l2a}
    		\fmfv{decor.shape=hexacross,decor.size=0.1w}{r1a}
    		\fmfv{decor.shape=hexacross,decor.size=0.1w}{r2a}
    		\fmfv{label={\scriptsize [time $\eta_1$]},label.angle=180}{v1}
    		\fmfv{label={\scriptsize [time $\eta_2$]},label.angle=0}{v2}
    		\fmfv{label={\scriptsize $\alpha$}}{l1}
    		\fmfv{label={\scriptsize $\beta$}}{l2}
    		\fmfv{label={\scriptsize $\gamma$}}{r1}
    		\fmfv{label={\scriptsize $\delta$}}{r2}
    	\end{fmfgraph*}
    }
    \hfill
    \mbox{}
    
    \caption{\label{fig:4pf-form-factors}
    Form-factors available for the four-point function.}

\end{figure*}

For example, for the four-point function we can write
\begin{equation}
\begin{split}
	\langle
		\delta \phi_\alpha(\vect{k}_1)
	&
		\delta \phi_\beta(\vect{k}_2)
		\delta \phi_\gamma(\vect{k}_3)
		\delta \phi_\delta(\vect{k}_4)
	\rangle_\tau
	= 
	\\ & 2\Re \Big[
		w^\ast_{\alpha \lambda}(k_1,\tau)
		w^\ast_{\beta \mu}(k_2,\tau)
		w^\ast_{\gamma \nu}(k_3,\tau)
		w^\ast_{\delta \pi}(k_4, \tau)
		I_{\lambda \mu \nu \pi}(\tau,
			\vect{k}_1,\vect{k}_2,\vect{k}_3,\vect{k}_4)
	\Big] ,
	\label{eq:4pf-structure}
\end{split}
\end{equation}
where the `vertex' integral $I_{\lambda \mu \nu \pi}$ receives
contributions at tree-level from graphs with the two topologies
shown in Fig.~\ref{fig:4pf-form-factors}.
As above, factors of $\ln \kfact$ from the external wavefunctions
of both diagrams can be
absorbed
by contracting 
with $\Gamma_{\alpha i}$. 
Comparison with the formulae of Refs.~\cite{Seery:2006js,Byrnes:2006vq}
shows that
form-factors are available to absorb factors of $\ln \kfact$
from $I_{\lambda \mu \nu \pi}$ which are proportional to
$k_1^{-3} k_2^{-3} k_3^{-3}$
or
$k_1^{-3} k_2^{-3} |\vect{k}_1 + \vect{k}_3|^{-3}$
(or their permutations).
These two possibilities correspond to
the contact and exchange graphs, respectively.
However, the form-factor for $k_1^{-3} k_2^{-3} |\vect{k}_1 + \vect{k}_3|^{-3}$
is built out of the form-factor used to absorb divergences in the
three-point function. Only the form-factor for the contact interaction
of Fig.~\ref{fig:4pf-contact} is `new'.
A similar pattern repeats for all higher $n$, with only the irreducible
$n$-point contact graph generating `new' divergences: all others must be
absorbed by form-factors which have already appeared in $m$-point
functions with $m < n$.
It is clear from inspection of Fig.~\ref{fig:4pf-exchange}
and its analogues for higher $n$-point functions
that this arrangement is plausible because all
diagrams except the contact graph are built out of lower-order
vertices whose divergences can be described in terms of lower-order
form-factors.

\section{The dynamical renormalization group}
\label{sec:drge}

In this section we revisit the analysis
of~\S\S\ref{sec:resummation}--\ref{sec:factorization}
from a slightly more formal viewpoint---that of the `dynamical
renormalization group'.
The original purpose of these methods was to determine
dynamical scaling laws for correlation functions near an
out-of-equilibrium critical point~\cite{cardy1996scaling,kamenev2011field}.
However, they have found various applications
to inflationary correlation
functions~\cite{Boyanovsky:1996rw,Boyanovsky:1997cr,Boyanovsky:1997xt,
Boyanovsky:2001ak,
Podolsky:2008du,Podolsky:2008qq}.
The approach developed here is similar to the
discussion of time-dependence generated from loop
corrections
given by
Burgess et~al.~\cite{Burgess:2009bs}.
More recently, Collins et~al.~suggested that the
dynamical
renormalization group method could be used to determine
time evolution
generated by integrating out heavy modes~\cite{Collins:2012nq}.
Our analysis demonstrates the steps which would be required to
achieve this for an arbitrary $n$-point function.

\paragraph{Role of `renormalizability'.}
To apply the renormalization group requires a guarantee
that all large logarithms can be absorbed into a finite number
of `renormalized' quantities.
When applied
to ultraviolet behaviour
this guarantee is provided either
by the criterion of renormalizability (in a strictly renormalizable theory),
or 
the fact that only a finite number of irrelevant operators
need to be kept in order to make predictions at
fixed accuracy (in an effective field theory).

In the present case
there is no notion of `renormalizability'
and
we do not have either of these guarantees.
The property which replaces renormalizability is factorization,
in the sense of~\S\S\ref{sec:resummation}--\ref{sec:factorization}.
In particular, what is required is a guarantee that
all divergences produced by `vertex integrals'
such as $I_{\lambda \mu \nu}$
and $I_{\lambda\mu\nu\pi}$ are proportional to a finite number
of combinations of the external momenta.
As we saw in~\S\ref{sec:separate-universe}, the separate universe
asserts that
all divergences generated by $I_{\lambda\mu\nu}$ produce 
the shape $k_1^{-3} k_2^{-3}$ (or its permutations),
and all divergences generated by $I_{\lambda\nu\nu\pi}$ produce the
shapes $k_1^{-3} k_2^{-3} k_3^{-3}$
or $k_1^{-3} k_2^{-3} |\vect{k}_1 + \vect{k}_3|^{-3}$
(or their permutations).

These properties
(and their analogues for higher $n$-point functions)
are implied by
the superhorizon structure of perturbation theory,
but it is difficult to see them emerge at the
level of Feynman diagrams.
A rigorous proof
is a `factorization theorem'.
Theorems of this kind provide the missing guarantee that only a finite
number of form-factors can be used to absorb all large logarithmically-enhanced
contributions.
They are an integral part of the apparatus
used to study perturbative
QCD processes such as jets and
deep inelastic scattering
\cite{Collins:1989gx,Brock:1993sz,Collins:2011zzd,Dissertori:2003pj,
Chiu:2011qc}.

\subsection{Two-point function}
\label{sec:twopf}

In this section our purpose is to prove a factorization theorem for
the vertex integral $I_{\lambda\mu\nu}$.
Before doing so, we
briefly
return to Eqs.~\eqref{eq:twopf-fixed} and~\eqref{eq:twopf-basic}
and repeat the analysis
of~\S\ref{sec:factorization} in a way which can be generalized
more easily to the
three-point function.

Ignoring terms which decay like positive powers of $|k\tau|$,
the coefficient $\Sigma_{\alpha\beta}$
defined in~\eqref{eq:factorized-2pf}
depends on $k$ and $\tau$ only through powers of
$\ln(-k_\ast \tau)$ and $\ln k/k_\ast$.
Therefore it can be interpreted as a Taylor series in
these logarithms,
expanded around the arbitrary horizon-crossing time for $k_\ast$.
The renormalization group
is a method
to reverse-engineer a function from the first few
terms in its Taylor expansion.
When
applied to $\Sigma_{\alpha\beta}$ this allows an all-orders reconstruction
from information about the lowest terms in the
perturbative series.
The term `dynamical' merely indicates that the ordinary renormalization-group
method is being applied to a
series expansion in time rather than energy.

\paragraph{Dynamical renormalization group analysis.}
To proceed we define a
`hard' contribution to $\Sigma_{\alpha\beta}$
which excludes any enhancement due to soft effects from 
$\ln|k_\ast \tau|$,
\begin{equation}
	\Sigma^{\hard}_{\alpha\beta} =
	H_\ast^2 \left(
		\delta_{\alpha\beta}
		+ 2 r_{\alpha\beta}^\ast
		- 2 M^\ast_{\alpha\beta} \ln \frac{2k}{k_\ast}
	\right) .
	\label{eq:sigma-hard}
\end{equation}
We interpret $\Sigma^{\hard}_{\alpha\beta}$ as a function of $k$
and $k_\ast$, but not $\tau$.
The full two-point function can
be written 
\begin{equation}
	\Sigma_{\alpha\beta}
	\simeq
		\Sigma^{\hard}_{\alpha\beta}
		- \Big[
			\Sigma^{\hard}_{\alpha\lambda} u_{\lambda\beta}^\ast
			+ \Sigma^{\hard}_{\beta\lambda} u_{\lambda\alpha}^\ast
		\Big]
		\ln(-k_\ast \tau)
	,
	\label{eq:sigma-hard-soft}
\end{equation}
where `$\simeq$' indicates that this expression is valid up to
$\NLO$ accuracy.
In particular, we have replaced
$H_\ast^2 \delta_{\alpha\beta}$
with $\Sigma_{\alpha\beta}$ in the factor multiplying
$\ln(-k_\ast \tau)$. Although there is a potential mismatch in
this exchange,
it would appear at \emph{next}-next-order and is therefore below
the accuracy to which we are working.
In the solution to the renormalization-group equation
this will translate to
an error below leading-logarithmic accuracy.%
	\footnote{\label{footnote:leading-log} The
	leading-logarithm
	approximation resums terms of the form
	$[\varepsilon_\ast \ln (-k_\ast \tau)]^n$ for all $n$,
	but not
	$\varepsilon [\varepsilon_\ast \ln (-k_\ast \tau)]^n$ or smaller.
	For a renormalization group equation of the schematic form
	\begin{equation}
		- \frac{\partial \Sigma}{\partial \ln \tau}
		= u \Sigma ,
	\end{equation}
	it can be proved by standard methods
	(see, eg., Ref.~\cite{Collins:1984xc})
	that resummation of the leading logarithms requires
	knowledge of $u$ to $\Or(\varepsilon)$; next-to-leading logarithms
	requires knowledge of $u$ to $\Or(\varepsilon^2)$, and so on.
	Mathematically, our methods will justify the separate universe
	method to leading-logarithm order.
	Although this is likely to be sufficient for observable
	inflation, subleading logarithms must become important
	over very long timescales.}

We now
interpret the right-hand side
of~\eqref{eq:sigma-hard-soft}
as a Taylor series expansion around
the arbitrary horizon-crossing time for $k_\ast$.
According to Taylor's theorem,
\begin{equation}
	\Sigma_{\alpha\beta}(k,\tau)
	=
		\Sigma_{\alpha\beta} |_{|k_\ast\tau| = 1}
		+ \left.
			\frac{\d \Sigma_{\alpha\beta}}{\d \ln \tau}
			\right|_{|k_\ast\tau| = 1}
			\ln (-k_\ast \tau)
		+ \cdots
	,
	\label{eq:twopf-taylor-expansion}
\end{equation}
and so
the coefficient of $\ln(-k_\ast \tau)$ is
the derivative of $\Sigma_{\alpha\beta}$ evaluated at $|k_\ast \tau| = 1$.
Therefore
\begin{equation}
	- \left.
		\frac{\d \Sigma_{\alpha\beta}}{\d \ln \tau}
		\right|_{|k_\ast\tau| = 1}
	=
		\Sigma^{\hard}_{\alpha\lambda} u_{\lambda\beta}^\ast
		+ \Sigma^{\hard}_{\beta\lambda} u_{\lambda\alpha}^\ast
	=
		\Sigma_{\alpha\lambda}^\ast u_{\lambda\beta}^\ast
		+ \Sigma_{\beta\lambda}^\ast u_{\lambda\alpha}^\ast .
	\label{eq:twopf-proto-rge}
\end{equation}
\emph{A priori}
this is a statement about
the value of the derivative
only at the fixed time $\tau = -k_\ast^{-1}$.
But precisely because $k_\ast$ is arbitrary,
the
function
$\d \Sigma_{\alpha\beta} / \d \ln \tau|_{|k_\ast \tau| = 1}$
obtained in this way
must be the same function of the horizon-crossing time for $k_\ast$
as
$\d \Sigma_{\alpha\beta} / \d \ln \tau$
is of $\tau$.
Therefore we can immediately promote it to
a differential equation valid for all $\tau$,
obtained by removing `$\ast$' from all terms
in~\eqref{eq:twopf-proto-rge}
\begin{equation}
	-\frac{\d \Sigma_{\alpha\beta}}{\d \ln \tau}
	=
	\Sigma_{\alpha\lambda} u_{\lambda\beta}
	+
	\Sigma_{\beta\lambda} u_{\lambda\alpha} .
	\label{eq:twopf-rge}
\end{equation}
In this equation
$\Sigma_{\alpha\beta}$ is a function of $k$ and $\tau$,
but $u_{\alpha\beta}$ is a function of $\tau$ only.
In the literature it is often rewritten
in terms of the inflationary e-folding
time $N = \int H \, \d t$
using $- \d \ln \tau = \d N$.

It might appear that~\eqref{eq:twopf-rge} could have been obtained by
differentiation of Eq.~\eqref{eq:sigma-hard-soft} with respect to $\tau$.
This procedure would produce an equation which is symbolically
identical, but where the right-hand side is
evaluated at $|k_\ast \tau| = 1$. To understand
why~\eqref{eq:twopf-rge} is valid for arbitrary $\tau$
it is necessary to construct~\eqref{eq:twopf-rge}
via comparison with Taylor's theorem, as described above.
This justifies the more heuristic method used to obtain
Eq.~\eqref{eq:twopf-backwards-rge}.

An explicit solution
of Eq.~\eqref{eq:twopf-rge} 
requires an initial condition.
According to~\eqref{eq:twopf-taylor-expansion} we can
obtain this initial condition
from the constant term in the Taylor expansion,
$\Sigma^{\hard}_{\alpha\beta}$.%
	\footnotemark%
	\footnotetext{Therefore
	only the lowest two Taylor terms are required to reconstruct
	$\Sigma_{\alpha\beta}$.
	However, these receive contributions from all orders in
	the slow-roll expansion.}%
	$^{,}$\footnotemark%
	\footnotetext{\label{footnote:factorization-scheme}
	In any renormalization-group procedure there is some
	arbitrariness in setting
	the initial condition~\eqref{eq:twopf-rge-ic},
	because
	we are free to treat some or all
	of the constant term $\Sigma^{\hard}_{\alpha\beta}$
	as an additive constant rather than the zero-order term in
	the Taylor expansion.
	This makes no difference in an exact calculation, but
	can influence the result at finite orders in perturbation theory.
	A prescription for this choice is called a
	\emph{factorization scheme}.
	For details, see eg. Ref.~\cite{Dissertori:2003pj}.
	We absorb the entirety of
	$\Sigma^{\hard}_{\alpha\beta}$
	as the initial condition.}
To obtain an accurate estimate using
Eq.~\eqref{eq:sigma-hard-soft}
we should set $k_\ast = k$
to remove large contributions from powers of $\ln k/k_\ast$, which yields%
	\footnote{At $|k\tau| = 1$
	there are contributions to $\Sigma_{\alpha\beta}$ from decaying
	power-law corrections which we have not written explicitly.
	(See also~\S\ref{sec:matching}.)
	These terms make no contribution
	to~\eqref{eq:twopf-rge-ic}
	because they do not form part of the Taylor series
	at late times.
	
	The contribution from $M_{\alpha\beta}$ could be removed
	by choosing $k_\ast = 2k$.
	For higher $n$-point functions, the analogous contribution
	would be removed by choosing $k_\ast = k_t$.
	To maximize the accuracy of each initial condition one should
	therefore evolve each $n$-point function from a marginally different
	initial time,
	but for the three- and four-point functions this effect will be
	negligible.}
\begin{equation}
	\Sigma_{\alpha\beta} =
	H_k^2 \left(
		\delta_{\alpha\beta}
		+ 2 r_{\alpha\beta}^k
		- 2 M_{\alpha\beta}^k \ln 2
		+ \cdots
	\right)
	\quad
	\text{at $|k \tau| = 1$} ,
	\label{eq:twopf-rge-ic}
\end{equation}
where `$\cdots$' indicates higher-order corrections beyond $\NLO$,
which we have not computed,
and a sub- or superscript `$k$' now indicates evaluation at the
horizon-crossing time for $k$.
Together,
Eqs.~\eqref{eq:sigma-hard-soft}, \eqref{eq:twopf-rge}
and~\eqref{eq:twopf-rge-ic}
provide an interpretation of the renormalization-group analysis.
The hard subprocess---here, creation of fluctuations near horizon
exit---depends only on the scale $k$, and therefore
fixed-order perturbation
theory can be used to obtain an accurate estimate
such as~\eqref{eq:twopf-rge-ic}.
The absence of other scales implies we can be confident that the
subleading terms represented by `$\cdots$' are smaller than those we
have calculated.
This subprocess is used as an initial condition for the
renormalization-group evolution, which evolves it to lower energies---here,
represented by the decreasing value of $k/aH$. This evolution is
accomplished by successively `dressing' the hard factor with the
cumulative effect
of soft processes.

Eq.~\eqref{eq:twopf-rge} is the dynamical renormalization group
equation for the two-point function.
It has already appeared in the literature as
a `transport' equation for
$\Sigma_{\alpha\beta}$,
as described
by Mulryne et~al.~\cite{Mulryne:2009kh,Mulryne:2010rp,Seery:2012sx}.
If desired, the methods of Ref.~\cite{Seery:2012sx} can be used
to convert~\eqref{eq:twopf-rge} into an
equivalent
equation for the
form-factor $\Gamma_{\alpha i}$ which coincides
with~\eqref{eq:twopf-backwards-rge}.
Therefore, the approach of this section is exactly equivalent to that
of~\S\ref{sec:factorization},
or
the standard $\delta N$ formula~\cite{Lyth:2005fi}.

\subsection{A factorization theorem for the bispectrum}
\label{sec:factorization-thm}

Let us return to the factorization properties of
$I_{\lambda\mu\nu}$.
We should interpret
divergences in this vertex integral
to mean that
three-body interactions continue to arbitrarily late
times, rather than being localized to the neighbourhood
of horizon-crossing.
Therefore the divergences
originate from a kinematic region where
all decaying modes become extinct.
In this region the wavefunctions $w_{\alpha\beta}$
lose their $\vect{k}$-dependent information and begin to evolve
in the same way.
It is this property which
limits the number of momentum configurations
which can be enhanced by divergences,
and ultimately leads to the possibility of factorization.
Mathematically,
this means that we
require only asymptotic information
about the dependence of each mode function on $k$ and $\tau$.%
	\footnote{In models containing extra characteristic scales,
	such as the horizon scale associated with a step or other
	sharp feature in the potential
	(see eg., Refs.~\cite{Adshead:2011bw, Adshead:2011jq}),
	this may no longer be true.
	In particular, it may happen that interactions continue
	outside the horizon until
	some characteristic time, and then switch off.
	In these models it need not be possible to analyse the
	divergence structure of the vertex integral using only
	asymptotic information, and extra momentum configurations may
	receive large enhancements.
	We do not consider such scenarios in this paper.}

\paragraph{Asymptotic behaviour.}
We will prove the factorization property in two parts.
The first step is to show that each mode function
has a `gap'
in its asymptotic expansion in the limit $|k\tau| \rightarrow 0$,
\begin{equation}
	\Psi_i(k, \tau) = \psi_i(k, k_\ast) \Big[
		A_i(k, k_\ast, \tau) + \Or( k \tau )^2
	\Big] .
	\label{eq:asymptotic-condition}
\end{equation}
Here, $\Psi_i(k,\tau)$ is a mode function for the scale $k$
and may depend on other labels which we collectively denote $i$.
The prefactor $\psi_i(k,k_\ast)$ is time-independent and
adjusted so that each Feynman propagator constructed
from $\Psi_i$ has the correct normalization.
If the calculation were carried to all orders in the
slow-roll expansion then $\Psi_i$ would be
independent of the reference
scale $k_\ast$,
but we allow $\psi_i$ and $A_i$ to include an explicit dependence
when truncated to any finite order.
In addition, $A_i(k,k_\ast,\tau)$ should be slowly varying,
by which we mean an arbitrary polynomial in
$\ln \tau$. 
Therefore there is a `gap' in the asymptotic
expansion caused by the absence of a term linear
in $k \tau$. 
If the elementary wavefunctions satisfy~\eqref{eq:asymptotic-condition}
then so do their cosmic-time derivatives.
Where required
we treat these as wavefunctions of different species, distinguished by
the label $i$.

For a background spacetime sufficiently close to de Sitter,
Eq.~\eqref{eq:asymptotic-condition} can be proved using the
methods of Weinberg's theorem~\cite{Weinberg:2005vy}.
Weinberg constructed asymptotic approximations to the mode function
of a \emph{massless} scalar field by making an expansion in powers of
$k^2$. The inclusion of a slowly varying mass promotes the
constants in Weinberg's computation to arbitrary polynomials
of logarithms.
Because the expansion is in powers of $k^2$ there is automatically
a gap in the asymptotic expansion of the growing mode.
If the decaying mode were to begin at relative order $k\tau$
this property could be lost.
However,
Weinberg's analysis shows that the decaying mode begins
at relative order $(k\tau)^3$, leading to the
estimate~\eqref{eq:asymptotic-condition}.

The same conclusion can be reached by studying the
behaviour of solutions to the field equation in the asymptotic
future $\tau \rightarrow 0$~\cite{Seery:2006tq}.
The field equation is
\begin{equation}
	\frac{\d^2 \phi}{\d \tau^2}
	+ 2 aH \frac{\d \phi}{\d \tau}
	+ m^2 a^2 \phi = 0 .
\end{equation}
Assuming quasi-de Sitter expansion this is an equidimensional equation
with solutions $\phi \sim \tau^\Delta$, where
\begin{equation}
	\Delta = \frac{3}{2} \pm \sqrt{ \frac{9}{4} - \frac{m^2}{H^2} } .
\end{equation}
In the massless case we conclude $\Delta = { 0, 3 }$.
As above,
corrections due to a slowly varying mass occur as logarithms
and subleading terms in the gradient expansion appear as powers of
$(k/a)^2$.

\subsubsection{Divergence structure of vertex integral}
\label{sec:vertex-divergences}
The second step uses~\eqref{eq:asymptotic-condition}
to show that the vertex integral produces divergences only
in very restricted combinations.
This point was emphasized by Weinberg~\cite{Weinberg:2006ac},
but here we give a more detailed analysis.
It is possible to regard the theorem proved in this section as
a sharper version of Weinberg's theorem---providing
control not only
over power-law divergences, but also
the way in which logarithmic divergences appear.
On the other hand,
Weinberg's theorem applies to all orders in the loop expansion
whereas our argument applies only to the bispectrum at tree-level.
It would be of considerable interest to strengthen
this result
to constrain the possible momentum combinations which can be
logarithmically enhanced at loop level.

A generic contribution to the vertex integration will
mix three `internal' wavefunctions,
evaluated at the time $\eta$ of the vertex,
and three `external' wavefunctions which are evaluated at the
time of observation $\tau$.
The labels for the internal and external parts need not agree.
For example, they may be mixed by an off-diagonal mass matrix
as in~\eqref{eq:propagator} or~\eqref{eq:propagator-final},
or the internal wavefunctions may be differentiated with respect
to time.
Also, depending on the assignment of $+$ and $-$ type vertices,
half of the wavefunctions will be conjugated with respect to the other
half. These details are irrelevant for the purposes of our discussion.
We denote the external labels $\{ e_1, e_2, e_3 \}$
and the internal labels $\{ i_1, i_2, i_3 \}$.
In conclusion, the late-time
behaviour of each diagram can be deduced from a sum of terms of the form
\begin{equation}
\begin{split}
	I_n = \mbox{} &
		\bigg(
			\prod_{m \in \{ 1, 2,3 \}} \psi_{e_m}(k_m) \psi_{i_m}(k_m)
			\Big[ A_{e_m}(k_m, \tau) + \cdots \Big]
		\bigg)
	\\ & \mbox{}
	\times
	\int_{-\infty}^{\tau} \frac{\d \eta}{\eta^{4-2n}}
	f_{e_1 e_2 e_3 i_1 i_2 i_3}(\eta)
	\Big[ A_{i_1}(k_1, \eta) + \cdots \Big]
	\Big[ A_{i_2}(k_2, \eta) + \cdots \Big]
	\Big[ A_{i_3}(k_3, \eta) + \cdots \Big] ,
\end{split}
\label{eq:I_n}
\end{equation}
where $f_{e_1 e_2 e_3 i_1 i_2 i_3}$ is also taken to be slowly
varying in the sense described above.
For perturbatively massless fields $n$ is a nonnegative integer
arising from powers of the scale factor $a \sim -1/\tau$.
We begin with four powers of $a$ from the integration measure $\sqrt{-g}$.
Further powers appear in combination with spatial gradients, which
each gradient producing a power of $1/a$.
Inverse gradients such as $\partial^{-2}$ produce positive
powers of $a$.
The
precise value of $n$
depends on the \emph{net} number of spatial gradients
carried by the operator giving rise to~\eqref{eq:I_n}.

\paragraph{More than two spatial gradients: $n \geq 2$.}
There are only
two choices which produce divergences in the vertex
integral: $n=0$ and $n=1$. We will study these in detail below.
For $n \geq 2$, the vertex integral in~\eqref{eq:I_n} is convergent.
[Clearly $I_n$ itself may still be divergent
because of $\ln \tau$ terms from the external wavefunction factors
$A_{e_m}(k_m, \tau)$.
Divergences of this type are not under
discussion in the present section.]

\subsubsection{Two spatial gradients}
\label{sec:two-gradients}
Now consider the case of two net spatial gradients.
The primitive degree of divergence of
the integral in~\eqref{eq:I_n} is
$\int^\tau \d \eta / \eta^2 \sim 1/\tau$.
Therefore
decaying parts of the internal wavefunctions contribute only
to the convergent part of the integral,
and decaying parts of the external wavefunctions produce
contributions which decay at least as fast as $\tau$.

\paragraph{Vertex integral.}
To proceed we isolate
the integral in~\eqref{eq:I_n},
which measures the cumulative effect of
three-body interactions.
For $n=1$ these interactions
are suppressed by a net-positive
number of spatial gradients $\sim k/a$,
and we should expect them to switch off in the
superhorizon limit
$k/a \rightarrow 0$.
Hence, we
anticipate that the integral produces no divergences.
The late-time limit is determined by
\begin{displaymath}
	\int^\tau \frac{\d \eta}{\eta^2} f_{e_1 e_2 e_3 i_1 i_2 i_3}(\eta)
	A_{i_1}(k_1, \eta) A_{i_2}(k_2, \eta) A_{i_3}(k_3, \eta)
	+ \text{convergent} .
\end{displaymath}
The integral has dimensions of wavenumber or inverse time.
Because $f$ and the $A_{i_m}$ are only slowly varying,
the only scale available with the appropriate dimensions is
$1/\tau$ itself. Therefore the divergent part of the integral
must scale like $1/\tau$ multiplied by a polynomial in logarithms.%
	\footnote{An alternative way to reach the same conclusion is
	to argue that $\int^\tau \d \eta \; \eta^{-2} \ln^m \eta
	\sim \tau^{-1}$ for any power $m$,
	which
	can be proved by induction.}
For the interactions encountered in typical inflationary
theories, Weinberg's theorem
guarantees that these `fast', power-law divergences
cancel~\cite{Weinberg:2005vy}.

Therefore, after cancellation of
the `fast' divergences,
two-gradient operators with $n=1$ behave in the same way as
operators with $n \geq 2$.
The vertex integral is dominated by three-body interactions which
occur around the time of horizon exit, with negligible contributions
from interactions occurring much later.
Any divergent terms in $I_1$ arise only from $\ln\tau$-dependent
terms appearing in the external wavefunction factors $A_{e_m}$.

\subsubsection{Zero spatial gradients}
\label{sec:zero-gradients}
The remaining case is $n=0$, generated by operators with
zero net spatial gradients.
In this case there is no suppression in the limit $k/a \rightarrow 0$,
and
therefore no expectation that these
three-body interactions should
switch off
at late times.

\paragraph{Vertex integral.}
As above, we first focus on the integral term in~\eqref{eq:I_n}.
There is no longer anything
to prevent a `slow', pure logarithmic divergence
in addition to the `fast'
terms involving inverse powers of $\tau$.
However, we show that
these slow divergences are generated by
$k$-dependent terms drawn from a single internal wavefunction.

First, consider the contribution from the growing modes of each
internal wavefunction. This can be written
\begin{displaymath}
	\int^\tau \frac{\d \eta}{\eta^4} f_{e_1 e_2 e_3 i_1 i_2 i_3}(\eta)
	A_{i_1}(k_1, \eta) A_{i_2}(k_2, \eta) A_{i_3}(k_3, \eta)
	.
\end{displaymath}
The argument of~\S\ref{sec:two-gradients} shows that this diverges
like $\tau^{-3}$
multiplied by a polynomial in
logarithms.
We will discuss these terms when we consider the external
wavefunction factors.
On the other hand, because of the `gap' in the
asymptotic expansion~\eqref{eq:asymptotic-condition},
the contribution from the $\Or(k\eta)^2$-suppressed
terms from two \emph{different} wavefunctions is
of the form
\begin{displaymath}
	\int^\tau \frac{\d \eta}{\eta^4} f_{e_1 e_2 e_3 i_1 i_2 i_3}(\eta)
	A_{i_1}(k_1, \eta) \times
	\Or(k_2 \eta)^2 \times
	\Or(k_3 \eta)^2
	= \text{convergent}
	.
\end{displaymath}
Therefore any pure logarithmic divergences must come from the
$\Or(k\eta)^3$ term associated with a \emph{single}
internal wavefunction $i_m$.
The integral has dimensions of $k^3$, and hence these pure logarithms
(which do not involve powers of $\tau$)
must be proportional to $k_m^3$,
because no other scales are available.
This is the distinctive `local' shape which we have been seeking.

\paragraph{External wavefunctions.}
In addition to these `local' pure logarithms,
the vertex integral will produce fast power-law divergences
proportional to
$\tau^{-3}$ and $\tau^{-1}$.
The `gap' in~\eqref{eq:asymptotic-condition} guarantees that there
are no terms proportional to $\tau^{-2}$.
The $\tau^{-1}$ terms can be ignored.
When multiplied by the growing modes $A_{e_m}$ from each external
wavefunction they generate contributions which are also of order
$\tau^{-1}$, and Weinberg's theorem guarantees that they must
cancel~\cite{Weinberg:2006ac}.%
	\footnote{For this conclusion, we require the stronger argument
	of Ref.~\cite{Weinberg:2006ac} which applies to operators
	proportional to $a^4 \sim 1/\tau^4$ in conformal time.
	To be clear, we would like to emphasize that Weinberg's theorem
	excludes only `fast' power-law divergences in correlation
	functions.
	As discussed explicitly in Ref.~\cite{Weinberg:2006ac},
	it does not restrict the appearance of `slow' logarithmic
	effects.}
When multiplied by decaying terms in each external wavefunctions
they yield contributions which decay at least as fast as $\tau$,
and therefore become negligible at late times.

The situation is different for the $\tau^{-3}$ divergences.
Weinberg's theorem likewise guarantees that, when multiplied by
the growing modes from each external wavefunction, their net contribution
must cancel.
But it is also possible for these divergences to promote decaying
terms from the external wavefunctions.
As we now explain, these also produce only `local' combinations of the
momenta. The argument is similar to that for the internal wavefunctions.

First, consider the case where the $\tau^{-3}$ divergence
combines with
decaying contributions from two different external wavefunctions.
The net contribution will be
$\tau^{-3} \times \Or(\tau^2) \times \Or(\tau^2) =
\Or(\tau)$. Therefore these contributions decay at late times
and become negligible.
It follows that pure logarithmic effects can be generated only by the
$\Or(k \tau)^3$ term from a single external wavefunction $e_m$,
and
on dimensional grounds will be proportional to $k_m^3$.
There is no need to keep track of any $\tau^{-1}$ contributions generated
in the same way because these are also controlled by Weinberg's theorem.

\subsubsection{Inverse spatial gradients}
\label{sec:inverse-gradients}
We can not yet conclude that the vertex integration produces only
divergences proportional to a pure power $k_i^3$
of one
of the external momenta,
because in Einstein gravity some modes of the metric are not
dynamical: instead, they are removed by constraints.
The process of solving these constraints can produce
inverse spatial gradients
$\partial^{-2}$~\cite{Maldacena:2002vr,Seery:2005gb}.
Working in ADM variables, the metric can be written
\begin{equation}
	\d s^2 = - N^2 \, \d t^2 + h_{ij} ( \d x^i + N^i \, \d t )
		( \d x^j + N^j \, \d t ) ,
	\label{eq:adm-decomposition}
\end{equation} 
where $N$ is the lapse function and $N^i$ is the shift vector.
Only the three-metric $h_{ij}$ carries
independent degrees of freedom. The lapse and shift are determined
by
constraints.
(See Appendix~\ref{subsec:3pfappendix}.)
To obtain the third-order action it is only necessary to
solve these constraints to first order~\cite{Maldacena:2002vr,Chen:2006nt}.
We set
$N_i = \partial_i \vartheta_1 + \beta_{1 i}$,
where $\beta_{1 i}$ is divergenceless.
It can be ignored for the purposes of the three-point function
because it does not contribute to the third-order action.
The solution for $\vartheta_1$ is~\cite{Seery:2005gb}
\begin{equation}
	\frac{4H}{a^2} \partial^2 \vartheta_1
	= - 2 \dot{\phi}^\alpha \delta \dot{\phi}_\alpha
	+ \Big(
		2 \ddot{\phi}^\alpha + \frac{\dot{\phi}^2}{H} \dot{\phi}^\alpha
	\Big)
	\delta\phi_\alpha .
	\label{eq:shift-vector}
\end{equation}
In principle the shift vector $N_i = \partial_i \vartheta_1$
can appear in operators with zero net
spatial gradients, generating 
enhanced
shapes formed from 
rational
functions of the external momenta.
(If present, these ratios of the external momenta would
appear as a prefactor in Eq.~\eqref{eq:I_n} which we have not
written explicitly.) 
Although this outcome is compatible with the conclusions of
\S\S\ref{sec:two-gradients}--\ref{sec:zero-gradients}
these shapes are not local and could not be reproduced by
separate-universe type formulae.
To demonstrate that only form-factors compatible with the
separate universe method are required we must show that
these enhanced non-local shapes are absent.

Whether $\vartheta_1$ can produce divergences in the vertex integral
depends on its asymptotic behaviour.
This is an important issue
beyond the renormalization-group approach we
are developing in this paper,
because the decay rate of the shift vector is important
in any attempt to justify the separate universe
method. Weinberg gave an argument using a broken-symmetry
approach~\cite{Weinberg:2008nf,Weinberg:2008si}.
More recently, Sugiyama, Komatsu \& Futamase
showed that the Einstein equations require the shift vector
to decay at late times during inflation~\cite{Sugiyama:2012tj}.
Up to next-order, it can be verified by explicit calculation
that~\eqref{eq:shift-vector} gives
\begin{equation}
	\frac{4H}{a^2} \partial^2 \vartheta_1 = \Or(k\eta)^2 .
\end{equation}
This gives $\vartheta_1 = \Or(1)$.
The general
analysis of Sugiyama et~al.~extends this conclusion to all orders
in the slow-roll expansion.

\paragraph{Absence of time-dependent terms generated by the shift.}
This decay rate is not sufficiently rapid to prevent
the generation of
time-dependent
terms from arbitrary operators involving the shift
vector $N_i$.
However, because the decay is exponentially fast in cosmic time,
it will prevent divergences in any operator which involves two
or more powers of $\partial^2 \vartheta_1 / a^2$.
Only operators linear in $\partial^2 \vartheta_1 / a^2$
can generate time dependence.%
	\footnote{In principle, terms such as $N_i N^i = \partial_i \vartheta_1
	\partial_i \vartheta_1 / a^2$ can generate time dependence, even though
	they are quadratic in $\vartheta_1$.
	However, such contractions are not generated in
	Einstein gravity combined with the scalar field theories
	we are considering.}
It can be shown that the second-order action is entirely independent of
$\vartheta$, and that the third-order action contains no terms
linear in $\partial^2 \vartheta_1 / a^2$.
Therefore the lapse function gives rise to no time-dependent
terms in the two- or three-point functions, and no
nonlocal momentum configurations can be enhanced by divergences.

\subsubsection{Factorization for the three-point function}
In conclusion, to all orders in the slow-roll expansion,
each diagram contributing to the three-point function
with external wavefunctions carrying labels $\{ e_1, e_2, e_3 \}$
will schematically be of the form
\begin{equation}
\begin{split}
	I_{e_1 e_2 e_3} =
	\Big[
		\prod_m \psi_{e_m}(k_m) & \psi_{i_m}(k_m)
	\Big]
	A_{e_1}(k_1, \tau)
	A_{e_2}(k_2, \tau)
	A_{e_3}(k_3, \tau)
	\\ &
	\mbox{} \times
	\Big[
		u(\tau) k_1^3
		+ v(\tau) k_2^3
		+ w(\tau) k_3^3
		+ \text{finite}
	\Big]_{e_1 e_2 e_3 i_1 i_2 i_3} ,
\end{split}
\label{eq:factorization-theorem}
\end{equation}
where $u$, $v$ and $w$ are slowly-varying functions of $\tau$
(and perhaps also $k$ or $k_\ast$)
given by arbitrary polynomials of logarithms.
The `finite' term is independent of $\tau$.
The precise form of $u$, $v$, $w$ and the finite piece depends on the
interaction under discussion.
The notation $[ \cdots ]_{e_1 e_2 e_3 i_1 i_2 i_3}$ indicates that both
the divergent and finite pieces produced by the vertex integral can
depend on the labels for the internal and external wavefunctions.
Eq.~\eqref{eq:factorization-theorem} is the formal statement of
the factorization theorem.%
	\footnote{As
	described in~\S\ref{sec:zero-gradients},
	the time-dependent `local' contributions
	$u$, $v$ and $w$ may include terms generated by
	promotion of decaying
	terms in an external wavefunction $\Psi_{e_m}$
	due to $\tau^{-3}$ divergences
	in the vertex integral.
	In~\eqref{eq:factorization-theorem} we have
	redefined these contributions by
	dividing out
	a factor of the
	corresponding external growing mode $A_{e_m}$.
	This is harmless overall, leading only to a redefinition of the
	associated source term.
	No such terms have yet been encountered in practical calculations,
	so there has been no need to keep track of this redefinition.
	The first term of this type would arise from subleading corrections
	to the $V''' \delta\phi^3$ vertex, which already contributes at
	next-order. Therefore the effect is \emph{next}-next-order
	in fixed-order perturbation theory
	and would contribute only to the next-leading-logarithm
	solution of the renormalization group equation.
	(See footnote~\ref{footnote:leading-log}
	on p.~\pageref{footnote:leading-log}.)}

As for the two-point function, perturbative calculations
using quantum field theory will not
provide us with the functional form of $u$, $v$, or $w$.
Instead, they generate terms in the Taylor series
expansion for these functions based at the arbitrary horizon-crossing
time for $k_\ast$. The full functions must be reverse-engineered
using the methods of the renormalization group.

Eq.~\eqref{eq:factorization-theorem} clearly exhibits
the Sudakov-like enhancements generated by 
divergent
logarithms.
In agreement with the discussion of~\S\ref{sec:separate-universe},
we see that
large Sudakov-like effects in each external wavefunction
can be absorbed into the form factor $\Gamma_{\alpha i}$.
This accounts for the first line in~\eqref{eq:deltaN-3pf}.
Also,
we now see that~\eqref{eq:factorization-theorem} guarantees
the vertex integral will produce
precisely the structure identified in~\S\ref{sec:separate-universe}
as a prerequisite
for the second line in~\eqref{eq:deltaN-3pf}.

To develop a renormalization-group approach to the three-point
function we will not pursue the factorization of this form-factor
directly. Instead we follow the approach of~\S\ref{sec:twopf}.
Specializing to a multiple-field model in which the external labels
$\{ e_1, e_2, e_3 \}$ correspond to the different species of
scalar fields, we see that
the lowest terms in the Taylor expansion
around the horizon-crossing time for $k_\ast$ are
\begin{equation}
\begin{split}
	I_{\alpha\beta\gamma} \approx I^\ast_{\alpha\beta\gamma}
	& \mbox{} - \big[
		u_{\alpha\lambda}^\ast I^\ast_{\lambda\beta\gamma}
		+ u_{\beta\lambda}^\ast I^\ast_{\alpha\lambda\gamma}
		+ u_{\gamma\lambda}^\ast I^\ast_{\alpha\beta\lambda}
	\big] \ln(-k_\ast \tau)
	\\ & \mbox{}
	+ \Big[
		\prod_m |\psi_\alpha(k_m)|^2
	\Big]
	A^\ast_{\alpha\lambda}(k_1)
	A^\ast_{\beta\mu}(k_2)
	A^\ast_{\gamma\nu}(k_3)
	\left. \frac{\d u_{\lambda\mu\nu}}{\d N} \right|_\ast
	k_1^3 \ln (-k_\ast \tau)
	\\ & \mbox{}
	+ \text{permutations}
	,
	\label{eq:3pf-factorization}
\end{split}
\end{equation}
where $\d u_{\lambda\mu\nu}/\d N$ is defined by
this expression, and
`permutations' includes the terms generated from the
Taylor expansion of $v(\tau) k_2^3$ and $w(\tau) k_3^3$
in~\eqref{eq:factorization-theorem}.
The Taylor coefficient $u_{\alpha\beta}$ associated with time-dependence
of the external wavefunction factors is already known.
Therefore
to reverse-engineer the full functional form of $I_{\alpha\beta\gamma}$
we require only an estimate of the derivative
$\d u_{\lambda\mu\nu} / \d N|_\ast$,
which
can be read off from the $\ln (-k_\ast \tau)$ terms in a
next-order computation of the three-point function.
We collect the details of this calculation in Appendix~\ref{appendix:3pf}.

\subsection{Renormalization group analysis for the three-point function}
\label{sec:threepf}

We can now use~\eqref{eq:3pf-factorization}
to apply the dynamical renormalization group to the three-point function.
The procedure is very similar to the two-point function analysis
in~\S\ref{sec:twopf}, although potentially complicated
because the three-point function depends on three distinct external
momenta $k_1$, $k_2$, $k_3$.
In addition to a hierarchy between the scale of these momenta and the
horizon scale $aH$, there may now be hierarchies among the
external momenta themselves.

\paragraph{Shape and scale effects.}
The three-point function enforces momentum conservation
$\vect{k}_1 + \vect{k}_2 + \vect{k}_3 = 0$
for the wavevectors $\vect{k}_i$ carried by its external lines.
Therefore the $\vect{k}_i$ can be regarded as forming a triangle.

In addition to the time-dependent logarithms $\ln|k_\ast \tau|$ which
we have studied
in~\S\S\ref{sec:resummation}--\ref{sec:factorization-thm},
it is now possible to generate logarithms involving ratios of the
form $\ln k_i/k_t$ and $\ln k_t/k_\ast$, where $k_t = k_1 + k_2 + k_3$
is the perimeter of the triangle.
(Logarithms such as $\ln k_i / k_\ast$ can be converted
to $\ln k_i / k_t + \ln k_t / k_\ast$.)
These effects were discussed by Burrage
et~al.~\cite{Burrage:2011hd,Ribeiro:2012ar},
who identified them with a response to changes in the size or shape
of the momentum triangle.
When any of these hierarchies become large, resummation of the
corresponding
logarithms will endow the three-point function
with a new, enhanced Sudakov-like structure.

We can regard $k_t$
as an analogue of the single scale $k$ in~\eqref{eq:twopf-basic}.
Indeed, the term $\ln 2k/k_\ast$
appearing in~\eqref{eq:twopf-basic} can be regarded as
$\ln k_t / k_\ast$, with $k_t = k_1 + k_2$ the `perimeter' of the
momentum 2-gon.
If the momentum $k_i$ associated with each side of the triangle is not
too different from $k_t$ then there is only one
relevant momentum scale and
situation is very similar to
that of the two-point function.
This is the `equilateral' configuration.
The converse situation occurs if one momentum is very much smaller than $k_t$.
In this case there are multiple hierarchies and computation of the three-point
function becomes more complex.
In this paper we focus only on the case where $k_i \sim k_t$ and there
is a single hierarchy.
We intend to return to the question of
multiple hierarchies in a future publication.

\paragraph{External wavefunctions.}
We divide the analysis into time-dependent terms arising from
external wavefunction factors, and those arising from the vertex integral.
The time-dependence associated with
external wavefunction factors is `unsourced', in the sense
that new contributions are not continuously generated.
(One way to regard the time-dependent
terms generated by external wavefunctions is as
a resummation of the infinite sequence of Feynman diagrams generated
by arbitrary insertions of the mass
operator $m_{\alpha\beta} \delta\phi_\alpha \delta\phi_\beta$
on each external line.)
In contrast,
time-dependence arising from the vertex integral
is actively `sourced' by ongoing
three-body interactions,
as described in~\S\ref{sec:factorization-thm}.

The factorization theorem~\eqref{eq:3pf-factorization} shows
how to deal with these unsourced terms.
We define a `bispectrum' $B_{\alpha\beta\gamma}$ by
\begin{equation}
	\langle
		\delta \phi_\alpha(\vect{k}_1)
		\delta \phi_\beta(\vect{k}_2)
		\delta \phi_\gamma(\vect{k}_3)
	\rangle
	=
		(2\pi)^3
		\delta(\vect{k}_1 + \vect{k}_2 + \vect{k}_3)
		\frac{B_{\alpha\beta\gamma}}{4 \prod_i k_i^3}
		.
	\label{eq:B-tensor-def}
\end{equation}
The factor $(4 \prod_i k_i^3)^{-1}$ is generated from
products of the $\psi_i$ which normalize each wavefunction
in~\eqref{eq:asymptotic-condition}.
The factorization theorem shows that
the `sourced' and `unsourced' pieces contribute additively to
$B_{\alpha\beta\gamma}$
and can therefore be considered separately.
Therefore
we ignore all `sourced'
contributions from the vertex integral,
which will be dealt with below.

Excluding the remaining
soft effects generated by powers of $\ln|k_\ast\tau|$
from each external wavefunction,
we obtain
a `hard' component $B^{\hard}_{\alpha\beta\gamma}$
analogous to~\eqref{eq:sigma-hard}.
Because there is no hierarchy between the external momenta,
the hard component can be computed using fixed-order perturbation theory.
Its lowest-order contribution
was calculated in Ref.~\cite{Seery:2005gb}.
Alternatively it may
be obtained by extracting the lowest-order terms
from Eqs.~\eqref{eq:LO-A-external}, \eqref{eq:LO-B-external}
and~\eqref{eq:LO-C-external}.
This contribution is analogous to the lowest-order term
$H_\ast^2 \delta_{\alpha\beta}$ in~\eqref{eq:twopf-rge-ic}.
In Appendix~\ref{subsec:3pfappendix}
we extend the results of Ref.~\cite{Seery:2005gb} by
computing the three-point function to next-order.
The subleading contributions from this calculation which are
not enhanced by soft logarithms are analogous to the next-order
term $r_{\alpha\beta}^\ast$ in~\eqref{eq:twopf-rge-ic}.
However,
for this discussion in this section we will not need an explicit
expression for $B^{\hard}_{\alpha\beta\gamma}$.

Comparison with~\eqref{eq:3pf-factorization}
shows that
the unsourced contribution to the bispectrum,
which we denote
$\hat{B}_{\alpha\beta\gamma}$, has the Taylor series expansion
\begin{equation}
	\hat{B}_{\alpha\beta\gamma}
	\simeq
		B^{\hard}_{\alpha\beta\gamma}
		-
		\Big[
			u_{\alpha\lambda}^\ast B^{\hard}_{\lambda\beta\gamma}
			+ u_{\beta\lambda}^\ast B^{\hard}_{\alpha\lambda\gamma}
			+ u_{\gamma\lambda}^\ast B^{\hard}_{\alpha\beta\lambda}
		\Big]
		\ln (-k_\ast \tau)
		+ \cdots .
	\label{eq:3pf-unsourced-taylor}
\end{equation}
We now apply the argument
of~\S\ref{sec:twopf}
to obtain the dynamical renormalization-group equation
\begin{equation}
	- \frac{\d \hat{B}_{\alpha\beta\gamma}}{\d \ln \tau}
	=
		u_{\alpha\lambda} \hat{B}_{\lambda\beta\gamma}
		+ u_{\beta\lambda} \hat{B}_{\alpha\lambda\gamma}
		+ u_{\gamma\lambda} \hat{B}_{\alpha\beta\lambda} .
	\label{eq:3pf-unsourced-rge}
\end{equation}
A suitable
initial condition can be determined from the fixed-order
perturbative expression for $B^{\hard}_{\alpha\beta\gamma}$.
As for the two-point function, the methods of Ref.~\cite{Seery:2012sx}
can be used to show that~\eqref{eq:3pf-unsourced-rge}
is equivalent to absorption
of the time-dependent logarithms in the form-factor $\Gamma_{\alpha i}$.

\paragraph{Vertex integral.}
We now return to the `sourced' contributions
generated by the vertex integral
which we introduced 
in Eq. \eqref{eq:3pf-factorization}.
Collecting terms from Appendix~\ref{subsec:3pfappendix}, we find
\begin{equation}
	B_{\alpha\beta\gamma}^{\text{div,vertex}}
	\supseteq
	- H_\ast^4
	u_{\alpha\beta\gamma}^\ast
	\ln (-k_\ast \tau)
	\sum_i k_i^3 ,
	\label{eq:3pf-div}
\end{equation}
where
a sub- or superscript `$\ast$'
denotes evaluation at the time $|k_\ast \tau| = 1$,
and $u_{\alpha\beta\gamma}$ is obtained by differentiating
the expansion tensor $u_{\alpha\beta}$, given
in Eq.~\eqref{eq:expansion-tensor},
\begin{equation}
	u_{\alpha\beta\gamma} = \partial_\gamma u_{\alpha\beta}
	=
	- \frac{V_{\alpha\beta\gamma}}{3H^2}
	+ \frac{\dot{\phi}_\alpha}{H} u_{\beta \gamma}
	+ \frac{\dot{\phi}_\beta}{H} u_{\alpha \gamma}
	+ \frac{\dot{\phi}_\gamma}{H} u_{\alpha \beta}
	- \frac{\dot{\phi}_\alpha \dot{\phi}_\beta \dot{\phi}_\gamma}{H^3} .
	\label{eq:u-three}
\end{equation}
It is symmetric under exchange of any two indices.
The superscript `div,vertex' is a reminder that Eq.~\eqref{eq:3pf-div}
includes only divergent terms from the vertex integral.
We see that
time-dependent logarithms multiply only the local shapes
$k_1^3$, $k_2^3$ and $k_3^3$,
as required by the factorization theorem.
The symmetries
of $u_{\alpha\beta\gamma}$ reduce these to the combination
$\sum_i k_i^3$.

To apply a renormalization-group analysis, we introduce a form-factor
for each shape which can be logarithmically enhanced.
In this case these are the local combinations,
\begin{equation}
	B_{\alpha\beta\gamma}^{\text{sourced}}
	=	
		a_{\alpha|\beta\gamma} k_1^3
		+ a_{\beta|\alpha\gamma} k_2^3
		+ a_{\gamma|\alpha\beta} k_3^3
		.
	\label{eq:3pf-form-factors}
\end{equation}
The form-factors $a_{\alpha|\beta\gamma}$ are symmetric under exchange of
$\beta$ and $\gamma$, but need have no other symmetries.
The full bispectrum is obtained by adding these sourced
contributions to the unsourced terms obtained
from~\eqref{eq:3pf-unsourced-rge}.

As usual,
exclusion of all terms enhanced by soft logarithms yields `hard'
components for the $a_{\alpha|\beta\gamma}$,
and
in the absence of other large hierarchies they can
also be computed using
fixed-order perturbation theory.
Comparison of~\eqref{eq:3pf-factorization},
\eqref{eq:3pf-div} and~\eqref{eq:3pf-form-factors}
shows that the $a_{\alpha|\beta\gamma}$ have a Taylor series expansion
\begin{equation}
	a_{\alpha|\beta\gamma}
	\simeq
		a^{\hard}_{\alpha|\beta\gamma}
		-
		\Big[
			u_{\alpha\lambda}^\ast a^{\hard}_{\lambda|\beta\gamma}
			+ u_{\beta\lambda}^\ast a^{\hard}_{\alpha|\lambda\gamma}
			+ u_{\gamma\lambda}^\ast a^{\hard}_{\alpha|\beta\lambda}
			+ u_{\alpha\lambda\mu}^\ast
				\Sigma^{\hard}_{\lambda\beta}
				\Sigma^{\hard}_{\mu\gamma}
		\Big]
		\ln(-k_\ast \tau)
		+ \cdots .
\end{equation}
We have replaced $H_\ast^4$ by a suitable product of
$\Sigma^{\hard}_{\alpha\beta}$ which preserves the index symmetries.
As in~\S\ref{sec:twopf} this may generate a small mismatch
at next-next-order,
which
translates to an error in the solution of the renormalization
group which is below leading-logarithm accuracy.
The renormalization-group prescription gives
\begin{equation}
	- \frac{\d a_{\alpha|\beta\gamma}}{\d \ln \tau}
	=
		u_{\alpha\lambda} a_{\lambda|\beta\gamma}
		+ u_{\beta\lambda} a_{\alpha|\lambda\gamma}
		+ u_{\gamma\lambda} a_{\alpha|\beta\lambda}
		+ u_{\alpha\lambda\mu} \Sigma_{\lambda\beta} \Sigma_{\mu\gamma} .
\end{equation}
This is the transport equation for $a_{\alpha|\beta\gamma}$ derived
in Ref.~\cite{Seery:2012sx},
where it was demonstrated that its solution
generates the form-factor $\partial^2 \phi_\alpha / \partial \phi_i^\ast
\partial \phi_j^\ast$.
Combining the sourced and unsourced
components, we reproduce the anticipated expression~\eqref{eq:deltaN-3pf}.

There is some arbitrariness in choosing an initial condition
for $a_{\alpha|\beta\gamma}$.
We are free to treat any local terms which appear
in the lowest-order correlation function $B^{\hard}_{\alpha\beta\gamma}$
\emph{either} as contributions to the unsourced part
$\hat{B}_{\alpha\beta\gamma}$,
\emph{or}
an initial condition for $a_{\alpha|\beta\gamma}$.%
	\footnote{For clarity,
	we note that this is not the same ambiguity
	discussed in footnote~\ref{footnote:factorization-scheme}
	on p.~\pageref{footnote:factorization-scheme},
	which was resolved by a choice of factorization scheme.}
Whichever choice we make the result is the same,
because the results of Ref.~\cite{Seery:2012sx} show that
this initial condition contributes additively to the final bispectrum.

\section{Applications}
\label{sec:applications}

In this section we briefly discuss two examples which illustrate the
utility of the renormalization group framework.

\subsection{Matching between quantum and classical eras}
\label{sec:matching}

In~\S\S\ref{sec:factorization}--\ref{sec:drge} we emphasized the separation
between `hard', model-independent processes by which inflationary
fluctuations are created, and `soft', model-dependent processes by
which they evolve.
The hard creation process is associated with loss of a decaying mode,
causing interference effects
to cease and the fluctuations to behave classically.
Usually,
the separate universe picture is considered as
a framework which can be used to evolve these classicalized
fluctuations~\cite{Starobinsky:1986fxa,
Sasaki:1995aw,Lyth:2005fi}.

\paragraph{Power-law corrections.}
Implicit in this point of view is a matching between the
quantum and classical parts of the calculation.
Beginning with
Polarski \& Starobinsky \cite{Polarski:1994rz,Polarski:1995jg},
several authors have noticed that this leads to a potential
ambiguity.
The issue was later studied
in more detail by Leach \& Liddle \cite{Leach:2000yw}
and has recently been revisited by Nalson et~al.~\cite{Nalson:2011gc}.

The problem can be stated simply.
In Eqs.~\eqref{eq:twopf-fixed} and~\eqref{eq:twopf-basic},
decaying power-law corrections which scale like positive powers of
$|k\tau|$ have been neglected.
These already occur at lowest order in slow roll, where the
equal-time two-point function at an arbitrary conformal time
$\tau$ can be written
\begin{equation}
	\langle
		\delta \phi_\alpha(\vect{k}_1)
		\delta \phi_\beta(\vect{k}_2)
	\rangle_\tau
	=
	(2\pi)^3
	\delta(\vect{k}_1 + \vect{k}_2)
	\frac{H_\ast^2}{2k^3}
	(1 + k^2 \tau^2) .
\end{equation}
The term $k^2 \tau^2$ is unity at horizon crossing but decays exponentially
fast outside the horizon, leading to evolution of the typical
fluctuation amplitude $\langle \phi^2 \rangle^{1/2}$ by a factor of
$\surd 2$
between horizon crossing and a few e-folds later.
By itself this is not significant because
one should not think of a
measurable
classical fluctuation with fixed
amplitude until the decaying mode is lost, which corresponds to
$k^2 \tau^2$ becoming negligible~\cite{Lyth:1984gv}.
However, it does demonstrate that
(from this point of view)
classical reasoning
can not be used
until
at least a few e-folds outside the horizon.
A similar issue will exist
for higher $n$-point functions.

\begin{figure}

	\hfill
	\subfloat[][$H$ at matching time\label{fig:matching-vary}]{
		\includegraphics[width=4.5cm]{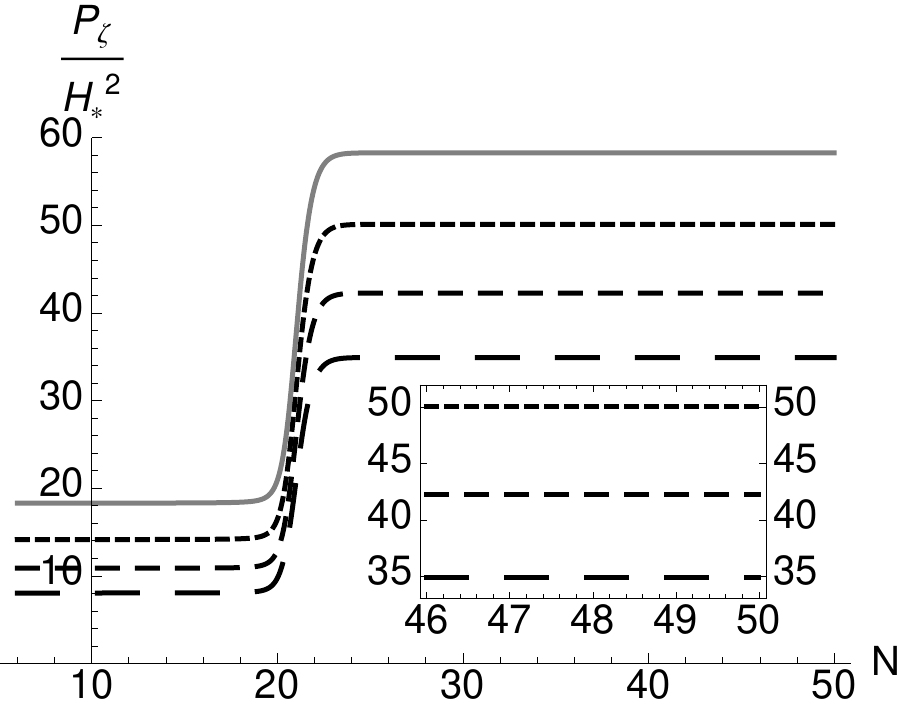}
	}
	\hfill
	\subfloat[][$H$ at horizon-crossing\label{fig:matching-fixed}]{
		\includegraphics[width=4.5cm]{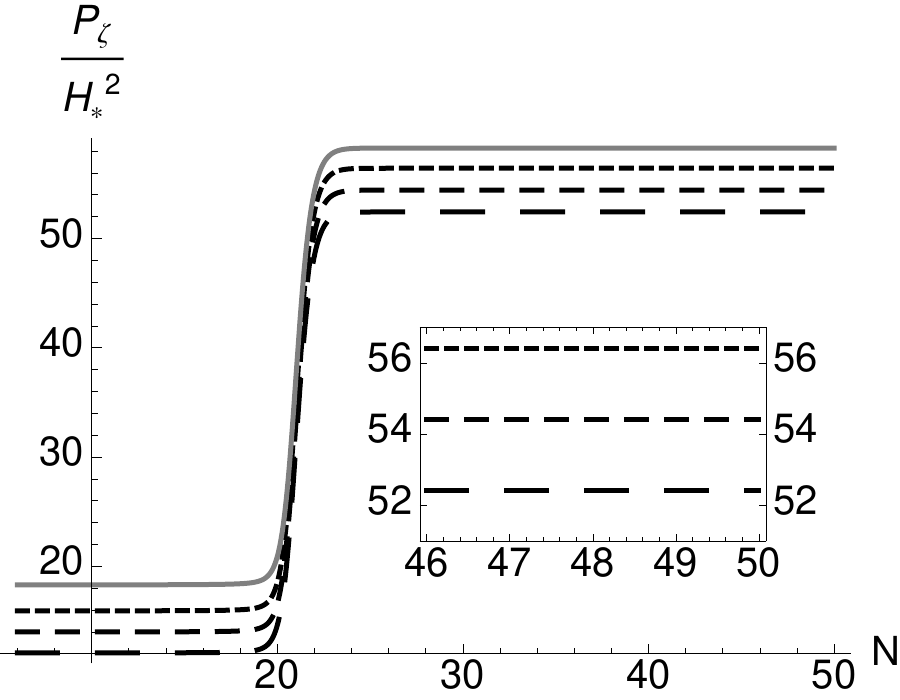}
	}
	\hfill
	\subfloat[][$\NLO$ initial condition\label{fig:matching-nlo}]{
		\includegraphics[width=4.5cm]{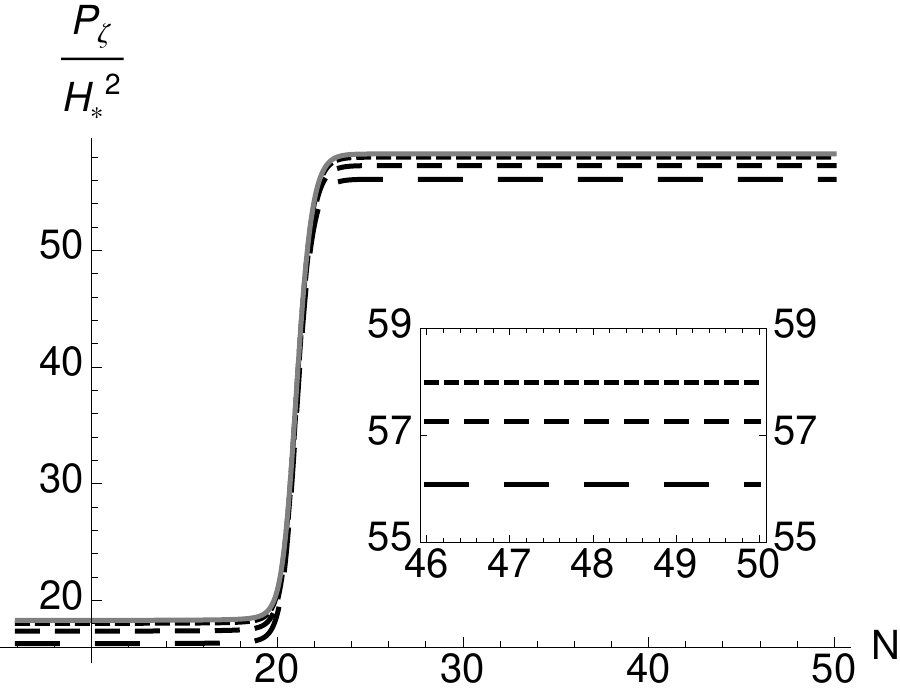}
	}
	\caption{Impact of different matching prescriptions on the
		final value of the power spectrum of $\zeta$ in the
		double-quadratic model~\eqref{eq:double-quadratic}.
		The quoted value is for the scale which passed outside
		the horizon with field values
		$\varphi = 8.2 \Mp$ and $\chi = 12.9 \Mp$,
		labelled $N=0$ on the figures.
		In each figure, the short dashed line corresponds to
		beginning the classical calculation at $N=2$ e-folds after
		horizon exit; the medium dashed line at
		$N=4$ e-folds; and the long dashed line at $N=6$ e-folds.
		The evolution generated by the dynamical renormalization
		group is the grey solid line.
		In Fig.~\ref{fig:matching-vary}, initial conditions
		are chosen using $\Sigma_{\alpha\beta} = H^2_N \delta_{\alpha\beta}$,
		with $H_N$ the value of the Hubble rate at the matching time.
		In Fig.~\ref{fig:matching-fixed} the matching is performed
		using $\Sigma_{\alpha\beta} = H_k^2 \delta_{\alpha\beta}$,
		where $H_k$ is the value of the Hubble rate at horizon
		crossing.
		In Fig.~\ref{fig:matching-nlo} the full $\NLO$
		initial condition~\eqref{eq:twopf-fixed} is used,
		setting $k_\ast = k$ to minimize the $\ln k/k_\ast$
		terms.
		In each plot the inset panel
		shows a close-up of the last four e-folds of evolution.
		\label{fig:matching}}

\end{figure}

\paragraph{Choice of matching surface.}
If the separate universe method is regarded in classical terms
we cannot apply it at the moment of horizon exit.
But
Eqs.~\eqref{eq:twopf-fixed} and~\eqref{eq:twopf-basic} also show that we
cannot wait too long before switching to the
classical calculation, otherwise this quantum initial condition will be
invalidated by the growing logarithm
$\ln(-k_\ast \tau)$.
At the earliest,
we could perhaps consider the fluctuations to be approximately classical
when they are $\sim 2$ e-folds outside the horizon, making
$k^2 \tau^2 \sim \e{-4} \approx 0.018$.
In Fig.~\ref{fig:matching} we show the effect of different matching
prescriptions in the double-quadratic model~\eqref{eq:double-quadratic},
using the $k$-scale, initial conditions and model
parameters of Fig.~\ref{fig:dqpowerspectrum}.
We plot the observable quantity $P_\zeta$, which can be obtained
from $\Sigma_{\alpha\beta}$ by a gauge
transformation~\cite{Sasaki:1995aw}.
We compute $\Sigma_{\alpha\beta}$ numerically using
Eq.~\eqref{eq:twopf-rge}
and making use of the slow-roll approximation.

In Fig.~\ref{fig:matching-vary} we
perform the matching at $N = 2$, $N = 4$ and $N = 6$
e-folds after horizon crossing.
Initial conditions for the classical evolution are set
using the lowest-order approximation
with $\Sigma_{\alpha\beta} = H^2_N
\delta_{\alpha\beta}$, where $H_N$ is the value
of the Hubble rate on the matching surface.
The decaying power-law term $k^2 \tau^2$ is ignored
in setting these initial conditions.%
	\footnote{In~\S\ref{sec:factorization-thm} we argued that the
	decaying mode began at $\Or(k\tau)^3$,
	and therefore this term must originate from the growing
	mode which dominates the subsequent classical evolution.
	Hence, its contribution could be included in a self-consistent
	calculation.
	This possibility was suggested by
	Polarski \& Starobinsky~\cite{Polarski:1994rz} and
	Lalak et~al.~\cite{Lalak:2007vi}.

	We have explained in~\S\ref{sec:twopf} that decaying
	terms of the form
	$k^2 \tau^2$ do not contribute the {\DRGE} evolution,
	and for this reason we have not retained it when setting
	initial conditions.}

In Fig.~\ref{fig:matching-fixed} the matching is also performed
at $N = 2$, $N = 4$ and $N = 6$ e-folds after horizon crossing, but
with the initial condition $\Sigma_{\alpha\beta} = H_k^2
\delta_{\alpha\beta}$ where $H_k$ is the Hubble rate at
the moment of horizon exit.
Finally, in Fig.~\ref{fig:matching-nlo} we use the full
$\NLO$ expression~\eqref{eq:twopf-fixed} to provide an initial condition
at each matching surface.
In each figure, the solid grey line shows the evolution computed
using the dynamical renormalization group (see below).

With the full $\NLO$ initial condition
(Fig.~\ref{fig:matching-nlo}) the discrepancy between the asymptotic
value of $P_\zeta$ computed using each prescription, and also
the {\DRGE} evolution,
is a few percent.
In Fig.~\ref{fig:matching-fixed} the discrepancy
between different matching prescriptions
is of order $3\%$--$7\%$.
The discrepancy between the $N=6$ prescription and the {\DRGE} evolution
is as large as $10\%$. 
In Fig.~\ref{fig:matching-vary} the discrepancies are very large,
between $10\%$ and $40\%$.

In Fig.~\ref{fig:matching-nlo}, the
expression~\eqref{eq:twopf-basic} is used to provide an estimate of
$\Sigma_{\alpha\beta}$ at the matching surface.
One might have expected
this estimate to be accurate because it is computed to next-order,
and the slow-roll parameters are individually small.
Therefore we might also have expected
the final difference in $P_\zeta$ to be
negligible,
because the separate universe method is supposed to be independent
of the time at which initial conditions are set.
The discrepancy arises because, with matching performed at $N > 0$,
the fixed-order expression is used to
evolve $\Sigma_{\alpha\beta}$
up to $N$ e-folds.
With the matching performed at $N = 0$, all-orders resummation is used.
Hence,
the non-negligible
difference between each line in Fig.~\ref{fig:matching-nlo}
measures the discrepancy between the resummed and fixed-order
calculations.
It would be even more significant if these first $N$ e-folds
coincided with stronger evolution, such as that occuring
near the spike in Fig.~\ref{fig:dqpowerspectrum}.

\paragraph{{\DRGE} analysis.}
As observations improve,
an ambiguity of order $10\%$ will certainly be important.
Therefore we would like to understand which of these prescriptions, if any,
is correct.
This question cannot be answered within a classical framework.
The resolution is that there is no requirement to divide the calculation
into `quantum' and `classical' eras joined by a matching condition.
Although this is a useful picture which guides our thinking,
the calculation can be carried out entirely within the framework of
quantum field theory and
yields a unique answer.
Indeed,
as we have seen in~\S\S\ref{sec:resummation}--\ref{sec:drge},
the relevant division is not into `quantum' and `classical'
effects, but `hard' and `soft' processes.
It simply happens that,
because
interference effects are absent in the soft superhorizon
evolution,
it can be described at tree-level
using the classical equations of motion.
In the quantum field theory calculation there is no arbitrary
matching between `quantum' and `classical' eras, but only
the floating factorization scale. 
Changes in the factorization scale leave all correlation functions
invariant.

The renormalization-group approach we have described allows
each correlation function to be calculated purely using the
methods of quantum field theory
and therefore provides a unique resolution of this matching
ambiguity.
The prescription can be extracted from Eqs.~\eqref{eq:twopf-rge}
and~\eqref{eq:twopf-rge-ic}.
Positive powers of $k\tau$ harmlessly decay and do not contribute
to the Taylor series in $\ln(-k_\ast \tau)$ at late times.
When reversing-engineering this Taylor expansion,
the renormalization group evolution should begin at the
expansion point $k_\ast$ and the initial condition should be the
constant term in the Taylor series.
As we argued in~\S\ref{sec:twopf}, to obtain the most accurate
answer we should evaluate~\eqref{eq:twopf-rge-ic}
at the horizon-crossing time for $k$.
Therefore we should simultaneously begin the renormalization-group
evolution at this time.
Hence the prescription provided by the {\DRGE} is to use the initial
condition $\Sigma_{\alpha\beta} \approx H_k^2 \delta_{\alpha\beta}$
but to begin the `classical' evolution at $N=0$.
If desired, higher-order corrections
can be retained to give a more accurate estimate
of the initial value. This has been
done to obtain the grey lines in Fig.~\ref{fig:matching},
although
the change due to inclusion of $\NLO$ terms in the {\DRGE} initial
condition is only of order $0.3\%$.

The prescription provided by the dynamical renormalization group
is incompatible with classical intuition, but by itself this is
not important because no classical phase of evolution is being invoked in
this calculation.

\subsection{Nontrivial field-space metric}
\label{sec:nontrivial-metric}

Elliston et~al.~\cite{Elliston:2012he} recently calculated the
inflationary bispectrum
generated by a $\sigma$-model Lagrangian with nontrivial field-space
metric.
The action is
\begin{equation}
	S = \frac{1}{2} \int \d^3 x \, \d t \; a^3 \Big[
		\Mp^2 R
		- \Gmetric_{\alpha\beta} \partial_a \phi^\alpha \partial^a \phi^\beta
		- 2 V
	\Big]
	\label{eq:sigma-model}
\end{equation}
where $V$ is a potential and $\Gmetric_{\alpha\beta}$ is an arbitrary
symmetric matrix which can be interpreted as a metric.

To preserve manifest covariance with respect to $\Gmetric_{\alpha\beta}$
it is helpful to
describe fluctuations using
a variable which obeys a covariant
transformation law~\cite{Saffin:2012et,Gong:2011uw}.
Gong \& Tanaka
gave a prescription for rewriting inflationary perturbation theory in terms
of such a variable, which they denoted $Q^\alpha$,
and computed the action for field fluctuations to
third-order~\cite{Gong:2011uw}.

\paragraph{Two-point function.}
The equal-time two-point function for $Q^\alpha$
had already been obtained
to leading order by Sasaki \& Stewart~\cite{Sasaki:1995aw}.
The analogue of~\eqref{eq:twopf-fixed} was
computed by Nakamura \& Stewart~\cite{Nakamura:1996da}.
There is a correction to the mass matrix
and a more complex structure due to preservation of manifest covariance,%
	\footnote{In Ref.~\cite{Nakamura:1996da} the parallel propagator
	matrices ${\Pimetric^\alpha}_i$ were omitted.
	Because the metric is determined by parallel transport,
	${\Pimetric^\alpha}_i {\Pimetric^\beta}_j \Gmetric^{ij}
	= \Gmetric^{\alpha\beta}$.
	Therefore, provided
	we carefully keep track of
	the evaluation point for the metric,
	these matrices are not needed explicitly at lowest-order.}
\begin{equation}
\begin{split}
	\langle
		Q^\alpha(\vect{k}_1)
	&
		Q^\beta(\vect{k}_2)
	\rangle
	\supseteq
	(2\pi)^3 \delta(\vect{k}_1 + \vect{k}_2)
	\frac{H_\ast^2}{2k^3}
	{\Pimetric^\alpha}_i {\Pimetric^\beta}_j
	\\
	& \times \left\{
		\Gmetric^{ij} \left[
			1 + 2 \varepsilon_\ast \left(
				1 - \EulerGamma - \ln \frac{2k}{k_\ast}
			\right)
		\right]
		+
		2 \umetric^{ij} \left[
			2 - \ln (-k_\ast \tau) - \ln \frac{2k}{k_\ast}
			- \EulerGamma
		\right]
	\right\} .
	\label{eq:2pf-metric}
\end{split}
\end{equation}
In accordance with our usual convention we have neglected
decaying power-law corrections.
Greek indices $\alpha$, $\beta$, {\ldots} label
the tangent space at field-space coordinate $\phi^\alpha(\tau)$,
whereas indices $i$, $j$, {\ldots} label
the tangent space at coordinate $\phi^\alpha(\tau_\ast)$, where
$\tau_\ast$ is the horizon-crossing time for a reference scale $k_\ast$.
The expansion tensor $\umetric_{ij}$ satisfies
\begin{equation}
	\umetric_{ij}
	=
	- \frac{V_{ij}}{3H^2}
	+ \frac{1}{3H^2} \frac{1}{a^3} \frac{\D}{\d t} \left(
		\frac{a^3}{H} \dot{\phi}_i \dot{\phi}_j
	\right)
	+ \frac{1}{3} \RiemannMetric_{imnj}
		\frac{\dot{\phi}^m}{H} \frac{\dot{\phi}^n}{H} .
	\label{eq:utensor-metric}
\end{equation}
The indices $i$, $j$ imply that~\eqref{eq:utensor-metric}
is evaluated at time $\tau_\ast$.
The same applies for $\Gmetric_{ij}$.
The derivative $\D / \d t$ is
$\dot{\phi}^\alpha \grad_\alpha$, where $\grad_\alpha$ is the
covariant derivative compatible with $\Gmetric_{\alpha\beta}$
and $\RiemannMetric_{\alpha\beta\gamma\delta}$ is
the corresponding Riemann tensor.
Finally, the parallel propagator ${\Pimetric^\alpha}_i$
expresses parallel transport along the inflationary trajectory
in field-space between $\phi^\alpha(\tau_\ast)$ and $\phi^\alpha(\tau)$,
\begin{equation}
	{\Pimetric^\alpha}_i =
	\TimeOrder \exp \Big(
		- \int_{\tau_\ast}^\tau \d \eta \; \Gamma^{\alpha'}_{\beta' \gamma'}
		\d \phi^{\beta'}
	\Big)
	\delta^{\gamma'}_i .
\end{equation}
Indices $\alpha'$, $\beta'$, {\ldots} label the tangent space associated
with the integration time $\eta$, and $\TimeOrder$ is the time-ordering
operator.
For all details we refer to Elliston et~al.~\cite{Elliston:2012he}.

\paragraph{Three-point function.}
The equal-time three-point function for $Q^\alpha$ was obtained
up to $\Or(\dot{\phi}/H)^2$
in Ref.~\cite{Elliston:2012he}. It is
\begin{equation}
	\langle
		Q^\alpha(\vect{k}_1)
		Q^\beta(\vect{k}_2)
		Q^\gamma(\vect{k}_3)
	\rangle
	\supseteq
	(2\pi)^3 \delta(\vect{k}_1 + \vect{k}_2 + \vect{k}_3)
	\frac{H_\ast^4}{4 \prod_i k_i^4}
	{\Pimetric^\alpha}_i {\Pimetric^\beta}_j {\Pimetric^\gamma}_k
	A^{ijk}(N) ,
	\label{eq:3pf-metric}
\end{equation}
where $A^{ijk}$ transforms as a three-tensor in the tangent
space at time $\tau_\ast$ and is defined by
\begin{equation}
\begin{split}
	A^{ijk} = \mbox{}
	&
	\frac{1}{\Mp^2} \frac{\dot{\phi}^i}{H} \Gmetric^{jk}
	\bigg(
		\frac{k_1}{2} \vect{k}_2 \cdot \vect{k}_3
		- 2 \frac{k_2^2 k_3^2}{k_t}
	\bigg)
	\\ & \mbox{}
	+ \frac{4}{3} {\RiemannMetric}^{i(jk)m} \frac{\dot{\phi}_m}{H}
	\bigg[
		k_1^3 \Big(
			\EulerGamma 
					+ \ln(-k_\ast \tau)+ \ln \frac{k_t}{k_\ast}
		\Big)
		- k_1^2 k_t
		+ \frac{k_1^2 k_2 k_3}{k_t}
	\bigg]
	\\ & \mbox{}
	+ \frac{1}{3} \grad^{(i} {\RiemannMetric}^{j | mn | k)}
		\frac{\dot{\phi}_m}{H} \frac{\dot{\phi}_n}{H}
	\bigg[
	-	k_1^3 \Big(
			\ln(-k_\ast \tau) + \ln \frac{k_t}{k_\ast} + \EulerGamma + \frac{1}{3}
		\Big) 
		+ \frac{4}{9} k_t^3 - k_t K^2
	\bigg]
	\\ & \mbox{}
	- \frac{4}{3} \grad^n {\RiemannMetric}^{i(jk)m}
		\frac{\dot{\phi}_m}{H} \frac{\dot{\phi}_n}{H}
	\bigg[
		\frac{k_1^3}{2} \Big(
			\ln^2(-k_\ast \tau) - \EulerGamma^2 + \frac{\pi^2}{12} - \Big[
				2 \EulerGamma + \ln \frac{k_t}{k_\ast}
			\Big] \ln \frac{k_t}{k_\ast}
		\Big)
	\\ & \mbox{}
	\hspace{38mm}
		+ k_1^2 k_t \Big(
			\ln \frac{k_t}{k_\ast} + \EulerGamma - 1
		\Big)
		- \frac{k_1^2 k_2 k_3}{k_t} \Big(
			\EulerGamma + \ln \frac{k_t}{k_\ast}
		\Big)
	\bigg]
	\\ & \mbox{}
	+ \text{cyclic} .
	\label{eq:A-metric}
\end{split}
\end{equation}
We have used $k_t = k_1 + k_2 + k_3$ and $K^2 = \sum_{i < j} k_i k_j =
k_1 k_2 + k_1 k_3 + k_2 k_3$.
As in~\eqref{eq:utensor-metric}, all time-dependent
quantities are evaluated at
the horizon-crossing time for $k_\ast$.

Eqs.~\eqref{eq:3pf-metric}--\eqref{eq:A-metric} are
considerably more complicated than their
flat field-space counterparts.%
	\footnote{The first line in~\eqref{eq:A-metric}
	covariantizes the lowest-order
	flat-space computation given in Ref.~\cite{Seery:2005gb}.
	If next-order corrections were to be kept they would covariantize the
	computation described in Appendix~\ref{appendix:3pf}.}
Growing terms involving $\ln(-k_\ast \tau)$ are already
present at lowest nontrivial order in $\dot{\phi}/H$.
At $\Or(\dot{\phi}/H)^2$ a double-logarithmic term involving
$\ln^2 (-k_\ast \tau)$
appears.
In flat field-space,
growing terms appear only
at $\Or(\dot{\phi}/H)^3$ and double logarithms do not occur until
even higher orders.
In combination with the propagator terms ${\Pimetric^\alpha}_i$ these
make the time dependence of~\eqref{eq:3pf-metric} appear
extremely complicated.

\paragraph{Dynamical renormalization group analysis.}
Eqs.~\eqref{eq:2pf-metric}, \eqref{eq:3pf-metric} and~\eqref{eq:A-metric}
present a case study in the use of 
renormalization group
techniques to extract a Sudakov-like structure from
logarithmic divergences.

The Taylor theorem on a curved manifold is
\begin{equation}
	{A^{\alpha \cdots}}_{\beta \cdots}(x) =
	{\Pimetric^\alpha}_{\alpha'}(x,x') \cdots
	{\Pimetric_\beta}^{\beta'}(x,x') \cdots
	\left[
		{A^{\alpha' \cdots}}_{\beta' \cdots}\Big|_{x'}
		+
		\frac{\D}{\d \lambda}
		{A^{\alpha' \cdots}}_{\beta' \cdots}
		\bigg|_{x'}
		\delta \lambda
		+
		\cdots
	\right] ,
	\label{eq:curved-taylor-theorem}
\end{equation}
where the omitted terms are $\Or(\delta\lambda^2)$.
To write~\eqref{eq:curved-taylor-theorem} we have
assumed a smooth path connecting $x$ and $x'$, with
${\Pimetric^\alpha}_{\alpha'}(x,x')$ the parallel propagator
along this path and $\D/\d \lambda$ the corresponding
parametric derivative.
As in~\S\ref{sec:drge}, the Taylor theorem~\eqref{eq:curved-taylor-theorem}
is our principal tool in the renormalization-group analysis.
The path connecting $x$ and $x'$ will be the inflationary trajectory,
and the parametric derivative along it will be $\D / \d N$.

The argument now parallels~\S\ref{sec:drge} precisely.
Comparing Eq.~\eqref{eq:2pf-metric} with~\eqref{eq:curved-taylor-theorem}
allows us to conclude
\begin{equation}
	\frac{\D}{\d N} \Sigma^{\alpha\beta}\bigg|_{-k_\ast \tau = 1} =
	2 H^2 \umetric^{\alpha\beta} \Big|_{-k_\ast \tau = 1} .
	\label{eq:2pf-metric-proto-rge}
\end{equation}
The arbitrariness of $k_\ast$ allows this to be promoted to a differential
equation.
Replacing $H_\ast$ by suitable copies of $\Sigma_{\alpha\beta}$,
as in~\S\ref{sec:drge}, we find
\begin{equation}
	\frac{\D \Sigma^{\alpha\beta}}{\d N}
	=
	{\umetric^\alpha}_\gamma \Sigma^{\gamma\beta}
	+ {\umetric^\beta}_\gamma \Sigma^{\gamma\alpha} .
	\label{eq:2pf-metric-rge}
\end{equation}

A similar analysis can be performed for the three-point function.
Eq.~\eqref{eq:A-metric} shows that all time-dependent logarithms
multiply only local momentum shapes, in agreement with the
factorization theorem of~\S\ref{sec:factorization-thm}.
(This is not altered by the presence of a nontrivial metric.)
Repeating the argument of~\S\ref{sec:threepf},
we conclude
\begin{equation}
	\frac{\D a_{\alpha|\beta\gamma}}{\d N}
	=
		{\umetric_\alpha}^\lambda a_{\lambda|\beta\gamma}
		+ {\umetric_\beta}^\lambda a_{\alpha|\lambda\gamma}
		+ {\umetric_\gamma}^\lambda a_{\alpha|\beta\lambda}
		+ {\umetric_\alpha}^{\lambda\mu}
			\Sigma_{\lambda\beta} \Sigma_{\mu\gamma} ,
	\label{eq:3pf-metric-rge}
\end{equation}
where $\umetric_{\alpha\beta\gamma}$ satisfies
\begin{equation}
	\umetric_{\alpha\beta\gamma} = \grad_{(\alpha} \umetric_{\beta\gamma)}
		+ \frac{1}{3} \left(
			\grad_{(\alpha} {\RiemannMetric}_{\beta|\lambda\mu|\gamma)}
				\frac{\dot{\phi}^\lambda}{H}
				\frac{\dot{\phi}^\mu}{H}
			- 4 {\RiemannMetric}_{\alpha(\beta\gamma)\lambda}
				\frac{\dot{\phi}^\lambda}{H}
		\right) .
	\label{eq:3pf-utensor}
\end{equation}

\paragraph{Jacobi equation.}
Elliston et~al.~deduced equations
equivalent to~\eqref{eq:2pf-metric-rge}
and~\eqref{eq:3pf-metric-rge}--\eqref{eq:3pf-utensor}
from the evolution of a covariant connecting vector
between two adjacent inflationary trajectories~\cite{Elliston:2012he},
and demonstrated that
(when solved perturbatively) these transport
equations reproduce the
`divergences' involving $\ln(-k_\ast \tau)$ and $\ln^2 (-k_\ast \tau)$
in~\eqref{eq:2pf-metric} and~\eqref{eq:3pf-metric}--\eqref{eq:A-metric}.
The analysis in this section shows that this agreement is not
restricted to the lowest powers of $\ln(-k_\ast \tau)$ but will extend to the
leading-logarithm terms at all orders.

Comparing the analysis of Elliston et~al.~\cite{Elliston:2012he}
with the discussion given in this section highlights
the advantages and disadvantages of the renormalization-group technique.
Eqs.~\eqref{eq:2pf-metric-rge}--\eqref{eq:3pf-utensor} involve combinations
of the Riemann curvature tensor
and its derivative which one could not expect to
obtain by accident. The analysis of Elliston et~al.~\cite{Elliston:2012he}
provides an explanation for the precise combinations which appear, and
shows how they can be understood in geometrical terms.
On its own, however, it does not demonstrate agreement to
higher orders in $\ln(-k_\ast \tau)$.
This is important, because as we argued
in~\S\ref{sec:resummation}
and was demonstrated explicitly in
Fig.~\ref{fig:matching-nlo},
contributions
from all powers of $\ln(-k_\ast \tau)$ are required to
obtain an accurate prediction.

In contrast,
by combining
the dynamical renormalization-group framework
with a factorization theorem,
we derived
a correct set of transport equations
directly from the linear $\ln(-k_\ast \tau)$ `divergences'.
(As we explained in footnote~\ref{footnote:leading-log}
on p.~\pageref{footnote:leading-log},
there is no need to take special measures to account for the
double-logarithmic term.)
Therefore the {\DRGE} procedure gives an algorithmic method
to obtain a correct evolution equation, valid to
leading-logarithm order.
But conversely it gives no information about the physical interpretation
of this evolution, in the same way that the successful calculation of
a negative $\beta$-function in QCD did not directly give an interpretation
of asymptotic freedom as a consequence of antiscreening due to
gluonic fluctuations.

In this case Elliston et~al.~were able to guess the correct
physical interpretation of Eqs.~\eqref{eq:2pf-metric-rge}
and~\eqref{eq:3pf-metric-rge}--\eqref{eq:3pf-utensor}.
But it is easy to imagine that, in more complex circumstances
(perhaps where the effect of time-dependence from loop corrections
is included), it would no longer be possible to
guess the correct
evolution equation using simple physical arguments.
In such cases the dynamical renormalization group provides
a straightforward and algorithmic approach.
In canonical models it requires the extra
complexity of a next-order calculation to obtain
the first contributions to the linear term in the Taylor expansion
[that is, the term proportional to $\ln(-k_\ast \tau)$].
For the nontrivial metric, the relevant new terms involving
the Riemann curvature appear already at lowest order.

\section{Summary and discussion}
\label{sec:discussion}

The separate universe picture is a powerful
method to compute the superhorizon evolution of inflationary
correlation functions. Our main result is a
derivation
of this  method from the renormalization 
group technique.
We have focused on the bispectrum, where the simplest
discussion can be given, but it is clear what would be involved
in extending the analysis to higher $n$-point functions.

The main steps in the argument are simple.
We first show that the vertex integral obeys a `factorization'
principle: to all orders in the slow-roll expansion,
time-dependent logarithms appear only in combination with a few
fixed functions of the external momenta.
The second step is to derive renormalization-group equations
for the coefficients of these fixed functions.
These equations turn out to be the transport equations
for the two- and three-point functions, which were shown
in Ref.~\cite{Seery:2012sx}
to be equivalent to the
separate-universe
Taylor expansion
introduced by
Lyth \& Rodr\'{\i}guez.
Therefore, in combination,
the factorization principle and RGE approach
reproduce the separate universe method.
Technically our argument demonstrates this in the leading-logarithm
approximation.

Our result suggests a systematic method to
incorporate more than the
lowest-order contribution to each correlation function.
Burgess et~al.~\cite{Burgess:2009bs}
studied time-dependent effects from loop-level corrections to the power
spectrum, making use of the same dynamical renormalization group
methods employed in~\S\ref{sec:drge}.
More recently,
Collins et~al.~\cite{Collins:2012nq} studied effective field theories
in a nonequilibrium or time-dependent setting, finding
nontrivial
contributions to the time-dependence after integrating out
a massive field.
To perform calculations
analogous to these for $n$-point functions with $n \geq 3$
during an inflationary phase
it would be necessary to proceed as
in~\S\ref{sec:drge}---first, proving a suitable factorization
theorem; and second, constructing the relevant renormalization-group
equation.
(For the two-point function,
factorization is usually a much more straightforward question
due to the absence of vertex integrals.)

The factorization property admits a clear physical interpretation
which is related to the usual division of separate-universe
calculations into a quantum contribution, generated near horizon-crossing,
followed by classical evolution.
The renormalization-group analysis
gives a sharper version of this division.
Fluctuations are created as they inflate beyond the Hubble length,
on a timescale which is rapid compared to their subsequent evolution
and which is insensitive to any model-dependent parameters.
Factorization divides the physics of the fluctuations into
two parts:
a `hard' model-independent subprocess associated with the creation event,
and `soft', model-dependent
interactions which subsequently dress the hard subprocess.
One way to understand why correlation functions factorize
into self-contained parts associated with these `hard' and `soft'
processes is that the disparity in timescales prevents
interference effects
between the two.

Because we work to tree-level, it follows that the soft physics can
be described using the classical equations of motion.
Had we chosen to compute beyond tree-level, 
the applicability of classical equations of motion might be lost
but the division into hard and soft physics would remain.
Therefore,
as a conceptual tool, the hard/soft division takes the place of the usual
quantum/classical separation, allowing correlation functions
to be calculated without invoking any classical phase at all.
Naturally, this does not absolve us of the obligation to explain why
the correlation functions obtained in this way can
subsequently
be re-interpreted
as the correlations of a classical stochastic process---%
that is, why we see fluctuations which behave classically.
The renormalization-group method has nothing new to say about
this difficulty.

In the course of our analysis we treated all fields as perturbatively
massless. (That is, mass terms are treated as interactions.)
This has two important consequences.
First, all $k$-dependence occurs as integer power laws
modified by logarithms.
For example, $k^{3 + \delta} \approx k^3 [ 1 + \delta \ln k +
\Or(\delta^2) ]$;
it is the infinite sum of logarithms which reproduces a
power-law scaling.
This restriction to
integer powers of $k$
was helpful in proving the factorization theorem
in~\S\ref{sec:factorization-thm},
which depended on the existence of a `gap' in the asymptotic expansion
of each wavefunction
in integer powers of $k$, Eq.~\eqref{eq:asymptotic-condition}.
In our applications, the coefficient $\delta$ is of order a slow-roll
parameter,
so in a strict mathematical
interpretation we are restricted to scenarios in which
the slow-roll parameters are small
throughout the evolution.
Second, the
appearance of time-dependent logarithmic `divergences' of the
form $\ln(-k_\ast \tau)$ is itself a consequence of the massless
approximation. Incorporation of masses would provide a cutoff which
prevents these terms growing unboundedly.

Finally, it should be remembered that
our analysis does not apply to all theories.
Scenarios which contain other distinguished scales need not
satisfy the conditions of the factorization theorem proved
in~\S\ref{sec:factorization-thm}.
To obtain this theorem
it was assumed that asymptotic information about the
wavefunctions provided all necessary data about the superhorizon
regime.
Where this assumption fails,
other combinations of the external momenta
could receive significant enhancements.
This corresponds to the well-known restriction that, in applying
separate-universe arguments, all relevant scales should first
be outside the horizon.

\end{fmffile}

\acknowledgments
We would like to thank
Enrico Pajer for discussions, and
Joseph Elliston,
Richard Holman and 
Andrew J. Tolley for comments on a draft version of this paper.
We would like to thank the referee for 
drawing our attention to Ref. \cite{Sasaki:1998ug}.

MD is financially supported by
Funda\c{c}\~{a}o para a Ci\^{e}ncia e a Tecnologia 
[grant SFRH/BD/ 44958/2008].
RHR acknowledges financial support
from Funda\c{c}\~{a}o para a Ci\^{e}ncia e a Tecnologia 
[grant SFRH/BD/35984/2007]
in the early stages of this work.
RHR also acknowledges the hospitality
of the University of Sussex.
DS acknowledges support from the Science and Technology Facilities Council
[grant number ST/I000976/1]
and the Leverhulme Trust.
DS acknowledges that
this material is based upon work supported in part by the
National Science Foundation under Grant No. 1066293 and
the hospitality of the Aspen Center for Physics.
The research leading to these results has received funding from
the European Research Council under the European Union's
Seventh Framework Programme (FP/2007--2013) / ERC Grant
Agreement No. [308082].

DS would like to thank Misao Sasaki for the
suggestion (made long ago at the workshop `Non-Gaussianity from
Inflation' held at DAMTP, Cambridge, in April 2006) that the
methods of Refs.~\cite{Gong:2001he,Gong:2002cx}
could be used
to determine the superhorizon evolution of the three-point function.

\vspace*{1cm}
\hrule

\appendix

\section{Three-point function at next-to-leading order}
\label{appendix:3pf}

In this Appendix we
collect the components of the three-point function up to $\NLO$ in 
the slow-roll expansion,
and briefly sketch some details of its calculation.
For full details of the `in--in' or `Schwinger--Keldysh'
method used to obtain
inflationary correlation functions we refer to the
literature~\cite{Maldacena:2002vr,Weinberg:2005vy,Seery:2007we,Chen:2010xka,
Koyama:2010xj}.

\subsection{Propagator}
\label{appendix:propagator}

We require the second- and third-order action for a collection of
light, canonically-normalized scalar fields coupled to gravity.
The second-order action has already been given in~\eqref{eq:quadratic-action}.
To obtain the propagator $G$, defined in conformal time by
\begin{equation}
	G^{\alpha\beta}(\vect{x}, \tau ; \tilde{\vect{x}}, \tilde{\tau})
	= \langle
		\delta \phi^\alpha(\vect{x}, \tau)
		\delta \phi^\beta( \tilde{\vect{x}}, \tilde{\tau})
	\rangle ,
\end{equation}
we should invert the quadratic structure appearing in the action.
That gives
\begin{equation}
	\left\{
		\delta_{\alpha\beta}
		\left(
			\frac{\partial^2}{\partial \tau^2}
			+ 2 \frac{\partial \ln a}{\partial \tau} \frac{\partial}{\partial \tau}
			- \partial^2
		\right)
		+ a^2 m_{\alpha\beta}
	\right\}
	G^{\beta\gamma}
	=
	- \frac{\im}{a^2}
	\delta^\gamma_\alpha \
	\delta(\vect{x} -  \tilde{\vect{x}})
	\delta(\tau - \tilde{\tau}) .
\end{equation}
Because spatial translations and rotations are unbroken by the
de Sitter background, $G$ can depend only on the relative
displacement $\vect{x} - \tilde{\vect{x}}$.
Diagonalizing this by passing to Fourier space,
\begin{equation}
	G^{\alpha\beta}(\tau,\tilde{\tau};\vect{x}-\tilde{\vect{x}})
	=
	\int \frac{\d^3 k}{(2\pi)^3} \ 
	G^{\alpha\beta}_k(\tau,\tau')
	\e{\im \vect{k}\cdot(\vect{x} - \tilde{\vect{x}})}
	,
\end{equation}
and introducing a dimensionless time $y= k \tau = - k/aH$, we find
\begin{equation}
	\left\{
		\delta_{\alpha\beta}
		\left(
			\frac{\d^2}{\d y^2}
			+ 2 \frac{a'}{a} \frac{\d}{\d y}
			+ 1
		\right)
		+ \frac{a^2}{k^2} m_{\alpha\beta}
	\right\}
	G^{\beta\gamma}_k
	=
	- \frac{\im}{k a^2} \delta^\gamma_\alpha \ \delta(y - \tilde{y}) ,
	\label{eq:propagator-eqn}
\end{equation}
where $\tilde{y} = k \tilde{\tau}$
and $a'$ now denotes $\d a / \d x$.
Eq.~\eqref{eq:propagator-eqn} can be transformed into a Bessel equation
by making the substitution $G^{\alpha\beta}_k = w^{\alpha\beta} \sqrt{-y} / a$.
After expanding $a$ to next-order (for details, see
Refs.~\cite{Lidsey:1995np,Burrage:2011hd}) we finally arrive at an equation
which can be solved perturbatively,
\begin{equation}
	\left\{
		\delta_{\alpha\beta}
		\left(
			\frac{\d^2}{\d y^2}
			+ \frac{1}{y} \frac{\d}{\d y}
			+ \left[
				1 
				- \frac{9/4}{y^2}
			\right]
		\right)
		- \frac{3 M_{\alpha\beta}}{y^2}
	\right\}
	w^{\beta\gamma}
	=
	- \frac{\im}{k a\sqrt{-y}}
	\delta^\gamma_\alpha \
	\delta(y-\tilde{y}) ,
	\label{eq:propagator-perturbative}
\end{equation}
where
$M_{\alpha\beta} = \varepsilon_\ast \delta_{\alpha\beta} + u_{\alpha\beta}$
was defined in Eq. \eqref{eq:M-def}.
It plays the role of an effective mass term, including contributions
from coupling to the background geometry.

\paragraph{Homogeneous solution.}
The technology to solve Eq. \eqref{eq:propagator-perturbative}
perturbatively in $M_{\alpha\beta}$ was developed by
Nakamura \& Stewart~\cite{Nakamura:1996da}
and
Gong \& Stewart~\cite{Gong:2001he,Gong:2002cx}.
We adopt their methods, with some small modifications noted below
to keep each wavefunction purely imaginary in the late-time limit.
The first step is to find a solution $v^{\beta\gamma}$ of the
homogeneous equation,
\begin{equation}
	\left\{
		\delta_{\alpha\beta}
		\left(
			\frac{\d^2}{\d y^2}
			+ \frac{1}{y} \frac{\d}{\d y}
			+ \left[
				1
				- \frac{9/4}{y^2}
			\right]
		\right)
		- \frac{3 M_{\alpha\beta}}{y^2}
	\right\}
	v^{\beta\gamma}
	=
	0 .
	\label{eq:v-eqn}
\end{equation}
To do so we work perturbatively in $M_{\alpha\beta}$,
writing $v^{\alpha\beta} = v_0^{\alpha\beta} + v_1^{\alpha\beta} + \cdots$,
with $v_n^{\alpha\beta} = \Or(M^n)$.
Up to $\Or(M)$ we find
\begin{subequations}
\begin{align}
	\delta_{\alpha\beta}
	\left(
		\frac{\d^2}{\d y^2}
		+ \frac{1}{y} \frac{\d}{\d y}
		+ \left[
			1
			- \frac{9/4}{y^2}
		\right]
	\right)
	v_0^{\beta\gamma}
	& = 0 ,
	\label{eq:v0-eqn}
	\\
	\delta_{\alpha\beta}
	\left(
		\frac{\d^2}{\d y^2}
		+ \frac{1}{y} \frac{\d}{\d y}
		+ \left[
			1
			- \frac{9/4}{y^2}
		\right]
	\right)
	v_1^{\beta\gamma}
	& =
	\frac{3}{y^2} M_{\alpha\beta} v_0^{\beta\gamma} .
	\label{eq:v1-eqn}
\end{align}
\end{subequations}

Eq.~\eqref{eq:v0-eqn} has solution in terms of Bessel functions
of order $3/2$. The linear combination with correct boundary
conditions to match the Minkowski-space wavefunctions in the
subhorizon limit $k/aH = |y| \rightarrow \infty$
is the Hankel function,
\begin{equation}
	v_0^{\alpha\beta} = \Umatrix^{\alpha\beta} H_{3/2}^{(2)}(-k\tau)
	\label{eq:v0-soln}
\end{equation}
where $\Umatrix^{\alpha\beta}$ is an arbitrary constant matrix.
Eq.~\eqref{eq:v1-eqn} can be solved using the
causal Green's function
$\Gamma(y,z)$
constructed by Gong \& Stewart~\cite{Gong:2001he,Gong:2002cx}.
This satisfies
\begin{equation}
	\frac{\d^2 \Gamma}{\d y^2}
	+ \frac{1}{y} \frac{\d \Gamma}{\d y}
	+ \left[
		1
		- \frac{9/4}{y^2}
	\right]
	\Gamma
	=
	\delta(y-z) ,
\end{equation}
and can be written explicitly in terms of the Hankel functions
\begin{equation}
	\Gamma(y,z) =
	-
	\frac{\im \pi z}{4} \times
	\begin{cases}
		0 & y < z \\
		H_{3/2}^{(2)}(-y) H_{3/2}^{(1)}(-y)
		- H_{3/2}^{(1)}(-y) H_{3/2}^{(2)}(-y) & y > z
	\end{cases}
	\quad
	.
\end{equation}
After integrating over $\Gamma$, the solution $v_1^{\alpha\beta}$ can be
written
\begin{equation}
	v^{\alpha\beta}_1 = {M^\alpha}_\gamma \ \Umatrix^{\gamma\beta} \ v_1 ,
	\label{eq:v1-soln}
\end{equation}
where $v_1$ satisfies~\cite{Gong:2001he,Gong:2002cx}
\begin{equation}
	v_1 = \im \sqrt{\frac{2}{\pi}} \frac{1}{(-y)^{3/2}}
	\left(
		2 \e{\im y}
		- (1 + \im y)\e{-\im x}
		\int_{-\infty}^y \frac{\d z}{z} \e{2\im z}
	\right) .
	\label{eq:v1-scalar}
\end{equation}
In writing~\eqref{eq:v1-soln} we have adjusted the position
of the indices on $M$
to match on both sides of the equation. However, since we are
generally working
with a flat field-space metric,
except in~\S\ref{sec:nontrivial-metric}, there is no distinction between
co- and contravariant indices.

The scalar part of the zero-order wavefunction $H_{3/2}^{(2)}(-y)$ has the
property that it is purely imaginary at late times,
\begin{equation}
	H_{3/2}^{(2)}(-y) \rightarrow
	\im \sqrt{\frac{2}{\pi}} \frac{1}{(-y)^{3/2}} + \Or(y) .
\end{equation}
However, in this limit $v_1$ is complex.
Assembling~\eqref{eq:v0-soln}, \eqref{eq:v1-soln} and~\eqref{eq:v1-scalar}
gives
\begin{equation}
	v^{\alpha\beta}
	\rightarrow \im \sqrt{\frac{2}{\pi}} \frac{1}{(-y)^{3/2}}
	\left(
		\Umatrix^{\alpha\beta}
		+
		{M^\alpha}_\gamma \ \Umatrix^{\gamma\beta}
		\Big[
			2 - \EulerGamma - \ln(-2y) - \im \frac{\pi}{2}
		\Big]
	\right) .
	\label{eq:v-late-limit}
\end{equation}
It is the term involving $\im \pi/2$ which is problematic.
This arises from the asymptotic expansion of the exponential integral
appearing in Eq.~\eqref{eq:v1-scalar}.
We will see 
below that to clearly exhibit
the cancellation of unwanted divergent terms in the calculation of
the three-point functions it is helpful to adjust
the phase of $v^{\alpha\beta}$
so that this term is removed.
Therefore we redefine $v_{\alpha\beta}$ by a pure phase
\begin{equation}
	v^{\alpha\beta} \rightarrow
	\tilde{v}^{\alpha\beta} =
	\exp( \im \theta^{\alpha\gamma} )
	v^{\gamma\beta} ,
\end{equation}
where $\theta^{\alpha\beta}$ is a constant real matrix.
This redefinition does \emph{not} change the propagator, which depends on
the combination $\vect{v}^\dag \cdot \vect{v}$
in which this phase cancels out.
To remove the imaginary term in Eq.~\eqref{eq:v-late-limit}
we should choose $\theta^{\alpha\beta} = (\pi/2) M^{\alpha\beta}$.
The net effect is to replace $v_1$ in Eq.~\eqref{eq:v1-scalar} with 
a redefined function
(cf. Refs.~\cite{Chen:2006nt,Burrage:2010cu,Burrage:2011hd})
\begin{equation}
	v_1 \rightarrow \tilde{v}_1
	=
	\im \sqrt{\frac{2}{\pi}}
	\frac{1}{(-y)^{3/2}}
	\left(
		2 \e{\im y}
		+ \im \frac{\pi}{2} (1 - \im y) \e{\im y}
		- (1 + \im y)\e{-\im y} \int_{-\infty}^y \frac{\d z}{z} \e{2\im z}
	\right) .
	\label{eq:v1-phase}
\end{equation}

\paragraph{Feynman propagator.}
We now return to Eq.~\eqref{eq:propagator-perturbative}
and use the homogeneous solution $v^{\alpha\beta}$ to construct
the Feynman Green's function $w^{\alpha\beta}$.%
	\footnote{Strictly, in the context of the Schwinger method this
	is the time-ordered $++$ correlation function.
	Its complex conjugate is the time-ordered $--$ correlation
	function.
	The $+-$ and $-+$ correlation functions can be extracted from
	the two alternatives in Eq.~\eqref{eq:propagator-final},
	bearing in mind that $-$ fields are later than $+$
	fields for the purpose of time ordering, no matter 
	at what physical time they are evaluated.}
In this section it is helpful to switch to matrix notation,
with $(\vect{v})_{\alpha\beta} = v^{\alpha\beta}$,
$(\vect{w})_{\alpha\beta} = w^{\alpha\beta}$ and so on.
In the absence of the mass matrix $\vect{M}$
the field fluctuations are uncorrelated, which implies
that we must choose $\vect{\Umatrix} = \vect{1}$.
Together,
the Feynman boundary conditions and hermiticity require
\begin{equation}
	\vect{u} =
	\begin{cases}
		\vect{v}^\dag(\tilde{y}) \cdot \vect{Y} \cdot \vect{v}(y) & y < \tilde{y} \\
		\vect{v}^\dag(y) \cdot \vect{Y} \cdot \vect{v}(\tilde{y}) & y > \tilde{y}
	\end{cases}
\end{equation}
where $\vect{Y}$ is an Hermitian matrix. We will take it to commute
with $\vect{v}$ and $\vect{v}^\dag$. This can be justified
(with our choice $\vect{\Umatrix} = 1$)
because
Eq.~\eqref{eq:v-eqn}
implies that both $\vect{v}$ and $\vect{v}^\dag$ are power series
in the Hermitian matrix $\vect{M}$.
The normalization condition in~\eqref{eq:propagator-perturbative}
likewise guarantees that $\vect{Y}$ is a power series in $\vect{M}$.
Therefore all these matrices commute.
Imposing this normalization condition, we conclude
\begin{equation}
	\vect{Y} \cdot
	\left(
		\vect{v}(\tilde{y}) \cdot \frac{\d \vect{v}^\dag(\tilde{y})}{\d \tilde{y}}
		-
		\vect{v}^\dag(\tilde{y}) \cdot \frac{\d \vect{v}(\tilde{y})}{\d \tilde{y}}
	\right) =
	- \frac{\im}{(-\tilde{y})^{1/2} k a(\tilde{y})} \times \vect{1} .
\end{equation}
The matrix in brackets $( \cdots )$ is the `Wronskian matrix'
for $\vect{v}$ and $\vect{v}^\dag$. Abel's theorem shows that it
can be written
\begin{equation}
	\vect{W}(\tilde{y}) =
	\vect{v}(\tilde{y}) \cdot \frac{\d \vect{v}^\dag(\tilde{y})}{\d \tilde{y}}
	- \vect{v}^\dag(\tilde{y}) \cdot \frac{\d \vect{v}(\tilde{y})}{\d \tilde{y}}
	= \vect{W}(\tilde{y}_0) \exp
	\left(
		- \int_{\tilde{y}_0}^{\tilde{y}} \frac{\d z}{z}
	\right)
	= \vect{W}(\tilde{y}_0) \frac{\tilde{y}_0}{\tilde{y}} ,
\end{equation}
where $\tilde{y}_0$ is arbitrary and $\vect{W}(\tilde{y}_0)$ is independent of
$\vect{M}$. With our choice $\vect{\Umatrix} = 1$, we find
\begin{equation}
	\vect{W}(\tilde{y}) = - \frac{4\im}{\pi (-\tilde{y})} \times \vect{1} .
\end{equation}
Therefore $\vect{Y}$ is Hermitian and commutes with $\vect{M}$,
in agreement with our earlier argument. In conclusion, the
Feynman propagator can be written
\begin{equation}
	\vect{G}_k = \frac{\pi}{4k}
	\frac{(-y)^{1/2} (-\tilde{y})^{1/2}}{a(y) a(\tilde{y})}
	\times
	\begin{cases}
		\vect{v}^\dag(\tilde{y}) \cdot \vect{v}(y) & y < \tilde{y} \\
		\vect{v}^\dag(y) \cdot \vect{v}(\tilde{y}) & y > \tilde{y}
	\end{cases}
	.
	\label{eq:propagator-final}
\end{equation}

\paragraph{Uniform expansion of $\vect{G}_k$ to next-order.}
We require the lowest-order and next-order
contributions in slow-roll to $\vect{G}_k$.
This can be obtained from the solutions $v_0^{\alpha\beta}$ and
$v_1^{\alpha\beta}$
together with next-order corrections to the scale factor $a$,
for which purpose we must Taylor expand around an arbitrary reference
scale $k_\ast$ as described in~\S\ref{sec:resummation}.
Denoting evaluation at the horizon-crossing time for $k_\ast$
by a subscript `$\ast$',
we find that
$\vect{G}_k$ behaves like the product of two effective wave-function
factors $\vect{w}$,
\begin{equation}
	\vect{G}_k =
	\begin{cases}
		\vect{w}^\dag(\tilde{y}) \cdot \vect{w}(y) & y < \tilde{y} \\
		\vect{w}^\dag(y) \cdot \vect{w}(\tilde{y}) & y > \tilde{y}
	\end{cases} ,
	\label{eq:propagator-wavefunctions}
\end{equation}
where $\vect{w} = \vect{w}_0 + \vect{w}_1 + \cdots$ satisfies
\begin{subequations}
\begin{align}
	\vect{w}_0(y) & = \frac{\im}{\sqrt{2k^3}} H_\ast ( 1 - \im y ) \e{\im y}
		\cdot \vect{1}
	\label{eq:w0}
	\\
\begin{split}
	\vect{w}_1(y) & = \frac{\im}{\sqrt{2k^3}} H_\ast \Big\{
		\varepsilon_\ast \Big[ \big( \ln\frac{k_\ast}{k} (-y) \big) - 1 \Big]
			(1 - \im y) \e{\im y} \cdot \vect{1}
		\\ & \hspace{2.2cm} \mbox{}
		+ \vect{M} \cdot \Big[
			2 \e{\im y}
			+ \im \frac{\pi}{2} (1 - \im y)\e{\im y}
			- (1 + \im y)\e{-\im y}
			\int_{-\infty}^{y} \frac{\d z}{z} \e{2\im z}
		\Big]
	\Big\} .
	\label{eq:w1}
\end{split}
\end{align}
\end{subequations}
[This result was used in the main text in Eq.~\eqref{eq:propagator}.]
Eqs.~\eqref{eq:w0} and~\eqref{eq:w1} are both
exact in $y$ but perturbative in
powers of slow-roll and mass $\vect{M}$.

\subsection{Asymptotic behaviour of wavefunctions}
In practical calculations we
require the late-time expansion of the elementary wavefunctions, $\vect{w}$,
in order to compute the contribution from external lines.
In the present case it is only necessary to compute these
up to $\Or(y^2)$. We find
\begin{equation}
	\vect{w}(y)
	= \dfrac{\im}{\sqrt{2k^3}} H_{\ast} \e{\im y}
		\Big\{
	    	\vect{1}
	    	+ \vect{A}_\ast
	    	- \vect{u} \ln \frac{k_\ast}{k} (-y)
	    	- \im y \big(
	    		\vect{1} +\vect{A}_\ast
	    	\big) 
	    	+ \im y \vect{u} \ln \frac{k_\ast}{k} (-y)
	    	+ \Or(y)^2
	    \Big\} \ ,
		\label{eq:wavefunctionktau}
\end{equation}
where 
\begin{equation}
	\vect{A}
	\equiv \varepsilon_\ast
		\Big(
			1
			- \EulerGamma
			- \ln \frac{2k}{k_\ast}
		\Big) \vect{1} 
		+ \vect{u}_\ast \Big(
			2
			- \EulerGamma
			- \ln \frac{2k}{k_\ast}
		\Big) ,
	\label{eq:defA}
\end{equation}
and $\vect{u}$ is defined in Eq. \eqref{eq:expansion-tensor}.

\subsection{Three-point function}
\label{subsec:3pfappendix}
To compute the three-point function we require
the action to third-order.
This calculation is well-documented in the literature;
see, for examples, Refs.~\cite{Maldacena:2002vr,Seery:2005gb}.
Fluctuations in the fields will mix with the metric,
which can be accommodated using the
ADM decomposition~\eqref{eq:adm-decomposition}
\begin{equation}
	\d s^2 = - N^2 \, \d t^2 + h_{ij} ( \d x^i + N^i \, \d t )
		( \d x^j + N^j \, \d t ) ,
\end{equation}
where the lapse $N$ and shift vector $N^i$ are determined
by constraint equations which follow from invariance under
temporal and spatial reparametrizations.
At the level of the background we have $N = 1$ and $N^i = 0$.
When perturbations are included there are order-by-order
corrections, discussed in~\S\ref{sec:inverse-gradients},
\begin{subequations}
\begin{align}
	N & = 1 + \sum_{n = 1}^\infty \alpha_n , \\
	N_i & = \sum_{n=1}^\infty \partial_i \vartheta_n +
		\text{solenoidal component} .
\end{align}
\end{subequations}
The solenoidal part of $N_i$ is important only
when computing the action at fourth-order and
above~\cite{Sloth:2006az,Seery:2006vu}.
(Spatial indices $i$, $j$, {\ldots} are raised and lowered using the
three-metric $h_{ij}$.)
Also, although in principle there could be contributions
to the third-order action from
$\alpha_n$ and $\vartheta_n$ with $n \leq 3$, the structure of the
constraints in the action guarantees that only $\alpha_1$ and
$\vartheta_1$ contribute~\cite{Maldacena:2002vr,Chen:2006nt}.
Specializing to the spatially flat gauge $h_{ij} = a^2 \delta_{ij}$,
we find%
	\footnote{\label{footnote:vartheta}Eq.~\eqref{eq:third-action}
	contains a term
	linear in $\partial^2 \vartheta_1 / a^2$, in apparent contradiction
	with the conclusions of~\S\ref{sec:inverse-gradients}.
	This discrepancy is only an appearance, however.
	Eq.~\eqref{eq:third-action} has been simplified using
	the constraints, which allow us to trade $\vartheta_1$
	for specific combinations of the other metric and
	scalar fluctuations. Therefore any appearance of nonlocality
	in~\eqref{eq:third-action} is fictitious and will disappear in the
	final answer.
	See the discussion on p.~\pageref{page:factorization-theorem}.}
\begin{equation}
\begin{split}
	S_{\phi\phi\phi} \supseteq
	\frac{1}{2}
	\int \d^3 x \, \d t \; a^3 \bigg\{
		&
		- \frac{1}{3} V_{\alpha\beta\gamma}
			\delta\phi^\alpha \delta\phi^\beta \delta\phi^\gamma
		- \frac{2}{a^2} \delta \dot{\phi}^\alpha
			\partial_j \vartheta_1 \partial_j \delta \phi_\alpha
		\\ & \mbox{}
		+ \alpha_1 \Big[
			- \frac{1}{a^2} \partial_j \delta \phi^\alpha
				\partial_j \delta \phi_\alpha
			- \delta \dot{\phi}^\alpha \delta \dot{\phi}_\alpha
			- V_{\alpha\beta} \delta \phi^\alpha \delta \phi^\beta
		\\ & \hspace{1.09cm} \mbox{}
			- \frac{1}{a^4} \partial_i \partial_j \vartheta_1
				\partial_i \partial_j \vartheta_1
			+ \frac{1}{a^4} \partial^2 \vartheta_1
				\partial^2 \vartheta_1
			+ \frac{2}{a^2} \dot{\phi}^\alpha \partial_j \vartheta_1
				\partial_j \delta \phi_\alpha
		\\ & \hspace{1.09cm} \mbox{}
			+ \alpha_1 \Big(
				2 \dot{\phi}^\alpha \delta \dot{\phi}_\alpha
				+ \frac{4H}{a^2} \partial^2 \vartheta_1
				+ \alpha_1 ( 6H^2 - \dot{\phi}^2 )
			\Big)
		\Big]
	\bigg\}
\end{split}
\label{eq:third-action}
\end{equation}
where $\dot{\phi}^2 = \dot{\phi}_\alpha \dot{\phi}_\alpha$
and
repeated indices in the lower position are summed using the
flat Euclidean metric $\delta_{ij}$.
Eq.~\eqref{eq:third-action}
is exact. It has been obtained using an expansion in the
amplitudes $|\delta\phi|$ but does not invoke an expansion
in slow-roll parameters.
The solution for $\alpha_1$ is
\begin{equation}
	\alpha_1 = \frac{\dot{\phi}_\alpha \delta \phi_\alpha}{2H} ,
\end{equation}
and the solution for $\vartheta_1$ was used in~\eqref{eq:shift-vector},
\begin{equation}
	\frac{4H}{a^2} \partial^2 \vartheta_1 =
		- 2 V_{\alpha} \delta\phi_\alpha
		- 2 \dot{\phi}_\alpha \delta \dot{\phi}_\alpha
		+ 2\alpha_1 (-6H^2+\dot{\phi}^2) .
\end{equation}

We wish to compute the three-point function up to terms which are
next-to-leading (`$\NLO$') in a fixed-order slow-roll expansion.
For this purpose we need only extract the $\LO$ and $\NLO$
terms in~\eqref{eq:third-action}.
The $\LO$ contributions were given in Ref.~\cite{Seery:2005gb},
where the corresponding three-point function was also obtained.
They are
\begin{equation}
	S^{\LO}_{\phi\phi\phi} \supseteq
	\frac{1}{2} \int \d^3 x \, \d t \; a^3 \bigg(
		\frac{1}{H} \dot{\phi}^\beta \delta\dot{\phi}^\alpha
			\partial_j \delta \phi_\alpha \partial_j \partial^{-2}
			\delta \dot{\phi}_\beta
		- \alpha_1 \Big[
			\delta\dot{\phi}^\alpha \delta\dot{\phi}_\alpha
			+ \frac{1}{a^2} \partial_j \delta\phi^\alpha
				\partial_j \delta\phi_\alpha
		\Big]
	\bigg)\ \ ,
	\label{eq:third-action-lo}
\end{equation}
(In Ref.~\cite{Seery:2005gb} the action was
quoted in a different but equivalent form.)
The new contributions
at $\NLO$ are
\begin{equation}
\begin{split}
	S^{\NLO}_{\phi\phi\phi} \supseteq
	\frac{1}{2} \int \d^3 x \, \d t \; a^3 \bigg(
		&
		g_{\alpha\beta\gamma} H^2 \delta\phi^\alpha \delta\phi^\beta
			\delta\phi^\gamma
		- \frac{1}{2} q^\alpha H \delta\dot{\phi}^\beta
			\partial_j \delta\phi_\beta \partial_j \partial^{-2}
			\delta\phi_\alpha
		\\ & \mbox{}
		- \frac{\dot{\phi}^\alpha \dot{\phi}^\beta \dot{\phi}^\gamma}
			{2H^2} \delta\phi_\alpha \partial_j \delta \phi_\beta
			\partial_j \partial^{-2} \delta\dot{\phi}_\gamma
		\\ & \mbox{}
		- \frac{\dot{\phi}^\alpha \dot{\phi}^\beta \dot{\phi}^\gamma}
			{8H^3} \delta \phi_\alpha \Big[
				\partial_i \partial_j \partial^{-2}
					\delta \dot{\phi}_\beta
				\partial_i \partial_j \partial^{-2}
					\delta \dot{\phi}_\gamma
				- \delta \dot{\phi}_\beta  \dot{\phi}_\gamma
			\Big]
	\bigg) ,
	\label{eq:third-action-nlo}
\end{split}
\end{equation}
where
the coefficient tensors $g_{\alpha\beta\gamma}$
and $q_\alpha$ satisfy
\begin{subequations}
\begin{align}
	g_{\alpha\beta\gamma} & =
		- \frac{1}{3} \frac{V_{\alpha\beta\gamma}}{H^2}
		- \frac{\dot{\phi}_\alpha}{2H} \frac{V_{\beta\gamma}}{H^2}
		+ \frac{3}{4}
			\frac{\dot{\phi}_\alpha \dot{\phi}_\beta \dot{\phi}_\gamma}{H^3}
	\\
	q^\alpha & =
		2 \frac{\ddot{\phi}^\alpha}{H^2}
		+ 2 \varepsilon 
		\frac{\dot{\phi}^\alpha}{H} = 2\dfrac{\dot{\phi}_\beta}{H} u^{\alpha\beta}.
\end{align}
\end{subequations}
(We are using $\varepsilon = \dot{\phi}^2 / 2H^2$, 
which is accurate at lowest order in the slow-roll approximation.)
In particular, $g_{\alpha\beta\gamma}$ is symmetric under the
exchange $\beta \leftrightarrow \gamma$, but not under exchanges
involving $\alpha$.
The three-point function for
the $V_{\alpha\beta\gamma}$ term was obtained by
Falk, Rangarajan \& Srednicki~\cite{Falk:1992sf},
and later recomputed by Zaldarriaga~\cite{Zaldarriaga:2003my}.
These
calculations enable the three-point function for $g_{\alpha\beta\gamma}$
to be deduced,%
	\footnote{The three-point functions obtained by
	Falk et al.~and Zaldarriaga
	contained a small typographical error, which was corrected
	in Ref.~\cite{Seery:2008qj}.
	The correct expression is given in Eq.~\eqref{eq:NLO-A}.}
but results for the remaining terms
in Eq.~\eqref{eq:third-action-nlo} have not yet been obtained.

\subsubsection{Next-to-leading interactions}
Terms in the $\NLO$ action~\eqref{eq:third-action-nlo} contribute
to the three-point function beginning at next-to-leading order in 
the slow-roll expansion.
Therefore, we need only retain the lowest-order contribution~\eqref{eq:w0}
to the propagator. Also we require only the lowest order estimate
$a = -(H_\ast\tau)^{-1}$ for the scale factor, where as usual
we have introduced an arbitrary reference scale $k_\ast$ which serves
as the basis for a Taylor expansion of all background quantities,
as explained in~\S\ref{sec:resummation}.

\paragraph{Notation.}
The full set of three-point functions is lengthy.
To simplify the presentation, it is helpful to factor out
a number of common elements. In particular, we define
each three-point function in terms of a quantity
$\mathcal{B}^{\alpha\beta\gamma}$, in which 
we extract a further
common factor of $H_\ast^4$ in comparison to 
$B^{\alpha\beta\gamma}$ defined in 
\eqref{eq:B-tensor-def}.
This must be symmetrized to produce the final result,
\begin{equation}
\begin{split}
	\langle
		\delta\phi_\alpha(\vect{k}_1)
		\delta\phi_\beta(\vect{k}_2)
		\delta\phi_\gamma(\vect{k}_3)
	\rangle_\tau
	= \mbox{}
	&
	(2\pi)^3 \delta(\vect{k}_1 + \vect{k}_2 + \vect{k}_3)
	\frac{H_\ast^4}{4 \prod_i k_i^3} \
	\mathcal{B}_{\alpha\beta\gamma}(k_1, k_2, k_3)
	\\ & \mbox{}
	+ [ \text{5 permutations} ] ,
\end{split}
\end{equation}
where the permutations are formed by simultaneously exchanging
the index and momentum pairs
$(\alpha, \vect{k}_1)$,
$(\beta, \vect{k}_2)$
and
$(\gamma, \vect{k}_3)$.
We also define
\begin{subequations}
\begin{align}
	k_t & = k_1 + k_2 + k_3 , \\
	K^2 & = \sum_{i < j} k_i k_j = k_1 k_2 + k_1 k_3 + k_2 k_3 ,
\end{align}
\end{subequations}
and set $\EulerGamma$ to be the Euler--Mascheroni constant,
$\EulerGamma \approx 0.577$.

We highlight terms involving $\ln(-k_\ast \tau)$
in {\color{forestgreen}green}.
For vertex and internal wavefunction corrections,
these contribute to the divergent part of the
three-point function~\eqref{eq:3pf-div}.
For external wavefunctions, they contribute to~\eqref{eq:3pf-unsourced-taylor}.

\paragraph{$\delta \phi^\alpha \delta\phi^\beta \delta\phi^\gamma$
operator}
\begin{equation}
	\mathcal{B}_{\alpha\beta\gamma} \supseteq \frac{g^\ast_{\alpha\beta\gamma}}{2}
	\bigg(
		\frac{4}{9} k_t^2
		- K^2 k_t
		- \boxed{\frac{1}{3} \Big( \sum_i k_i^3 \Big)}
			\Big[  \boxed{\ln(-k_\ast \tau)}
			+ \ln \frac{k_t}{k_\ast}
			+ \EulerGamma
			+ \frac{1}{3}
		\Big]
	\bigg)
	\label{eq:NLO-A}
\end{equation}

\paragraph{$\delta\dot{\phi}^\beta \partial_j \delta\phi_\beta
\partial_j \partial^{-2} \delta\phi_\alpha$ operator}
\begin{equation}
	\mathcal{B}_{\alpha\beta\gamma} \supseteq
	\frac{q_\gamma^\ast \delta_{\alpha\beta}}{4}
	\frac{k_1^2}{k_3^2}
	(\vect{k}_2 \cdot \vect{k}_3)
	\Big(
		k_t
		- \frac{k_2 k_3}{k_t}
		- \boxed{k_1} \Big[
			\boxed{\ln(-k_\ast \tau) } + \ln \frac{k_t}{k_\ast} + \EulerGamma
		\Big]
	\Big)
	\label{eq:NLO-B}
\end{equation}

\paragraph{$\delta\phi_\alpha \partial_j \delta\phi_\beta
\partial_j \partial^{-2} \delta \dot{\phi}_\gamma$ operator}
\begin{equation}
	\mathcal{B}_{\alpha\beta\gamma} \supseteq
	\frac{\dot{\phi}^\ast_\alpha \dot{\phi}^\ast_\beta
	\dot{\phi}^\ast_\gamma}{4 H_\ast^3}
	(\vect{k}_2 \cdot \vect{k}_3)
	\Big(
		k_t
		- \frac{k_1 k_2}{k_t}
		- \boxed{k_3} \Big[
			\boxed{\ln(-k_\ast \tau)} + \ln \frac{k_t}{k_\ast} + \EulerGamma
		\Big]
	\Big)
	\label{eq:NLO-C}
\end{equation}

\paragraph{$\delta\phi_\alpha \delta \dot{\phi}_\beta \delta\dot{\phi}_\gamma$
operators}
\begin{equation}
	\mathcal{B}_{\alpha\beta\gamma} \supseteq
	- \frac{\dot{\phi}_\alpha^\ast \dot{\phi}^\ast_\beta
	\dot{\phi}^\ast_\gamma}{16 H_\ast^3}
	\frac{1}{k_t} \left( 1 + \frac{k_1}{k_t} \right)
	\big( [\vect{k}_2 \cdot \vect{k}_3]^2 - k_2^2 k_3^2 \big)
	\label{eq:NLO-D}
\end{equation}

\subsubsection{Leading-order interactions}
The $\LO$ operators which appear in~\eqref{eq:third-action-lo}
contribute to the three-point function at \emph{both}
lowest-order and next-order.
The next-order terms come from use of~\eqref{eq:w1} to correct the
wavefunctions associated with each internal or external line,
or from subleading terms in the Taylor expansion of background
quantities such as $H$ and $\dot{\phi}^\alpha$, including
factors of $H$ which appear in the scale factor.

\paragraph{Notation.}
These sources of
next-order corrections were identified
and described by Chen et~al.~\cite{Chen:2006nt}.
Subsequently,
rules for explicit evaluation of the integrals which appear
at intermediate stages of the computation
were given by Burrage et~al.~\cite{Burrage:2010cu,Burrage:2011hd},
to which we refer for all technical details.
For our present purposes we require only the functions
\begin{subequations}
\begin{align}
	J_0(k_3) & = \frac{1}{\vartheta_3} \ln ( 1 - \vartheta_3 )
	\label{eq:J0}
	\\
	J_1(k_3) & = \frac{1}{\vartheta_3^2} \Big(
		\vartheta_3 + \ln [1 - \vartheta_3]
	\Big) ,
	\label{eq:J1}
\end{align}
where $\vartheta_3$ is defined by
\begin{equation}
	\vartheta_3 = 1 - \frac{2k_3}{k_t} .
	\label{eq:vartheta}
\end{equation}
\end{subequations}
In writing Eqs.~\eqref{eq:J0}--\eqref{eq:vartheta} we have used the
convention that, although each expression depends on all $k_i$,
only the asymmetrically occurring momentum is written explicitly.
As described in Refs.~\cite{Burrage:2010cu,Burrage:2011hd},
the $J_n(k_3)$ approach finite values
in the limit $\vartheta_3 \rightarrow 0$
which corresponds to either of the (physical) squeezed configurations
$k_1 \rightarrow 0$ or $k_2 \rightarrow 0$.

To describe contributions from the late-time limit of~\eqref{eq:w1},
which are generated by external wavefunction factors,
we use Eq. \eqref{eq:defA}
and the combination
\begin{equation}
	t_\alpha =
	3 \varepsilon \frac{\dot{\phi}_\alpha}{H}
	+ \frac{\ddot{\phi}_\alpha}{H^2}
	= \frac{q_\alpha}{2} + 2 \varepsilon \frac{\dot{\phi}_\alpha}{H} .
\end{equation}
As above, the following expressions should be symmetrized over
simultaneous exchange of
the index and momentum pairs
$(\alpha, \vect{k}_1)$,
$(\beta, \vect{k}_2)$
and
$(\gamma, \vect{k}_3)$.

\begin{itemize}
	\item $\delta \phi_\beta \delta \dot{\phi}^\alpha
	\delta \dot{\phi}_\alpha$ operator
\end{itemize}

\subparagraph{$\LO$ plus $\NLO$ from external wavefunctions}
\begin{equation}
\begin{split}
	\mathcal{B}_{\alpha\beta\gamma}
	\supseteq
	- \frac{1}{4} \frac{k_2^2 k_3^2}{k_t}
	\frac{\dot{\phi}^\lambda_\ast}{H_\ast}
	\bigg(
		1 + \frac{k_1}{k_t}
	\bigg)
	\bigg(
		\delta_{\alpha\lambda} \delta_{\beta\gamma}
		&
		+
		\Big[
			A_{\alpha\lambda}^\ast(k_1) - \boxed{u_{\alpha\lambda}^\ast \ln(-k_{\ast} \tau)}
		\Big]
		\delta_{\beta\gamma}
		\\ & \mbox{}
		+
		\Big[
			A_{\beta\gamma}^\ast(k_2) + A_{\beta\gamma}^\ast(k_3)
			- \boxed{2 u_{\beta\gamma}^\ast  \ln(-k_{\ast} \tau)}
		\Big]
		\delta_{\alpha\lambda}
	\bigg)
	\label{eq:LO-A-external}
\end{split}
\end{equation}

\subparagraph{$\NLO$ from vertex}
\begin{equation}
	\mathcal{B}_{\alpha\beta\gamma}
	\supseteq
	- \frac{1}{4} \frac{k_2^2 k_3^2}{k_t}
	\delta_{\beta\gamma}
	\bigg\{
		2 \varepsilon_\ast \frac{\dot{\phi}_\alpha}{H_\ast}
		\Big(
			1 + \frac{k_1}{k_t}
		\Big)
		+ t_\alpha^\ast
		\bigg[  -
			\frac{k_1}{k_t}
			+
			\Big(
				\EulerGamma + \ln \frac{k_t}{k_\ast}
			\Big)
			\Big(
				1 + \frac{k_1}{k_t}
			\Big)
		\bigg]
	\bigg\}
\end{equation}

\subparagraph{$\NLO$ from internal wavefunctions}
\begin{equation}
\begin{split}
	\mathcal{B}_{\alpha\beta\gamma}
	\supseteq
	\mbox{}
	&
	\mbox{} - \frac{1}{4}
	\frac{\dot{\phi}_\alpha^\ast}{H_\ast}
	\bigg\{
	\boxed{u^\ast_{\beta\gamma}}
		\Big[
			\boxed{(k_2^3 + k_3^3)}
			\Big(
				\boxed{\ln(-k_\ast \tau)} + \ln \frac{k_t}{k_\ast} + \EulerGamma
			\Big)
			- k_t (k_2^2 + k_3^2)
		\Big]
	\\
	& \hspace{1.6cm} \mbox{}
		- 3 \varepsilon_\ast \delta_{\beta\gamma}
		\frac{k_2^2 k_3^2}{k_t}
		\Big(
			\EulerGamma + \ln \frac{k_t}{k_\ast}
		\Big)
		\Big(
		1 + \frac{k_1}{k_t}
		\Big)
	\\
	& \hspace{1.6cm} \mbox{}
		- M_{\beta\gamma}^\ast
		\frac{k_2^2 k_3^2}{k_t}
		\Big[
			J_0(k_2)
			+ J_0(k_3)
			+ \frac{k_1}{k_t} \big\{ J_1(k_2) + J_1(k_3) \big\}
		\Big]
	\bigg\}
	\\
	& \mbox{} + \frac{1}{4}
	\frac{k_2^2 k_3^2}{k_t}
	\frac{\dot{\phi}^\lambda_\ast}{H_\ast}
		\bigg\{\delta_{\beta\gamma}
		\bigg[M_{\alpha\lambda}^\ast
		\Big(
			J_0(k_1) - \frac{k_1}{k_t} J_1(k_1)
		\Big)
		- 2	u^\ast_{\alpha\lambda}\bigg]
	+\delta_{\alpha\lambda}\bigg[2 \varepsilon_\ast \delta_{\beta\gamma}-\frac{k_1(k_2+k_3)}{k_2k_3}u^\ast_{\beta\gamma}\bigg]
	\bigg\}
\end{split}
	\label{eq:LO-A-internal}
\end{equation}

\begin{itemize}
	\item $\delta \dot{\phi}^\alpha \partial_j \delta \phi_\alpha
	\partial_j \partial^{-2} \delta \dot{\phi}_\beta$ operator
\end{itemize}

\subparagraph{$\LO$ plus $\NLO$ from external wavefunctions}
\begin{equation}
\begin{split}
	\mathcal{B}_{\alpha\beta\gamma}
	\supseteq
	\frac{1}{2} k_2^2 \frac{\vect{k}_1 \cdot \vect{k}_3}{k_t}
	\frac{\dot{\phi}^\lambda_\ast}{H_\ast}
	\bigg( 1 + \frac{k_3}{k_t} \bigg)
	\bigg( \delta_{\alpha\lambda} \delta_{\beta\gamma}
	& \mbox{}
	+ \Big[
		A_{\alpha\lambda}^\ast(k_1) - \boxed{ u^\ast_{\alpha\lambda} \ln(-k_\ast \tau)}
	\Big]
	\delta_{\beta\gamma}
	\\
	& \mbox{}
	+
	\Big[
		A_{\beta\gamma}^\ast(k_2)
		+ A_{\beta\gamma}^\ast(k_3)
		- \boxed{2 u_{\beta\gamma}^\ast \ln(-k_\ast \tau)}
	\Big]
	\delta_{\alpha\lambda}
	\bigg)
	\label{eq:LO-B-external}
\end{split}
\end{equation}	

\subparagraph{$\NLO$ from vertex}
\begin{equation}
	\mathcal{B}_{\alpha\beta\gamma}
	\supseteq
	\frac{k_2^2 (\vect{k}_1 \cdot \vect{k}_3)}{2 k_t}
	\delta_{\beta\gamma}
	\bigg\{
		2 \varepsilon_\ast \frac{\dot{\phi}^\ast_\alpha}{H_\ast}
		\Big(
			1 + \frac{k_3}{k_t}
		\Big)
		-
		t^\ast_\alpha
		\bigg[
			\frac{k_3}{k_t} -
			\Big(
				\EulerGamma + \ln \frac{k_t}{k_\ast}
			\Big)
			\Big(
				1 + \frac{k_3}{k_t}
			\Big)
		\bigg]
	\bigg\}
\end{equation}

\subparagraph{$\NLO$ from internal wavefunctions}

\begin{equation}
\begin{split}
	\mathcal{B}_{\alpha\beta\gamma}
	\supseteq
	\frac{\vect{k}_1 \cdot \vect{k}_3}{k_1^2}
	\frac{\dot{\phi}^\lambda_\ast}{2H_\ast}
	& \bigg( \dfrac{k_1^2 k_2^2}{k_t} \Big[
	-\delta_{\alpha \lambda } M_{ \beta \gamma}^\ast   
				\big[J_0(k_2)+J_0(k_3) +\frac{k_3}{k_t}
				\big\{ J_1(k_2) - J_1(k_3) \big\}  \big] \\
		& \hspace*{1.3cm}	-\delta_{ \beta \gamma} M_{\alpha \lambda}^\ast 
							\big[J_0(k_1)+\frac{k_3}{k_t} J_1(k_1) \big]
	\Big]-k_t \big\{ k_1^2 \delta_{\gamma \lambda} u_{\alpha\beta}^\ast +k_2^2 \delta_{\beta\gamma} u_{\alpha\lambda}^\ast \big\} 
	\\
& \mbox{} 
-\bigg[  \EulerGamma +\ln\Big(\frac{k_t}{k_\ast}\Big)+\boxed{\ln(-k_\ast\tau)} \bigg]
\boxed{C^\ast_{\alpha\beta\gamma\lambda}}\\
	&\mbox{} +\frac{1}{k_t} \Big(\gamma^{0\ast}_{\alpha\beta\gamma\lambda}-\frac{1}{k_t} \big\{\gamma^{1\ast}_{\alpha\beta\gamma\lambda} +\delta^{1\ast}_{\alpha\beta\gamma\lambda} \big\} \Big)
	-\frac{1}{k_t} \Big(\EulerGamma+ \ln \frac{k_t}{k_\ast}\Big) \delta^{0\ast}_{\alpha\beta\gamma\lambda} 
\big(	1+\dfrac{k_3}{k_t}\big)\Big\}\bigg)
\end{split}
\label{eq:LO-B-internal}
\end{equation}
where we have defined
\begin{equation}
	\left\{
	\begin{split}
C^\ast_{\alpha\beta\gamma\lambda} &=	
	-k_1^3 u_{\beta\gamma}^\ast \delta_{\alpha\lambda}
	-k_2^3 \delta_{\beta\gamma} u_{\alpha\lambda}^\ast\\
\gamma^{0\ast}_{\alpha\beta\gamma\lambda} &\mbox{} = -2k_1^2 k_3^2 \varepsilon_\ast \delta_{\beta\gamma} \delta_{\alpha\lambda}
	+k_1^2k_2k_3 \delta_{\alpha\lambda}u_{\beta\gamma}^\ast +
	k_2^2k_3^2 \delta_{\alpha\lambda} (\varepsilon_\ast \delta_{\beta\gamma} +2u_{\beta\gamma}^\ast) +
	k_1 k_2^2 k_3 \delta_{\beta\gamma} u_{\alpha\lambda}^\ast\\
\gamma^{1\ast}_{\alpha\beta\gamma\lambda} &= k_2^2 k_3(k_3^2+k_1^2) \varepsilon_\ast \delta_{\alpha\lambda} \delta_{\beta\gamma}\\
\delta^{0\ast}_{\alpha\beta\gamma\lambda} &=	3 k_1^2 k_2^2 \varepsilon_\ast \delta_{\alpha\lambda} \delta_{\beta\gamma}\\
\end{split}
\right.
\end{equation}

\begin{itemize}
	\item $\delta \phi_\beta (\partial_i \delta \phi_\alpha)^2$ operator.
	
	This operator is simplified by our choice of phase
	convention~\eqref{eq:v1-phase}.
	With the unmodified wavefunction~\eqref{eq:v1-scalar},
	fast power-law divergences are generated by the internal wavefunctions
	which cancel against other power-law divergences arising from
	their external counterparts.
	
	With the redefined phase in~\eqref{eq:v1-phase},
	all power-law divergences cancel locally, in the sense that we
	do not need to group terms from different parts of the calculation.
	In practice this is a considerable convenience.
\end{itemize}

\subparagraph{$\LO$ plus $\NLO$ from external wavefunctions}

\begin{equation}
\begin{split}
	\mathcal{B}_{\alpha\beta\gamma}
	\supseteq
	\dfrac{\vect{k_2} \cdot \vect{k_3}}{4} & \bigg(
			 -\frac{\dot{\phi}^\ast_\lambda}{H_\ast}\frac{1}{k_t} \bigg(-k_t^2+K^2+\frac{k_1k_2k_3}{k_t} \bigg)
			\bigg[ \delta_{\alpha\lambda} \delta_{\beta\gamma}
			+ 
			\big\{A_{\alpha\lambda}^\ast (k_1)-\boxed{u^\ast_{\alpha\lambda} \ln(-k_\ast \tau)} \big\} \delta_{\beta\gamma}\\
			&\hspace{4.2cm} +
			\big\{A_{\beta\gamma}^\ast (k_2)+A_{\beta\gamma}^\ast (k_3)-\boxed{2u^\ast_{\beta\gamma} \ln(-k_\ast \tau)} \big\}
			\delta_{\alpha\lambda}\bigg]
			\bigg)
	\label{eq:LO-C-external}
\end{split}
\end{equation}

\subparagraph{$\NLO$ from vertex}
\begin{equation}
	\begin{split}
	\mathcal{B}_{\alpha\beta\gamma}
	\supseteq
	\frac{ (\vect{k}_2 \cdot \vect{k}_3)}{4 k_t}
	\delta_{\beta\gamma}
	&  \bigg(
		2 k_t^2 \varepsilon_\ast \frac{\dot{\phi}^\ast_\alpha}{H_\ast}+k_t^2 t^\ast_\alpha \big[\EulerGamma+\ln \frac{k_t}{k_\ast}-1 \big]
		-2 \varepsilon_\ast \frac{\dot{\phi}^\ast_\alpha}{H_\ast} \Big[K^2+\frac{k_1k_2k_3}{k_t}  \Big]	
			+t^\ast_\alpha \frac{k_1k_2k_3}{k_t} \\
		& -t^\ast_\alpha \Big[\EulerGamma+\ln \frac{k_t}{k_\ast}\Big]
			\Big[K^2+\frac{k_1k_2k_3}{k_t}  \Big]			
	\bigg)
	\end{split}
\end{equation}

\subparagraph{$\NLO$ from internal wavefunctions}
\begin{equation}
	\begin{split}
	\mathcal{B}_{\alpha\beta\gamma}
	\supseteq
	\frac{ (\vect{k}_2 \cdot \vect{k}_3)}{4} \frac{\dot{\phi}^\ast_\lambda}{H_\ast}
	&  \bigg(			
		-\delta_{\beta\gamma} M^\ast_{\alpha\lambda}
		\bigg[
			k_t \big(1-\ln \frac{k_t}{2k_1}\big) -2k_1\big(\EulerGamma-1+\ln \frac{2k_1}{k_\ast}\big)\\
		& \hspace{2cm}	+ \frac{1}{k_t} (k_1k_2+k_1k_3-k_2k_3) J_0(k_1) +\frac{k_1k_2k_3}{k_t^2} J_1(k_1)
		\bigg]\\
		&-\delta_{\alpha\lambda} M^\ast_{\beta\gamma}
		\bigg[
			k_t \big(2-\ln \frac{k_t^2}{4k_2 k_3}\big) -2k_2\big(\EulerGamma-1+\ln \frac{2k_2}{k_\ast}\big)\\
			&\hspace*{2cm}-2k_3\big(\EulerGamma-1+\ln \frac{2k_3}{k_\ast}\big) \\
		&\hspace*{2cm}	+ \frac{1}{k_t} \big\{ k_1k_2+k_2k_3-k_1k_3	\big\}J_0(k_2) \\
		&\hspace*{2cm}+\frac{1}{k_t} \big\{ k_2k_3+k_1k_3-k_1k_2 
			\big\}J_0(k_3) \\
			& \hspace*{2cm}	+
			\frac{k_1k_2k_3}{k_t^2} \big\{J_1(k_2)+J_1(k_3)\big\}
		\bigg] \\
		&+ D^\ast_{\alpha\beta\gamma\lambda} 
		+\frac{1}{k_t} \bigg(E^\ast_{\alpha\beta\gamma\lambda} -\frac{F^\ast_{\alpha\beta\gamma\lambda}}{k_t} \bigg)
		- \big( D^\ast_{\alpha\beta\gamma\lambda}+G^\ast_{\alpha\beta\gamma\lambda}\big) 
							\Big[\EulerGamma +\ln\frac{k_t}{k_\ast} \Big]\\
		& +L^\ast_{\alpha\beta\gamma\lambda} \bigg[
									k_t \Big(1-\EulerGamma-\ln \frac{k_t}{k_\ast}\Big) -\frac{k_1k_2k_3}{k_t^2} +
									\frac{1}{k_t} \Big(\EulerGamma+\ln \frac{k_t}{k_\ast}\Big) \Big(K^2+\frac{k_1k_2k_3}{k_t} \Big)
											\bigg]
		\bigg)
	\end{split}
\end{equation}
where we have defined
\begin{equation}
	\left\{
	\begin{split}
	D^\ast_{\alpha\beta\gamma\lambda} &=	
						3k_t \varepsilon_\ast \delta_{\alpha\lambda} \delta_{\beta\gamma} 
						+2k_t\delta_{\beta\gamma} u^\ast_{\alpha\lambda}
						+4 k_t\delta_{\alpha\lambda} u^\ast_{\beta\gamma}
						\\
	E^\ast_{\alpha\beta\gamma\lambda} &= \delta_{\beta\gamma} \big\{
															k_1(k_2+k_3) \varepsilon_\ast \delta_{\alpha\lambda} -k_2k_3 (\varepsilon_\ast \delta_{\alpha\lambda} +2u^\ast_{\alpha\lambda})
														\big\}	\\
														&\hspace*{0.45cm} + \delta_{\alpha\lambda} \big\{
																(K^2+k_2k_3) \varepsilon_\ast \delta_{\beta\gamma} -k_1(k_2+k_3)
																						(\varepsilon_\ast \delta_{\beta\gamma} +2u^\ast_{\beta\gamma})
														\big\}\\
						F^\ast_{\alpha\beta\gamma\lambda} &=-3 k_1k_2k_3 \varepsilon_\ast \delta_{\alpha\lambda} \delta_{\beta\gamma}\\
	G^\ast_{\alpha\beta\gamma\lambda} &=	k_1 \varepsilon_\ast \delta_{\alpha\lambda} \delta_\beta\gamma
			-2(k_2+k_3) \delta_{\beta\gamma}  u^\ast_{\alpha\lambda}
			-(k_1+k_t) \delta_{\alpha\lambda} (\varepsilon_\ast \delta_{\beta\gamma} +2u^\ast_{\beta\gamma})\\
	L^\ast_{\alpha\beta\gamma\lambda} &=3  \varepsilon_\ast \delta_{\alpha\lambda} \delta_{\beta\gamma}	
\end{split}
\right.
\end{equation}

\paragraph{Relation to factorization theorem.}
\label{page:factorization-theorem}
It may appear that only Eqs.~\eqref{eq:NLO-A}
and~\eqref{eq:LO-A-internal} satisfy
the factorization theorem proved in~\S\ref{sec:factorization-thm}.
In Eqs.~\eqref{eq:NLO-B}--\eqref{eq:NLO-C}
and~\eqref{eq:LO-B-internal}
there are divergences which do not appear to multiply local
combinations of the external momenta.
[As explained in~\S\ref{sec:factorization-thm}, there is no
requirement for time-dependence associated with \emph{external}
wavefunctions, as in Eqs.~\eqref{eq:LO-A-external},
\eqref{eq:LO-B-external} and~\eqref{eq:LO-C-external},
to appear in local combinations.]
This is caused by the appearance in~\eqref{eq:third-action}
of a term linear in $\partial^2 \vartheta_1 / a^2$.
As noted in footnote~\ref{footnote:vartheta}
on p.~\pageref{footnote:vartheta},
this term appears because the constraints have been used to
simplify the third order action. Therefore the apparent nonlocality
in this term is cancelled by a delicate combination of other
operators. Eqs.~\eqref{eq:NLO-B} and~\eqref{eq:LO-B-internal}
are the lowest-order and next-order parts of this operator
and combine to give a local result when symmetrized.
The same is true for~\eqref{eq:NLO-C},
in which $\partial^2 \vartheta_1 / a^2$
appears as a proxy for the operator
$\alpha_1^2 \dot{\phi}_\alpha \delta \dot{\phi}_\alpha$
which can only produce local divergences.
We could have avoided this complex series of cancellations by
working with an unsimplified action, at the expense of greater
algebraic complexity overall.

Note that the vertex integral arising from
the $\delta \phi_\alpha (\partial_i \delta\phi_\beta)^2$ operator
yields no time-dependent logarithms.
This is because the interaction is suppressed
by spatial gradients, in agreement with the
analysis of~\S\ref{sec:factorization-thm}. 

\renewcommand{\baselinestretch}{1}
\bibliographystyle{JHEPmodplain}
\bibliography{references}

\end{document}